\documentclass[12pt,preprint]{aastex}

\shorttitle{X-ray stars in the Orion Nebula}
\shortauthors{Feigelson et al.}

\slugcomment{To appear in the Astrophysical Journal}

\begin{document}

\title{X-ray emitting young stars in the Orion Nebula}

\author{
Eric D. Feigelson\altaffilmark{1},
Patrick Broos\altaffilmark{1},
James A. Gaffney III\altaffilmark{1},
Gordon Garmire\altaffilmark{1},
Lynne A. Hillenbrand\altaffilmark{2},
Steven H. Pravdo\altaffilmark{3},
Leisa Townsley\altaffilmark{1},
Yohko Tsuboi\altaffilmark{1}}

\altaffiltext{1}{Department of Astronomy \& Astrophysics,
525 Davey Laboratory, Pennsylvania State University, University Park
PA 16802}
\altaffiltext{2}{Department of Astronomy, MS 105-24, California
Institute of Technology, Pasadena CA 91125}
\altaffiltext{3}{Jet Propulsion Laboratory, MS 306-438, 4800 Oak
Grove Drive, Pasadena CA 91109}

\begin{abstract}

The Orion Nebula Cluster and the molecular cloud in its vicinity have
been observed with the ACIS-I detector on board the {\it Chandra X-ray
Observatory} with 23 hours exposure in two observations.  We detect
1075 X-ray sources, most with sub-arcsecond positional accuracy.
Ninety-one percent of the sources are spatially associated with known
stellar members of the cluster, and an additional 7\% are newly
identified deeply embedded cloud members.  This provides the largest
X-ray study of a pre-main sequence stellar population and covers the
initial mass function from brown dwarfs up to a 45 M$_\odot$ O star.
Source luminosities span 5 orders of magnitude from $\log L_x \simeq
28.0$ to 33.3 erg s$^{-1}$ in the $0.5-8$ keV band, plasma energies
range from 0.2 to $>$10 keV, and absorption ranges from $\log N_H <
20.0$ to $\sim 23.5$ cm$^{-2}$.  Comprehensive tables providing X-ray
and stellar characteristics are provided electronically.

We examine here the X-ray properties of Orion young stars as a function
of mass; other studies of astrophysical interest will appear in
companion papers.  Results include: (a) the discovery of rapid
variability in the O9.5 31 M$_\odot$ star $\theta^2$A Ori, and several
early B stars, inconsistent with the standard model of X-ray production
in small shocks distributed throughout the radiatively accelerated
wind; (b) support for the hypothesis that intermediate-mass mid-B
through A type stars do not themselves produce significant X-ray
emission; (c) confirmation that low-mass G- through M-type T Tauri
stars exhibit powerful flaring but typically at luminosities
considerably below the `saturation' level; (d) confirmation that the
presence or absence of a circumstellar disk has no discernable effect
on X-ray emission; (e) evidence that T Tauri plasma temperatures are
often very high with $T \geq 100$ MK, even when luminosities are modest
and flaring is not evident; and (f) detection of the largest sample of
pre-main sequence very low mass objects showing flaring levels similar
to those seen in more massive T Tauri stars and a decline in magnetic
activity as they evolve into L- and T-type brown dwarfs.

\end{abstract}


\section{Introduction \label{intro_sec}}

Two decades ago, the first imaging X-ray telescope pointed at nearby
star forming  regions and discovered X-ray emission from low-mass
pre-main sequence (PMS) stars at levels far above those seen from
typical main sequence stars \citep{Feigelson81a, Feigelson81b,
Walter81, Montmerle83}.   Initially seen in T Tauri stars with roughly
solar masses and characteristic ages of $\sim 10^6$ yr, the X-ray
excess was later found to extend from protostars with ages $\sim
10^4-10^5$ yr \citep{Koyama96} to post-T Tauri stars with ages $\sim
10^7$ yr \citep{Walter88}, and from substellar masses
\citep{Neuhauser99} to intermediate-mass Herbig Ae/Be stars
\citep{Zinnecker94} as well as high-mass OB stars \citep{Harnden79} .
The typical PMS star was found to have X-ray luminosities $10^1-10^4$
above those typically seen in older main sequence stars.  The hot
plasma temperatures in the range $0.5-5$ keV and frequent high
amplitude variability on timescales of hours suggested that the strong
X-ray emission was due to elevated levels of magnetic reconnection
flaring rather than a quiescent coronal process.

The Orion Nebula (= Messier 42), a blister H{\sc II} region on the near
edge of the Orion A giant molecular cloud, is illuminated by the
richest PMS stellar cluster in the nearby Galaxy.  Known variously as
the Orion Id OB association, the Trapezium and  the Orion Nebula
Cluster (ONC), it has $>2000$ stellar members with masses ranging from
$<$0.05 M$_\odot$ to nearly $50$ M$_\odot$ OB stars in the Trapezium
\citep{Hillenbrand97, Hillenbrand00}.  The ONC is a unique laboratory
where the entire intial mass function of a young stellar cluster can be
examined in a uniform fashion by an imaging X-ray telescope.  While
star formation appears to have ceased in the ONC itself $\sim 1$ Myr
ago, the molecular cloud behind the ONC is actively forming stars with
dozens of likely protostars both around the Becklin-Neugebauer object
\citep{Gezari98} and elsewhere in the cloud \citep{Lada00}.

The ONC was the first star forming region to be discovered in the X-ray
band with non-imaging instruments \citep{Giacconi72, denBoggende78,
Bradt79} which found that the emission was extended.  But these early
observations could not discriminate whether the X-rays were produced by the
Trapezium OB stars, the lower- mass T Tauri stars or a diffuse plasma.
Observations with the {\it Einstein}, {\it ROSAT} and {\it ASCA}
satellites resolved dozens of individual stars \citep[e.g.][]{Ku79,
Pravdo81, Gagne95, Geier95, Yamauchi96}, but could identify X-rays from
only a modest fraction of the cluster population due to sensitivity,
resolution and bandwidth limitations. Due to crowding and absorption by
the molecular material, high-resolution X-ray imaging at energies $> 2$
keV is critical for the study of the ONC and other rich  star forming
regions.  The {\it Chandra X-ray Observatory} provides these
capabilities and gives a greatly improved view of the ONC field 
\citep{Garmire00}.

While the X-rays from low-mass T Tauri stars are recognized to arise
from magnetic reconnection flares, a variety of astrophysical questions
remain concerning the origins of X-ray emission as a function of
stellar mass.  Among young massive OB stars, {\it Einstein} and {\it
ROSAT} investigations found that X-ray luminosity scales with the
bolometric luminosity as $L_x \propto 10^{-7} L_{bol}$ for stars
earlier than B1.5 \citep{Harnden79, Pallavicini81, Berghofer97}.  The
X-ray emission mechanism here is thought to be quite different from
that in lower mass PMS stars, produced in shocks arising from
instabilities in their radiatively driven stellar winds
\citep[e.g.][]{Lucy80, Owocki99}.  As these high-mass stars generally
exhibited little X-ray variability, the emission was generally
interpreted to arise in a myriad of little shocks, although the
possibility of high-amplitude variations from large shocks is discussed
by \citet{Feldmeier97}.  Recent {\it Chandra} and {\it XMM-Newton}
grating spectroscopic studies confirm the basic scenario of X-ray
production in the extended wind with broadened lines produced at
different depths within the wind \citep{Waldron01, Schulz00, Kahn01},
although evidence is growing that magnetically confined hot plasma
either near the surface or in the wind plays a significant role
\citep{Babel97, Waldron01}.

The nature of the transition between OB wind-generated X-rays and T
Tauri flare-generated X-rays has not been well-established.  Late B and
A type stars have neither strong winds nor outer convective zones where
magnetic fields may be amplified by dynamo processes, and are thus
predicted to be X-ray quiet.  Nonetheless, a substantial number of
these stars have been detected with a wide range of X-ray luminosities
\citep{Schmitt85, Caillault89, Berghofer94a, Simon95, Cohen97}.  Much,
but perhaps not all, of this emission can be attributed to the presence
of low-mass T Tauri star companions to the intermediate mass stars.
The X-ray emission is often stronger from the youngest
intermediate-mass stars, Herbig Ae/Be stars,  which exhibit accretion
and outflows from their circumstellar disks \citep{Zinnecker94}.  In at
least one case, high-amplitude flaring is clearly present
\citep{Hamaguchi00}.

The magnetic flaring model of X-ray emission from lower mass T Tauri
stars has been generally interpreted as enhanced solar-type activity
where fields generated by a magnetic dynamo in the stellar interior
erupt and undergo violent reconnection at the stellar surface
\citep[e.g. reviews by][]{Walter91, Feigelson91, Feigelson99}. This
solar activity model is supported by extensive multiwavelength evidence
of enhanced chromospheric lines, rotationally modulated cool starspots,
photospheric Zeeman measurements, and circularly polarized radio
continuum flares in T Tauri stars.  However, in support of arguments
for a magnetic coupling between PMS stars and their circumstellar
disks, the X-ray flares (particularly in protostars) have been
alternatively attributed to reconnection in long star-disk magnetic
field lines \citep[e.g.][]{Shu97, Montmerle00}.  The astrophysical
processes giving rise to PMS magnetic fields are also uncertain.
Possibilities include a standard $\alpha-\omega$ dynamo as in main
sequence stars, a turbulent dynamo particular to fully convective
stars, fossil fields inherited from the star forming process, or dynamo
processes in the circumstellar disk.  Addressing these open issues
regarding the origins of the enhanced X-ray emission from PMS stars is
a principal goal of the present effort.  \citet{Feigelson99} give a
comprehensive review of both the observational evidence and theoretical
interpretations of magnetic flaring in lower mass PMS stars.

A handful of the lowest mass objects in young stellar clusters, PMS
brown dwarfs, have been detected in the X-ray band at the limit of
sensitivity of the {\it ROSAT} and {\it Chandra} satellites
\citep{Neuhauser99, Garmire00, Imanishi01, Preibisch01}.  One nearby
older brown dwarf has also been seen with {\it Chandra} during a flare
\citep{Rutledge00}, and another object exhibited a powerful flare in
the optical band \citep{Liebert99}.  Their X-ray behavior appears
qualitatively similar to that seen in PMS stars, which is consistent
with the expectation that the internal conditions of PMS brown dwarfs
do not differ much from the conditions within PMS M-type stars.

The present paper is the second in a series based on {\it Chandra}
observations with the ACIS detector obtained during the first year of
the {\it Chandra} mission. \citet[][Paper I]{Garmire00} gave an
overview of the initial exposure and discussed the nature of sources in
the Becklin- Neugebauer/Kleinman-Low region.  The present study
analyzes the full dataset (\S \ref{obs_sec}), presenting comprehensive
tables and notes of individual source properties and counterparts (\S
\ref{tables_sec}), providing a global X-ray view of the young stellar
population (\S \ref{demog_sec}), and examining the X-ray properties of
young stars across the initial mass function (\S \ref{IMF_sec}).  Paper
III \citep{Feigelson02a} focuses on the magnetic activity of $\simeq 1$
M$_\sun$ analogs of the PMS Sun and discusses the implications for
energetic particle radiation onto solar nebula solids as evidenced in
ancient meteorites.   Paper IV \citep{Feigelson02b} discusses the
absence of the relationship between X-ray emission and stellar rotation
expected from a solar-type magnetic activity and dynamo model.  Paper V
(Gaffney et al., in preparation) presents detailed spectral and
variability studies of the brighest ONC sources.  The reader is also
referred to {\it Chandra} ONC studies using the ACIS spectroscopic
array \citep{Schulz00, Schulz01} rather than the ACIS imaging array
used here.

\section{Observations and data analysis \label{obs_sec}}

\subsection{Instrumental setup \label{instrum_sec}}

The ONC was observed with the Advanced CCD Imaging Spectrometer (ACIS)
detector on board the {\it Chandra X-ray Observatory}
\citep{Weisskopf01} on 12 October 1999 and 1 April 2000 (Table
\ref{obs_table}).  The two images are shown at low resolution in Figure
\ref{fullfields_fig}a and b. The principal results arise from the
imaging array (ACIS-I) which consists of four abutted $1024 \times
1024$-pixel front-side illuminated charge-coupled devices (CCDs)
specially designed for X-ray studies\footnote{Detailed descriptions of
the ACIS instrument and its operation can be found on-line at
\url{http://www.astro.psu.edu/xray/docs/sop} and
\url{http://www.astro.psu.edu/xray/docs/cal\_report}.}.  Each CCD chip
subtends 8.3\arcmin\/ and, after chip gaps and satellite dithering are
taken into account, the ACIS-I image covers about $17\arcmin \times
17\arcmin\/$ on the sky.  The aimpoint of the array for both
observations is $5^h 35^m 15.0^s$, $-5^\circ 23\arcmin 20\arcsec\/$
(J2000), 22\arcsec\/ west of  the brightest member of the Trapezium,
$\theta^1$C Ori.

The instrument configuration during the two observations differed in a
number of respects.  First, the focal plane temperature was -110
$^\circ$C during the October 1999 observation, but was reduced to -120
$^\circ$C before the April 2000 observation.  Consequently, the earlier
observation suffers a higher charge transfer inefficiency (CTI) during
readout of the CCD chips every 3.2 s.  The CTI of the front-side
illuminated ACIS chips deteriorated early in the {\it Chandra} mission
due to charged particle bombardment during passage through the Earth's
magnetospheric radiation belts \citep{Prigozhin00}.  Second, the
satellite roll angle rotated by 183$^\circ$ between the two
observations, resulting in an offset of the arrays and different
orientations of the readout trailed events.

In addition to the ACIS-I data, the S2 and S3 chips in the
spectroscopic array (ACIS-S) were also operational.  These data are
less useful:  the telescope point spread function (PSF) is considerably
degraded far off-axis; the chips cover different regions of the Orion
cloud due to the roll angle change; the chips differ in background
levels and CTI characteristics due to their construction (S3 is
back-side illuminated while S2 is front-side illuminated); and the S3
chip suffered a hardware failure during the October 1999 observation
(FEP0 electronics board problem).  The ACIS-S results will not be
reported here.

\subsection{Exposure times \label{expos_sec}}

The satellite was pointed at the ONC for about 26 hours during the two
observations (Table \ref{obs_table}).  Time series were generated to
locate periods of missing or bad data.  The October 1999 dataset
suffered many brief periods of missing data due to the saturation of
the telemetry by false events generated by the FEP0 electronics
failure.   We use the exposure averaged over the entire array in the
analysis here and ignore the $\sim 1$\% scatter in chip-dependent
exposures. The April 2000 observation had only 6.4 s of telemetry
dropouts and was thus a virtually uninterrupted datastream.

Two other exposure time corrections are applied.  First, exposures are
reduced by 1.3\% because 0.04 s of each 3.24 s frame was devoted to
chip readout, during which source photons appear as faint readout
trails parallel to the chip axis.  The readout trails for the brightest
Trapezium stars are clearly evident in the images.  Second, 1.5\%
(1.0\%) of the time was eliminated due to software-generated glitches
in the aspect solution in the October (April) datasets.  With these
corrections, the net exposure time averaged over the array is about 23
hours (Table \ref{obs_table}).

The effective exposure time for a given source at some distance from
the aimpoint is this array-averaged exposure time corrected for
telescope vignetting, for chip-dependent telemetry dropouts and, for
sources lying near chip edges, for satellite dithering motions that
move the source on and off the chip.  Thus, each source is accompanied
by an independently calculated effective exposure time which is
incorporated into the auxiliary response file (arf) generated for each
source during spectral analysis (\S \ref{spec_sec}).

\subsection{Data selection \label{select_sec}}

Our data analysis starts with Level 1 processed event lists provided by
the pipeline processing at the {\it Chandra} X-ray Center, which
includes all events telemetered by the ACIS detector.  The earlier
dataset was processed with ASCDS version R4CU4 in October 1999, and the
later dataset with ASCDS version R4CU5UPD2 in April 2000.  Our data
reduction methodology uses codes and functionalities provided by a
variety of software systems:  the CIAO package (version 1.0 and 2.0,
\url{http://asc.harvard.edu/ciao}) for {\it Chandra} data analysis
produced by the {\it Chandra} X-ray Center at the Smithsonian
Astrophysical Observatory; FTOOLS programs for FITS file manipulation
produced by the HEASARC at NASA Goddard Space Flight Center (version
5.0,
\url{heasarc.gsfc.nasa.govdocs/software/ftools/ftools\_menu.html}); the
TARA package for interactive ACIS data analysis developed by the ACIS
Team at Pennsylvania State University (version 5.8,
\url{http://www.astro.psu.edu/xray/docs}); and the ds9 data
visualization application developed at Smithsonian Astrophysical
Observatory (version 1.9, \url{http://hea-www.harvard.edu/RD/ds9}).
Technical notes on the software procedures used here can be found at
\url{http://www.astro.psu.edu/xray/axaf/recipes}.

We first removed an artificial random offset of $\pm 0.25$\arcsec\/
introduced to the location of each event during Level 1 processing, as
it produces an unnecessary blurring of the point spread function.
Second, the energy and grade of each event was corrected with a
procedure that models the CTI characteristics of each chip
\citep{Townsley00, Townsley02}.  This not only corrects the trend of
decreasing gain as one proceeds from the edge towards the center of
each chip, but also accounts for changes in splitting of charge between
adjacent pixels due to CTI.  An important effect of the latter
correction is the improvement in signal and reduction in noise at high
energies towards the center of the ACIS-I array, particularly for the
first observation obtained at -110 $^\circ$C.

Two cleaning operations were conducted to remove spurious events from
the image.  First, a temporal cleaning operation was conducted to
remove `cosmic ray afterglows' produced by high energy Galactic cosmic
rays.  Although the charge deposited immediately after a cosmic ray hit
is almost always rejected by on-board processing, in some cases the
central pixel will release residual charge over $10-30$ s.  We consider
the arrival of two or more events at the same chip pixel within 30 sec
to be the signature of afterglows.  When these spurious charges
resemble X-ray hits, they are included in the telemetry as real
events.  While these constitute only $\simeq 2$\% of the background
events in a typical ACIS-I observation, they can combine with ordinary
background events to produce spurious weak sources.  A sensitive source
detection algorithm can find up to dozens of spurious sources due to
cosmic ray afterglows if they are not removed.

However, two or more photons from celestial sources will sometimes
arrive at the same chip pixel within 30 sec by chance.  (Recall that
the distribution of times between adjacent events of a Poisson process
decreases exponentially with the lag time, so the probability of
closely spaced events is not small.)  Examination of the spatial
distribution of events flagged as cosmic ray afterglows indicates that
$2\%-10\%$ of counts are flagged from real sources in the Orion field
with intensities ranging from $0.001-0.2$ counts s$^{-1}$.  We found
that the incorrect removal of these true source photons can be avoided
by removing only flagged events lying more than 3\arcsec\/ from
identified sources.

Second, `hot columns' of spurious events are removed.  These arise from
several flickering pixels in both the imaging and framestore regions of
the CCD chips, and cosmic ray hits in the frame store area along
amplifier boundaries.  The location and grade classifications of these
false events are known and are easily removed.  Third, we select events
by their `grades' and their energy to remove most of the remaining
events arising from charged particles and detector noise.  We choose
events that exhibit `standard {\it ASCA} grades' (0, 2, 3, 4 and 6)
after CTI correction and events outside the energy range $0.5-8.0$ keV
are eliminated.  Fourth, the two exposures were merged into a single
image after positional alignment, as described in \S \ref{ids_sec}.

The images resulting from this data selection procedure are shown in
Figure \ref{fullfields_fig} for the two individual exposures, and in
Figures \ref{guide_fig}-\ref{image_fig} for the merged field.  The
background levels are very low (except near the very bright Trapezium
stars).  On average, there is only one background event in a $3\arcsec
\times 3\arcsec\/$ region in the merged image (Table \ref{obs_table}).

\subsection{Source detection \label{detect_sec}}

Sources were located in the image using a wavelet transform detection
algorithm implemented as the {\it wavdetect} program within the CIAO
data analysis system \citep{Dobrzycki99, Freeman02}.  We found that the
default threshold probability of $1 \times 10^{-6}$ omitted a
considerable number of weak sources having stellar counterparts, while
noise frequently triggered the algorithm for a threshold of $1 \times
10^{-4}$.  Because of the highly crowded field and the elevated
background in the central region of the ONC, the simulation that showed
that a threshold probability of $1 \times 10^{-6}$ gives one false
detection per $10^{6}$ pixels \citep{Freeman02} is inapplicable.  We
therefore opted to use a threshold of $1 \times 10^{-5}$ and to examine
each {\it wavdetect} source by hand, as described below.  

The {\it wavdetect} program is very successful in detecting sources
down to faint count limits across the entire ACIS array, despite the
changes in PSF and variations in background due to chip gaps or
overlapping arrays.  However, failures or errors of several types
occasionally occur.  First, when the threshold is set to obtain
maximum sensitivity near the field center, some false triggers of noise
occur far off axis.  In particular, we find that a Poissonian upward
fluctuation in background noise adjacent to a downward fluctuation
sometimes produces a false trigger.  Second, the program sometimes
consolidates closely spaced sources easily resolved by eye; this
failure begins for source separations $\leq 2.5$\arcsec\/ on-axis.
Third, the program naturally triggers spurious sources on read-out
trails of strong sources.  Fourth, as the source detection (in contrast
to source consolidation) is done without knowledge of the varying PSF,
the algorithm can locate sources smaller than the point spread function
far off-axis or larger than the PSF near the axis.  Fifth, the
sensitivity to sources rapidly deteriorates as one approaches within
$\sim 2$\arcmin of $\theta^1$ C Ori due to the elevation of background
from the O star's PSF wings.  Sixth, as with any source detection
algorithm, performance near the detection threshold is erratic and the
eye can locate untriggered faint concentrations similar to those that
were triggered.  Seventh, {\it wavdetect} computes the source counts in
a cell region which is not accurately scaled to the local PSF.
Off-axis photometry obtained by {\it wavdetect} is thus not always
reliable.

We address most of these problems by careful visual inspection of the
image with the locations of {\it wavdetect} sources marked.  Sources
which appear spurious (clearly noise or read-out trails) are omitted,
marginal sources are flagged in the table notes, and missing sources
(close doubles and marginal sources missed by the algorithm) are
added\footnote{The following sources were added by hand: \#148, 189,
202, 223, 241, 382, 384, 408, 420, 454, 504, 609, 614, 842 and 862.} .
Several dozen sources were adjusted in some way; these adjustments are
explained in the table notes.  A final examination was made of sources
with extracted counts (\S \ref{extract_sec}) very close to the
estimated background level.  None of these decisions was based on the
presence or absence of counterparts at other wavelengths.

The result of this entire source detection process is 1075 sources for
the merged Orion ACIS-I field, which are illustrated in Figure
\ref{image_fig} and tabulated in Table \ref{src_list_table}.

\subsection{Source positions and stellar counterparts \label{ids_sec}}

The {\it wavdetect} software provides an estimate of the source
position using a simple average of event positions in pixels in its
source region, where the conversion between pixel and celestial
locations is based on the satellite aspect solution.  These celestial
locations can be further corrected to match the stellar {\it Hipparcos}
frame of reference by associating ACIS sources with stars of known
position.  We proceeded as follows:

{\bf Boresight alignment} ~~ An absolute translational error is
frequently present in the {\it Chandra} aspect solution at a level
around $1\arcsec-2$\arcsec\/.  For the present dataset, this boresight
error was corrected using twenty-two sources with $>200$ counts in each
exposure, lying in the inner 3\arcmin\/ of the field, and having an
unambiguous optical or near-infrared counterpart. We found that the
October 1999 field required a translation of 0.6\arcsec\/ to the SE and
the April 2000 field required a translation of 1.9\arcsec\/ to the NE
to match the stellar positions in the 2MASS/ACT/Tycho reference frame
\citep{Hillenbrand00}.  When the {\it wavdetect} positions from the
merged image are considered, the residual offsets of the optical and
X-ray positions have a standard deviation of only 0.1\arcsec\/, so the
formal uncertainty of the frame alignment is only $0.1/\sqrt{22-1}=
0.02$\arcsec.  The alignment was later checked using $\simeq 600$
ACIS/optical positional comparisons lying within 4\arcmin\/ of the
field center, and was found to be excellent.

{\bf Stellar counterparts} ~~  After the exposures were aligned and
merged, source positions were obtained from {\it wavdetect} then
compared to a catalog of confirmed or likely ONC members.  This catalog
consists of several thousand stars based on a complete $V<20$ ONC
sample \citep[][with positions corrected in Hillenbrand \& Carpenter
2000]{Hillenbrand97, Hillenbrand98}, a deep $JHK$ survey of the inner
$5\arcmin \times 5\arcmin\/$ \citep{Hillenbrand00}, and a 2MASS
variability survey of the region \citep{Carpenter01}.  All ACIS/star
associations with offsets $\phi < 5$\arcsec\/ were initially
considered.  Outliers with large offsets were individually examined,
and were typically found to have multiple counterparts, weak X-ray
sources below the completeness limit, or positions far off-axis where
the PSF is broad.  About 50 cases of multiple counterparts were found,
typically two members of a visual binary lying within
$1\arcsec-2$\arcsec\/ of an ACIS source.  We generally associated the
source with the brighter member of the binary, but recognize that this
may be incorrect in some cases and may produce a bias in later study
(e.g. a $L_x - L_{bol}$ correlation plot).  These cases are noted in
table footnotes.  For sources without counterparts in the catalogs
listed above, we also searched the USNO A-2.0, 2MASS survey and SIMBAD
databases.  Two new photospheric counterparts (a 2MASS survey source
and the mid-infrared source IRc5) and 1 new radio counterpart were
found.

After culling unreliable sources and flagging multiple counterparts, we
find that 755 (70\% of 1075) ACIS sources have $V<20$ counterparts
\citep[][604 of these are in the early lists of Parenago 1954 and/or
Jones \& Walker 1988]{Prosser94, Hillenbrand97}, 218 (20\%) have JHK
but no optical counterparts \citep{Hillenbrand00, Carpenter01}, 1 has
only a mid-infrared counterpart, and 101 (9\%) have no photospheric
counterpart.  The sources with no counterparts will be discussed in
detail in Section~\ref{noid_sec}.

{\bf Astrometric accuracy} ~~ The {\it wavdetect} centroid can, in
principle, produce systematically incorrect positions for off-axis
sources due to anisotropies in the {\it Chandra} mirror PSF.  For
example, a point source $5\arcmin-10$\arcmin\/ off-axis will exhibit
both a 2\arcsec\/ asymmetric cusp and a $5\arcsec-10$\arcsec\/
elliptical halo whose orientation depends on location in the detector
\citep[see Figure 4.9,][]{Chandra00}. It is difficult to predict the
amplitude of this systematic error, as the asymmetries may be partially
erased by the merging of two exposures with opposite roll angles and by
wavelet processing.

Figure \ref{offsets_fig} shows the offset $\phi$ between ACIS and
stellar positions as a function of off-axis angle
$\theta$\footnote{Throughout this study, we calculate $\theta$ from
$\theta^1$C Ori rather than the aimpoint of the {\it Chandra} mirrors,
which differ by 0.3\arcmin.  This permits correction for the high
background caused by the wings of its PSF.}.  Potential matches falling
to the left of the dashed line were rejected as false.  In the inner
region where the PSF is flat and constant in shape, the median offsets
are quite small: 0.25\arcsec\/ for $\theta < 1$\arcmin\/ rising to
0.5\arcsec\/ around $\theta \simeq 4$\arcmin. However, the mean offset,
and scatter about that mean, increase considerably as $\theta$
increases to $8\arcmin-12$\arcmin\/ such that offsets of
$2\arcsec-5$\arcsec\/ are not uncommon towards the edge of the field.
Matches with $\phi > 3$\arcsec\/ were carefully considered; any
ambiguity in the source identification is noted with an `id' flag in
Table~\ref{src_list_table}.  While some of the largest offsets can be
attributed to very weak off-axis sources where the centroid is strongly
affected by background contamination, half of the 15 sources with $\phi
> 3$\arcsec\/ contain $>1000$ counts.

Systematic trends in the offsets are seen.  Specifically, the
Star$-$ACIS offset in right ascension runs from $\simeq -1$\arcsec\/
near the NE corner of the field to $\simeq +1$\arcsec\/ near the SW
corner.  Similarly, the Star$-$ACIS offset in declination runs from
$\simeq -1$\arcsec\/ near the NW corner to $\simeq +1$\arcsec\/ near
the SE corner.  These systematic offsets have two possible causes: a
0.083\% error in the ACIS pixel size (not recognized until late-2001),
and an interaction between the wavelet transform and asymmetries in the
off-axis PSF.  We have not attempted to correct these positions here.

\subsection{Photon extraction \label{extract_sec}}

We extract counts for source analysis from circular regions centered on
{\it wavdetect} source positions.  The extraction of source photons in
an optimal and reproducible way is not simple due to the non-Gaussian
shape and strong variation in the PSF across the field.  The behavior
of the {\it Chandra} PSF as one proceeds off-axis is complex:  the
shape is nearly circular and centrally condensed in the inner $\theta <
5$\arcmin, but broadens rapidly with increasing asymmetries for
$5\arcmin < \theta < 12$\arcmin.  The PSF core and wing components do
not evolve homologously so that the curves of full-width half maximum
and various encircled energy fractions (e.g., 50\%,  95\%) as a
function of off-axis angle are not parallel.  Extraction from a large
region (e.g.  the radius enclosing 99\% of the PSF) guarantees capture
of more source photons but also includes more background events which,
for weak off-axis sources, can dominate the signal.  Extraction from a
small region reduces background effects but loses events that can
improve statistics in later spectral and variability analysis.  In any
case, the estimate of the source flux must account both for the loss of
events from the PSF wings and the addition of background events.

For most sources, we chose to extract events from the 95\% encircled
energy radius as a function of off-axis angle, based on the PSF of a
1.49 keV monochromatic source\footnote{These 95\% encircled energy
radii were calculated using the CIAO program mkpsf and are consistent
with those obtained with the detailed raytraces using the SAOSAC model
for the {\it Chandra} mirror assembly (P. Zhao, private
communication).  The radius in arcsec is reasonably well-approximated
by the quadratic function $R(95\%EE) = 2.05 -0.55 \theta + 0.18
\theta^2$, where $\theta$ is the off-axis angle in arcmin.  The 99\%
encircled energy radii used for bright sources is approximately $R(99\%
EE) = 8 + 0.2 \theta$ and the 50\% radii used for nondetections in \S
\ref{limits_sec} is approximately $R(50\%EE) = 0.43 - 0.10 \theta +
0.050 \theta^2$.  \label{radius_xtr_footnote}}.  For very bright
sources with $>1000$ counts, the extraction radius was increased to
around the 99\% curve, as the benefit of increased source photons
exceeds the increase in background.  The radius was reduced below the
95\% curve for nearly 200 sources principally due to source crowding.
For each source, we extract counts $C_{xtr}$ in the total $0.5-8$ keV
band from within radius $R_{xtr}$.  These are the events used in all
later spectral and variability analysis.  For all cases, we calculate
the corresponding PSF fraction $f_{PSF}$ using the CIAO program mkpsf.
The resulting distribution of extracted counts as a function of
off-axis angle is shown in Figure \ref{cts_theta_fig}.

The number of background counts in the extraction circle is estimated
to be 
\begin{equation}
B_{xtr} = B(\theta) \times \pi R_{xtr}^2
\end{equation}
where $B$ is in counts (arcsec)$^{-2}$, $R_{xtr}$ is in arcsec, and
off-axis angle $\theta$ is in arcmin.  This background level is
approximately constant over most of the ACIS field with the values
given in Table \ref{obs_table}.  In the inner $\theta < 3$\arcmin\/ the
background is substantially elevated by the PSF wings of $\theta^1$C
Ori.  We estimate that $\sim 450,000$ photons were incident onto the
detector from $\theta^1$C Ori; the next brightest sources are $\sim 20$
times fainter and their PSF wings are much less important (see \S
\ref{bright_sec}).  Background levels were measured manually at several
dozen source-free locations in the inner region and an empirical fit to
the wings of the $\theta^1$C Ori point spread function gives
\begin{eqnarray}
\log B(0.1\arcmin < \theta < 1.0\arcmin) &=& 0.6 \theta^{-0.6} - 2.0, \nonumber \\
\log B(1.0\arcmin < \theta < 3.0\arcmin) &=& -0.25 \theta - 1.15.
\end{eqnarray}
Note that this background fit is not very accurate in the inner $\theta
< 0.5$\arcmin\/ as the steep slope to the $\theta^1$C Ori PSF, wings
and readout trails from other strong sources, and the slightly
displaced chip gaps from the two observations together produce spatial
variations in the background levels.  Despite these complications, the
background is relatively unimportant for the great majority of ONC
sources.

The count rate $CR$ for each source can be calculated from  
\begin{equation}
{\rm CR (ct~ks^{-1})} = (C_{xtr} - B_{xtr}) / (f_{PSF} E_{eff})
\end{equation}
where the effective exposure time $E_{eff}$ is given in Table 
\ref{obs_table}.  Except for sources near the detector edges, $E_{eff} = 
82.8$ ks. Confidence intervals of these count rates are dominated by 
statistical uncertainties of the extracted counts when $C_{xtr} <
500$ counts.  Systematic uncertainties in the other quantities are 
estimated to be 5\% or less and dominate for the strongest sources.

\subsection{Variability analysis \label{variability_sec}}

Lightcurves were constructed for all sources.  For the stronger sources,  
binsizes were chosen to give roughly 20 bins in the lightcurve.  An 
extraordinary variety of behaviors were found including: constant 
sources; constant within each observation but different between 
observations ($\sim 6$ months separation); slowly variable within one or 
both observation, consistent perhaps with rotational modulation of 
longitudinal structures on the stellar surface; and rapidly variable
phenomena reminiscent of magnetic reconnection flares on the Sun and
other late-type stars.  The reader can view examples of such variations
in our companion study of PMS solar analogs \citep{Feigelson02a} and
in the ACIS-S study by \citet{Schulz01}.

No simple quantity for tabulation adequately describes the variety of 
phenomenology seen.  The non-parametric Kolmogorov-Smirnov test, 
for example, does not provide a consistent distinction between variable 
and constant sources because of the $10^4$ range in count rates and 
$10^1$ range in accessible timescales.  Parametric modeling that 
accounts for each source's counting statistics, such as Bayesian Block 
models \citep{Scargle98}, would be useful but are beyond the scope of 
this study.  

We thus provide only the average count rates in each of the two 
observations and a simple subjective classification of the variations.
Lightcurves illustrating the four classes are shown in Figure 
\ref{var_class_fig}:
\begin{description}

\item [Constant]  The source is approximately constant in all 
available observations, though for weak sources this is not a strong 
constraint.

\item [Long-term variation] No variation is seen within an observation, 
but the average count rates in the October 1999 and April 2000 
observations differ at the $>3 \sigma$ level of significance; that is,  
\begin{equation}
|(C_{xtr}(Oct) - 1.2 C_{xtr}(Apr)| > 3  \sqrt{C_{xtr}(Oct) + 1.2^2 
C_{xtr}(Apr)},
\end{equation}
where the constant 1.2 is the ratio of the exposure times 
(Table \ref{obs_table}).  Note that most sources designated `Flare' or 
`Possible flare' also show long-term variation.  

\item [Possible flare]  A variation is seen within an observation, 
but with lower signal-to-noise ratio so that quantitative descriptions are 
not readily obtained.  Notes are not provided in most cases.  

\item [Flare]  A highly significant variation on timescales of hours is
present within one or both observations.  In these cases, the
lightcurve is briefly described in a note to Table
\ref{src_prop_table}.  When the entire event lies within the
observation, the note gives peak count rate $CR_p$, quiescent $CR_q$,
rise and decay timescales.  However, often the events extend beyond the
$\sim 12$ hour exposures and only partial information is available.  A
wide variety of flare morphologies are seen $-$ there is no `typical'
flare.

\end{description}

\subsection{Spectral analysis and absorption estimates
\label{spec_sec}}

The spectrum of each source was evaluated by fitting simple optically
thin thermal plasma models to the pulse height distribution of the
extracted photons.  Background contribution and variability were
ignored in the fitting procedure.  These omissions, combined with the
diversity of observed spectra  and both systematic and statistical
instrumental  uncertainties, mean that this spectral analysis quite
likely does not reflect the complexity of the astrophysical
phenomena.  We therefore limit our objectives here to a basic
measurement of the time-averaged temperature(s) of the emitting plasma,
an estimate of the intervening interstellar column density from the
soft X-ray absorption, and evaluation of time-averaged broad-band
fluxes and luminosities.  For the faintest sources, our objectives are
further limited to a single estimate of luminosity.

Spectral fitting was performed using the XSPEC code \citep{Arnaud96},
version 10, assuming a uniform plasma with 0.3 times solar elemental
abundances.  Continuum and emission line strengths were evaluated using
the MEKAL code \citep{Mewe91}; soft X-ray absorption was modeled using
atomic cross-sections of \citep{Morrison83}.  The choice of sub-solar
abundances was based on fits of {\it ASCA} CCD-resolution spectra of
PMS stars \citep[][see however Kastner et al.\ 1999 for a case with
solar abundances]{Tsuboi98, Hamaguchi00, Tsuboi00}.  Best-fit models
were found by $\chi^2$ minimization by comparing models with extracted
events in the range $0.5-8$ keV. The events are grouped into bins of 5
photons (except for the weakest sources). Free parameters of the model
are the plasma energy $kT$, equivalent hydrogen column density of
absorbing interstellar material $\log N_H$,  and a normalization factor
adjusting the model to the total count rate.  Astrophysical models are
convolved with an auxiliary response file (arf) describing the
telescope and ACIS detector effective area as a function of energy and
location in the detector, and a response matrix file (rmf) describing
the spectral resolution of the detector as a function of energy.  The
arf file includes source-specific reductions in exposure times due to
off-axis telescope vignetting and (for a few sources) the effects of
chip gaps or bad CCD columns convolved with satellite aspect
dithering.

A problem arises here: only standard arf and rmf files from CIAO
version 2 were available at the time of this analysis, which do not
take into account the improvements in gain correction, hard-band
sensitivity, and spectral resolution provided by the CTI correction
applied to the individual events (\S \ref{select_sec}).  This
discrepancy is evident in XSPEC plots comparing source and model
spectra; for example,  in strong sources, the data have sharper line
features than the models and the data$-$model residuals show
corresponding correlations.  A potentially important source of
systematic bias in the spectral fits is our use of standard CIAO curves
of quantum efficiency $vs.$ photon energy (incorporated into the arf
files) which do not take into account the improved recovery of hard
energy photons from our CTI correction procedure. This can result in
overestimation of plasma energies at high temperatures.  The importance
of this bias is difficult to evaluate, as it is significant only for
the October 1999 observation obtained with detector temperature
-110$^\circ$C and for sources near the detector center.  Altogether,
the spectral results reported here thus cannot be considered definitive
and further analysis with improved calibration methods is warranted.
The broad-band luminosities (\S \ref{lum_sec}) are not significantly
affected by these problems.

After construction of weighted average arf and rmf files from -110
$^\circ$C and -120 $^\circ$C calibration, each source spectrum was
fitted with an isothermal plasma with interstellar absorption.
Satisfactory fits to the broad-band shape of the spectrum were obtained
for about 80\% of the sources using this one-temperature plasma fit.
A two-temperature plasma model was introduced for spectra with poor
fits, which gave satisfactory results for about 10\% of the sources.
However, although the model fits were adequate, the two-temperature
results were sometimes astrophysically unrealistic: a very strong soft
component ($E \sim 0.2$ keV) would sometimes be introduced  with a high
absorption incompatible with the known visual absorption of the star.
We thus do not report the $N_H$ values from two-temperature fits, and
warn that the two-temperature $kT$ values may not accurately reflect
the astrophysical plasma.

The model fits for some sources were still poor for two reasons.
First, additional broad-band components, usually a soft ($<1$ keV) excess
or a hard ($>4$ keV) excess, were sometimes present which were not
included in the model.  This usually occurred because the value of
$\chi^2$ for the overall fit was satisfactory despite an apparent
misfit of the spectral shape.  In such cases, mentioned individually in
the notes to Table \ref{src_prop_table}, the derived broad-band
luminosities will be systematically underestimated.  Second, the
broad-band spectral shape may be well-modeled but narrow spectral
features in emission or absorption appear in the data that are not
present in the model.  The lines can be attributed to a variety of
ionized elements (e.g., neon, silicon, sulfur, argon, iron) and may
arise from plasmas with unusual elemental abundances.  For example,
recent spectral studies of flares in magnetically active stars in the
solar neighborhood using the {\it XMM-Newton} satellite show dramatic
variations in neon and iron abundances on timescales of hours during
magnetic flares \citep{Brinkman01, Gudel01, Drake01}.  From visual
inspection of the spectra and source$-$model residuals, we have flagged
about 15\% of the sources as likely cases of plasma with narrow
spectral features. Note that, in some of these cases, the plasma
abundances may prove to be normal when improved arf/rmf calibration
files and spectral models are used.

The individual $kT$ and $N_H$ values for weak sources are unreliable:
they have large statistical uncertainty, and sometimes more parameters
than independent spectral bins.  Sources with $C_{xtr} < 30$ are
omitted from scientific analysis.  The reported spectral fits for faint
sources, however, are not meaningless.  For these faint sources, the
$\chi^2$ fitting process give a solution that passes exactly through
the binned spectrum ($\chi^2 = 0.0$).  The solutions represent
non-unique spline-like fits to the event energy distribution, and are
thus useful for the calculation of broad-band X-ray luminosities.  We
report the spectral parameters in the tables, even for these faint
sources, so that future researchers can reproduce our luminosity
values, even though the fit parameters are not individually reliable
for astrophysical analysis.

The derived spectral parameters inherit a statistical uncertainty, and
perhaps a bias, from the $\chi^2$ fitting process.  We therefore
performed several thousand simulations with sources of known spectra and
differing count rates to estimate the statistical uncertainties of
$\log N_H$, kT and the broad-band luminosities (\S \ref{lum_sec}).  We
used the {\it fakeit} utility in the XSPEC package for simulations with
$20 \leq \log N_H \leq 22.5$ cm$^{-2}$, $1 \leq kT \leq 10$ keV, and
$10 \leq C_{xtr} \leq 1000$ counts.  The results show a systematic
tendency for the fitting process to systematically underestimate the
energies of $kT = 10$ keV sources by $\simeq 10$\% (1000 cts) to
$\simeq 50$\% (30 cts).  Two effects may contribute to this bias:  the
sparse photon population of the uppermost channels due to the rapid
decline in telescope effective area at high energies; and the
incompatibility between the CTI-corrected data and the uncorrected
arf/rmf files used here. Another bias occurs when $\log N_H$ values are
below $\simeq 20.5$ cm$^{-2}$.  The fitted values are often
ill-determined in these cases because we consider data only above 0.5
keV.

The statistical uncertainties of spectral parameters are estimated as
follows from these simulations.  The standard deviations of fitted
plasma energy values range roughly from $\Delta(kT)/kT \simeq 60$\% (30
cts) to 30\% (100 cts) and 10\% (1000 cts).  Column density
uncertainties range from $\Delta(\log N_H) \simeq 0.7$ (30 cts) to 0.3
(100 cts) and 0.1 (1000 cts).   Due to the nonlinearity of the models
and data, correlated errors are naturally present.  Broad-band
luminosity values $\log L_t$ exhibit standard deviations ranging from
35\% (10 cts) to 25\% (30 cts), 15\% (100 cts) and 4\% (1000 cts).
These are only somewhat larger than the optimal $\sqrt{C_{xtr}}$, even
for sources with as few as 10 counts.  All of these results are not
substantially affected by use of the likelihood ratio ({\it C})
statistic instead of the $\chi^2$ statistic.

We conclude from this simulation analysis that all broad-band
luminosities derived here are reliable, but that the individual
spectral parameters ({\it kT} and $\log N_H$) are unreliable for the
faintest sources.  As noted above, we include these parameters in Table
3 even for faint sources so that others may reproduce our luminosity
results.  To further emphasize that the spectral parameters of faint
sources should not be considered accurate, a warning is given in the
table notes via a faint source flag.  We thus confine discussion of
spectral properties to the 790 sources (74\% of 1075) with $C_{xtr}
\geq 30$ cts, for which estimated errors are much smaller than the
parameter ranges.  Finally, we note that even well-determined spectral
parameters may not have a clear astrophysical meaning, as they often
are the sums of photons from different stages of flare evolution and/or
different physical structures in the active young stellar system.

Figure \ref{spec_fig} shows three spectra that exemplify some of the
characteristics of ACIS ONC spectra.  The top panel shows a fairly
typical spectrum: 91 counts are extracted and successfully modeled with
a plasma at $kT = 1.1$ keV with moderate absorption of $\log N_H =
20.8$ cm$^{-2}$.  The counterpart is a very young ($t \leq 0.1$ Myr)
low mass (0.2 M$_\odot$) star with little visual absorption, a 3.2 day
rotational period, and a possible X-ray flare.  The middle panel shows
a deeply embedded faint source: 17 counts are extracted and are modeled
with a $kT = 2.8$ keV plasma with high absorption of $\log N_H = 22.8$
cm$^{-2}$.  The spectral fit (not shown) passes directly through the
data values as there are no degrees of freedom (3 fit parameters and 3
bins).  These spectral parameters are not individually reliable and
are only used to infer the luminosity of the source.  The counterpart
here is IRc 3 = Source i, an important luminous and massive member of
the BN/KL embedded cluster responsible for at least some of the
molecular outflows in the region.  The bottom panel shows a bright
source with high signal-to-noise across the spectrum: 3469 counts are
extracted and modeled with only partial success using a $kT = 2.1$ keV
plasma and moderate absorption of $\log N_H = 21.5$ cm$^{-2}$.  The
spectrum shows strong excess emission around $0.8-1.0$ keV
(attributable perhaps to Fe-L, Ne IX and Ne X lines) and perhaps also
around $3-4$ keV.  The counterpart here is a 2 M$_\odot$ star with age
around 1 Myr, rotational period of 9.2 days, $A_V = 3.2$ mag
absorption, and an infrared excess $\Delta(I-K)=0.7$ mag indicating a
circumstellar disk.

The reliability of one aspect of the spectral analysis is subject to
independent test.  Figure \ref{avnh_fig} compares the column absorption
$\log N_H$ (in cm$^{-2}$) obtained here with the visual absorption
$A_V$ (in mag) obtained by \citet{Hillenbrand97} from optical
spectroscopy and photometry.  The dashed curve shows the conversion
relationship $N_H = 2 \times 10^{21} A_V$ for interstellar material
with a standard gas-to-dust ratio \citep{Savage79}.  It suggests that
visual absorption may be systematically slightly higher than expected
from the X-ray absorption for a standard gas-to-dust ratio.  The
scatter is consistent with expected errors in most cases, but some
$\log N_H$ values are completely incorrect with values around $\log N_H
\simeq 22$ when $A_V \simeq 0$ or $\log N_H < 20$ when $A_V \geq 2$.
Gross errors of this type may arise from our use of simplistic
one-temperature plasma models, or incorrect $A_V$ values due to optical
veiling or spectral typing errors.  We conclude that the log$N_H$
values derived here should only be used as a rough guide to the true
absorption to each source.

\subsection{Broad-band luminosities \label{lum_sec}}

Despite the difficulties of producing spectral fits of a quality
necessary for detailed study of plasma properties, the fits provide
reliable broad-band luminosities.  Effectively, the spectral model is
treated here as a spline fit to the data, and we integrate under the
fitted curve to obtain the broad-band fluxes in the soft $0.5-2$ keV
and hard $2-8$ keV bands.  The fluxes are converted to luminosities
assuming a distance of 450 pc to the ONC, although we recognize that
this distance is not precisely established and could be as high as 480
pc.

For sources with $\geq 30$ counts, we provide four log luminosity
values: soft band log$L_s$ covering $0.5-2$ keV, hard band log$L_h$
covering $2-8$ keV, total band log$L_t$ covering $0.5-8$ keV, and the
total band after correction of absorption log$L_c$.  These values have
complementary uses: $L_s$ allows comparison with earlier measurements
from the {\it Einstein} and {\it ROSAT} satellites, but is most
vulnerable to differences in absorption between sources; $L_h$ is
nearly unaffected by absorption but measures only very hot
flare-generated plasma; $L_t$ most closely represents all of the
emission directly observed by {\it Chandra}, but underestimates the
true luminosity due to soft-energy absorption; and $L_c$ attempts to
correct for the absorption. Note that the wide dispersion and
not-infrequent errors in log$N_H$ values (Figure \ref{avnh_fig})
suggests that $L_c$ values should be used with considerable caution.
For faint ($C_{xtr} < 30$ counts) sources, only $L_t$ is reported.

All luminosity values have been corrected for the contribution of
background emission and the counts in the wings of the PSF by scaling
the luminosity derived from the spectral fit by the factor $(C_{xtr} -
B_{xtr})/(C_{xtr}f_{PSF})$ (see \S \ref{extract_sec}).  These
corrections are small, $< 0.1$ in log$L$, for most sources.  The formal
uncertainties to the luminosities are also estimated to be small for
most sources,  roughly $\pm0.1$ in log$L$, though higher for the
weakest sources with poor statistics and those with inaccurate spectral
fits (see footnotes to Table \ref{src_prop_table}).  However, the {\it
astrophysical} accuracy of the luminosities, in the sense of the
reproducibility of the measurements in independent observations, is
much worse due to intrinsic variability.  About half of the sources
showed significant variability during our two $\sim$12-hour
observations, with many of these changing in brightness by more than a
factor of two or $> 0.3$ in log$L$. It is therefore reasonable to
expect scatter of several tenths in scatter plots involving log$L$
values due in part to uncertainties in the analysis but mostly to
source variability.

The reader can calculate an approximate hardness ratio for each source
using $(\log L_h - \log L_s)$ or $(\log L_h - \log L_t)$.  However, we
caution that the hardness ratio is difficult to interpret here as the
effects of different absorptions and different intrinsic spectral
hardnesses are intermingled.  A bias in this hardness ratio will be
present in sources far off-axis as we do not correct for the energy
dependence of the telescope vignetting.  An absorption-corrected soft
X-ray luminosity can also be calculated from $L_c - L_h$, as the
correction rarely has significant effect above 2 keV.  But we caution
again that the corrected soft-band luminosity is particularly sensitive
to errors in $\log N_H$ (Figure \ref{avnh_fig}).

\subsection{Very bright sources and photon pileup \label{bright_sec}}

$\theta^1$C Ori (source \#542), the massive O6 star dominating the
Trapezium at virtually all wavelengths, suffers very significant pileup
in the CCD detector.  That is, several photons arrived in the same pixel
during a 3.2~s frame resulting in an incorrect evaluation of the photon
energy and exclusion by the on-board event processor.  Most of the real
events from $\theta^1$C Ori were lost in this way; we estimate that
$\simeq 450,000$ source photons were incident on the detector.  We
recover from this problem in an approximate fashion by extracting
$\simeq 9,000$ events from a $2\arcsec-4$\arcsec\/ annulus around the source
which contains $1\%-3$\% of the PSF encircled energy (the exact fraction
varies with photon energy), creating a special arf file appropriate for
this annulus using the IDL program {\it xpsf.pro} (G. Chartas, private
communication), and calculating a corrected spectral fit.  Several
other sources, notably the massive stars $\theta^1$A Ori (\#498) and
$\theta^2$A Ori (\#828) and the late-type stars JW 567 (= MT Ori,
\#626) and P 1771(\#243) suffer mild photon pileup and the resulting
luminosities are likely underestimated by $10-30$\%.  The spectral fits
and variability measurements may also be affected in complex ways.
Bright off-axis sources like KM Ori (\#77) are less affected because
the photons are distributed over many pixels by the broadened PSF.
Pileup warnings are provided in Table \ref{src_prop_table} notes
whenever $CR_1 > 100$ or $CR_2 > 100$ cts ks$^{-1}$.  

\subsection{Completeness and reliability of the catalog
\label{spurious_sec}}

To assess the level of contamination of the ACIS ONC catalog from
spurious sources, we examined the faintest 50 sources with
$(C_{xtr} - B_{xtr})/f_{psf} \leq 10$ in some detail:

{\bf Spatial distribution} ~~   Three spatial components are seen: half
are clustered around the Trapezium region, a group of $\sim$13 are in
the BN/KL region, and $\sim$15 are distributed randomly in the field.
Naturally they avoid a strong concentration in the inner 1-2\arcmin\/
as the background from $\theta^1$C Ori precludes finding very faint
sources there.  The observed pattern makes sense if the sources are
nearly all real: some ONC members, some embedded members, and some
extragalactic sources.  In particular, it suggests that an ultradeep
observation will see many more sources in the embedded BN/KL cluster.

{\bf Counterparts} ~~  One-third (16/50) do not have stellar counterparts,
compared to 8\% in the full sample.  This does not necessarily mean
that all faint sources lacking counterparts are spurious; due to the
$L_x$-$L_{bol}$ correlation, the fainter sources are less likely to
appear in flux limited optical and infrared catalogs.

{\bf Counterpart offset} ~~  Of the 34 sources with listed stellar
counterparts, in all but 8 the counterpart offset lies within the
dense concentration in Figure \ref{offsets_fig}.  This, we believe, is
strong evidence that these sources are real Orion stars.  The eight
outliers (\# 137, 198, 303, 545, 827, 930, 943, and 1073) with offsets
ranging from 0.9$\arcsec\/ to 3.1$\arcsec\/ are the strongest
candidates for spurious sources in the catalog.  If drawn randomly from
the full sample, only $2-3$ sources should be present with these large
offsets.

This last test is the clearest indication that several spurious sources
are likely present in the catalog.  When a few additional spurious sources
without stellar counterparts or with slightly higher count rates are 
considered, we estimate that $\simeq 10$ of the faintest sources, or 1\%
of the entire source catalog, are likely spurious.  

\subsection{Detection limit of the catalog \label{limits_sec}}

Given the complexities of the {\it wavdetect} algorithm applied to the
spatially varying PSF (\S \ref{detect_sec}), the subjective nature of
our corrections to the source list (\S \ref{detect_sec}), the
position-dependent extraction radii and background levels (\S
\ref{extract_sec}), and the wide range of source spectra (\S
\ref{spec_sec}),  it is not simple to establish an astronomically
useful detection limit.  One procedure might be to set an extraction
circle onto the image at the location of a specific source, determine
the Poissonian 99th-percentile upper limit in counts, divide by the
effective exposure time at that location in the image, and convert to a
luminosity limit using a spectral model.  But, as there are hundreds of
faint or embedded ONC members which are undetected in our image, we
provide here a more general method that can be applied to any location 
of interest.

First we determine the count limit of the {\it wavdetect} source 
detection algorithm (\S \ref{detect_sec}).  Figure \ref{cts_theta_fig} 
shows the distribution of extracted source counts as a function of 
off-axis angle $\theta$.  From careful examination of the image, we 
are very confident that all sources brighter than the plotted curve, 
\begin{equation}
C_{lim}(\theta) = 9 + 0.16 \theta + 0.28 \theta^2
\end{equation}
in the full band are detected.  Note that the sources found below this
curve are still reliable; we just can not be sure that all such sources
have been found.

For sources of marginal significance, one empirically achieves a better 
signal-to-noise ratio by evaluating the source counts (or limits
thereof) using an extraction radius with 50\% of the encircled energy
rather than the usual 95\% extraction radius; this is most likely due
to backgroun variations.  The limiting
background-corrected source count rate is then
\begin{equation}
CR_{lim}(\theta) = [C_{lim}(\theta) (0.95/0.50) -
\pi R_{xtr, 50\%}(\theta)^2B(\theta)] /(f_{PSF}f_{vig}E_{eff})
\end{equation}
where $R_{xtr}, 50\%$ is the 50\% extraction radius given in footnote
\ref{radius_xtr_footnote}, $B(\theta)$ is given by the background fits
in \S \ref{extract_sec}, $f_{PSF}=0.50$, $f_{vig} \simeq 1.00 - 0.014
\theta$ is an approximate correction for telescope vignetting at 1.5
keV, and $E_{eff}$ is the exposure time in ks given in Table
\ref{obs_table}.

While the limiting count rate calculated in this fashion is accurate to
about $\pm 30$\%, there is considerably more uncertainty in converting
this to a limiting astrophysical luminosity given the wide range in
spectral shapes and foreground absorptions.  If we assume a plasma
energy of $kT = 3$ keV, then the conversions between $CR_{lim}$ (in
counts ks$^{-1}$) and $\log L_{t,lim}$ (in erg s$^{-1}$ in the full
$0.5-8$ keV band) can be approximately expressed as
\begin{equation}
\log L_{t,lim} \simeq 28.9 + \log CR_{lim} + 0.3 
(\log N_H - 20.0) ~ {\rm erg~s}^{-1}.
\end{equation}
For stars with absorptions measured from optical or infrared measurements, 
$N_H$ can be estimated from the relationship $N_H = 2 \times 10^{21} A_V$ 
cm$^{-2}$ (see Figure \ref{avnh_fig}).  We caution that this $L_{t,lim}$ 
value for a given star could be seriously in error if the intrinsic 
spectrum differs from the assumed 3 keV plasma or if the absorption 
estimate is inaccurate.

The result of these computations is that undetected stars with 
negligible interstellar absorption have upper limits of $\log L_t < 
28.0$ erg s$^{-1}$ in the inner region of the detector (except close 
to $\theta^1$C Ori) and $\log L_t < 28.6$ erg s$^{-1}$ near the 
edge of the field.  At a given $\theta$, the limiting observed 
luminosity rises by $\log L_t \simeq 0.5$ if $A_V \simeq 1-2$ 
compared to $A_V = 0$.  

Statistical study of ONC subpopulations, such as the measure of X-ray
luminosity functions, requires consideration of both X-ray
nondetections of catalogued stars (in statistical parlance, censored
bias) and on incompleteness of the catalogued sample (truncation
bias).  The techniques of survival analysis provide strategies for
treating censoring, but it is more difficult to overcome truncation
biases \citep{Feigelson90, Feigelson92}.  The sample of 1576 stars with
$V<20$ by \citet{Hillenbrand97}, for example, should be virtually
complete for ONC members with masses $M  > $0.1 M$_\odot$ and absorptions 
$A_V < 2.5$ mag.  The ACIS observation detects X-rays from nearly this
entire sample:  only 8 stars from \citet{Hillenbrand97} with $M > 0.7$
M$_\odot$ and high membership probabilities inferred from proper motion
measurements are absent from the X-ray source tables (JW 62, 108, 407,
479, 531, 593, 608, and Parenago 1772).  Ninety-two additional stars
with smaller masses are absent.  Statistical analysis of ACIS results
based on the $V<20$ sample should therefore be reliable if one avoids
stars with low stellar masses and high absorption.  Statistical
analysis of other samples, such as brown dwarfs and deeply embedded
protostars, may be subject to considerable bias.

\section{Source list and properties \label{tables_sec}}

The database of sources found in the merged Orion fields is provided in
Tables \ref{src_list_table}  and \ref{src_prop_table} which appear in
their entirety in the electronic edition.  The first of these large
tables gives source positions, stellar identifications, and
multiwavelength stellar properties while the second table gives source
count rates, luminosity, spectral and variability information.  Some
stellar properties like mass and age are given to higher precision than
we believe is scientifically warranted.  This is done to reduce the
number of overlapping points in scatter plots.  Specifics regarding
table entries follow.
\begin{description}

\item [Table \ref{src_list_table}, column 1]  Source name in the form
CXOONC Jhhmmss.s-ddmmss ({\it Chandra X-ray Observatory} Orion Nebula
Cluster).  These names supercede those given by \citet{Garmire00},
which often differ in the last digit.

\item [Columns 2-3]  Source position in decimal degrees in epoch
J2000.  The field is aligned to the 2MASS/ACT/Tycho reference frame to
within $\pm 0.1$\arcsec\/ (\S \ref{ids_sec}), and individual source
positional accuracies vary between about $0.1\arcsec - 3$\arcsec\/ 
depending on the signal strength and off-axis distance.  See table notes 
(`x'  in column 17) for cases of crowding, location on bright source 
readout trail, or other issue regarding the X-ray image. 

\item [Column 4] Distance from the cluster center in arcminutes,
measured from $\theta^1$C Ori.  This quantity is useful for evaluating
point spread function and completeness effects (\S \ref{extract_sec}
and \S \ref{detect_sec}).

\item [Column 5] Detection in previous X-ray studies of the Orion
Nebula:  a = {\it Einstein Observatory} \citep{Ku79, Ku82, Gagne94}; b
= {\it ROSAT} Position Sensitive Proportional Counter \citep{Geier95,
Lohmann00}; c = {\it ROSAT} High Resolution Imager \citep{Gagne95}; d =
{\it ASCA} satellite \citep{Yamauchi96}; e = {\it Chandra X-ray
Observatory} ACIS-I \citep{Garmire00}; and f = {\it Chandra X-ray
Observatory} ACIS-S3 \citep{Schulz01}.  The associations between the
older lower-resolution sources and CXOONC sources are sometimes
uncertain due to confusion.

\item [Column 6] Stellar identification of the X-ray source (\S
\ref{ids_sec}):  P = \citet{Parenago54}; JW = \citet{Jones88}; PSH =
\citet{Prosser94}; H = \citet{Hillenbrand97}; HC =
\citet{Hillenbrand00};  CHS = \citet{Carpenter01}.  JW designations are
preferentially listed when available.  See table notes (`id' in column
17) for cases of multiple counterparts and for Greek letter labels
(e.g., $\theta^1$G Ori).

\item [Column 7]  Offset $\phi$ between the X-ray and stellar source in
arcseconds.  Star positions from \citet{Hillenbrand00} are
preferentially adopted when multiple values are available.  Potential
uncertainties in source identification due to large offsets are noted
(`id' in column 17).

\item [Column 8-10] Effective surface temperature log$T_{eff}$,
bolometric luminosity log$L_{bol}$, and visual absorption $A_V$ from
\citet{Hillenbrand97} and subsequent additions and updates to the
database.  These stellar properties, derived from spectroscopy of
$V<20$ stars and V band photometry, locate the star on the HR diagram.

\item [Column 11-12] Logarithm of the stellar mass (in M$_\sun$) and
age (in years) obtained from the HR diagram  location and the PMS
evolutionary tracks of \citet{DAntona97}.  These values are updated
from those given by Hillenbrand (1997) using older tracks.  Note that
considerable debate exists over the accuracy of PMS evolutionary tracks
and systematic errors may be present.  In particular, it is difficult
to distinguish stellar ages log$t < 5.5$ due to uncertainties in
initial conditions \citep{Stahler83}.

\item [Column 13] K band (2.2 $\mu$m) excess, $\Delta (I-K)$, over the
value expected for a photosphere with temperature log$T_{eff}$, based
on the infrared photometry of \citet{Hillenbrand98} and
\citet{Hillenbrand00}.  Values of $\Delta (I-K) > 0.3$ are widely
considered to indicate warm dust in a circumstellar disk.

\item [Column 14] Additional stellar properties:  FIR = possible
counterpart mid- and far-infrared source (see footnote for details); HH
= Herbig-Haro objects or their host star from Hubble Space Telescope
(HST) images \citep{Bally98, Bally00, Bally01}; L = L band (3.5$\mu$m)
excess interpreted as protostellar candidate \citep{Lada00}; N = N band
(10 $\mu$m) excess interpreted as truncated disk \citep{Stassun01}; pd
= proplyd and/or disk imaged by the HST in emission and/or silhouette
\citep{Odell94, Odell96, Bally00, Bally01}; r = radio continuum source
detected at centimeter \citep{Felli93} or millimeter \citep{Mundy95}
wavelengths; and wc = star with wind collision front \citep{Bally00}.
Identifications are based on positional coincidences consistent with
the X-ray positional accuracies shown in Figure \ref{offsets_fig}, and
may not always represent physical associations.

\item [Column 15] Rotational period obtained from photometric
modulations of starspots.  Uncertain or multiple periods are discussed
in the table notes (`p' in column 17).

\item [Column 16] Source of the rotational period: C =
\citet{Carpenter01}; H = \citet{Herbst00} and \citet{Herbst01}; S =
\citet{Stassun99}.

\item [Column 17] Footnote indicator: x = X-ray image issue; id =
stellar identification issue; and p = rotational period issue.

\end{description}

\begin{description}

\item [Table \ref{src_prop_table}, column 1] Source number from Table
\ref{src_list_table}.

\item [Columns 2-5] Quantities associated with event extraction from
the full band ($0.5-8$ keV) image described in \S \ref{extract_sec}:
total extracted counts $C_{xtr}$ and estimated background counts
$B_{xtr}$ in a circle of radius $R_{xtr}$ centered on the source
position given in Table \ref{src_list_table}, and the fraction of the
point spread function $f_{PSF}$ encircled by $R_{xtr}$ at the source
location in the ACIS field.

\item [Columns 6-7] Average source count rates $CR_1$ during the
October 1999 and $CR_2$ during the April 2000 observations.  $CR$ is
defined in \S \ref{extract_sec}.  Location-dependent exposure
variations are not included in these values.

\item [Column 8] Variability class defined in \S
\ref{variability_sec}:  Const (constant); LT Var (long-term
variability); Pos flare (possible flare); and Flare.

\item [Columns 9-10] Spectral parameters from one-temperature plasma
models, when the fit to the source spectrum is satisfactory (\S
\ref{spec_sec}).  log$N_H$ (in cm$^{-2}$) is the equivalent hydrogen
column density of intervening interstellar material producing soft
X-ray absorption, and $kT$ (in keV) is the energy of the plasma.  See
\S \ref{spec_sec} regarding the reliability of these values.  For faint
sources, these quantities are highly uncertain and are used only as
rough characterizations of spectral shape.

\item [Column 11]  Lower and upper plasma energies (in keV) for sources
fit with two-temperature plasma models.  Again, these values are only
suggestive for faint sources.

\item [Column 12] Flag indicating the presence of spectral features
indicative perhaps of specific elemental abundances enhanced over the
assumed 0.3 times solar levels.

\item [Columns 13-16] X-ray luminosities of the source assuming a
distance of 450 pc averaged over both observations: $L_s$ = soft-band
($0.5-2$ keV) luminosity; $L_h$ = hard-band ($2-8$ keV) luminosity;
$L_t$ = total band ($0.5-8$ keV) luminosity; and $L_c$ = total band
luminosity corrected for the estimated interstellar absorption.  These
values are corrected for all telescope and detector efficiencies
convolved with the spectral model indicated in columns $9-11$.
For faint sources, only $L_t$ is given.

\item [Column 17] Footnote indicator: f = faint source warning
(spectral parameters are only used as a spline fit for obtaining
$L_t$); p = photon pileup warning; s = spectral issue; v = variability
issue.  The table notes give details for spectral and variability
issues, in particularly describing temporal variations when the 
Variability Class is `Flare'. 

\end{description}

\section{Demographics of the X-ray population \label{demog_sec}}

\subsection{Sources without stellar counterparts \label{noid_sec}}

Before examining the broad X-ray properties of the ONC and its
molecular cloud environs, we seek to establish the level of
contamination by extraneous sources.  Table \ref{noid_table} presents
the 101 CXOONC sources that have no detection in the available optical
and near-infrared catalogs (\S \ref{ids_sec}).  Most of these
unidentified sources are heavily absorbed with $\log N_H \geq 22.0$
cm$^{-2}$ and thus lie behind or deeply embedded within the Orion
molecular cloud.  

Some of these must be members of an extragalactic (mainly active
galactic nuclei) or background Galactic source population seen through
the cloud \citep{Garmire00}.  However, Figure \ref{noid_fig} shows that
most are too clustered towards the field center for an isotropic
extragalactic population.  This is further confirmed by comparison of
the source fluxes with the extragalactic $\log N - \log S$
distribution.  From CO surveys, we estimate that the depth through the
cloud ranges from $\log N_H \simeq 22$ cm$^{-2}$ near the edges of the
ACIS field to $\log N_H \simeq 23$ cm$^{-2}$ near the center. Thus
emission from extragalactic sources will be absorbed below $2-4$ keV at
different locations in the field.  From the $2-10$ keV extragalactic
$\log N - \log S$ curve derived with ACIS-I from the northern Hubble
Deep Field, we estimate that extragalactic sources can not account for
more than 25 heavily absorbed sources present in the ONC image, and
nearly all of these will have $C_{xtr} < 25$ counts.  As most of the
faint sources in Table \ref{noid_table} have $25- 100$ counts (plus 14
that lie in the range $100 < C_{xtr} < 3700$) and are concentrated
towards the field center, we estimate that there are only $10-15$
extragalactic sources in the entire field.  The contribution of
background Galactic sources can not be confidently estimated as their
$\log N- \log S$ distribution has not been reported at these faint flux
levels for this Galactic latitude.  Altogether, the contamination from
extragalactic or background Galactic X-ray sources probably accounts
for $\simeq 20$ sources or $\simeq 2$\% of the full ONC source sample.

The $\simeq 80$\% of the ACIS sources without catalogued stellar
counterparts which are not contaminants must be new young stars
associated with the Orion cloud.  In column 8 of Table
\ref{noid_table}, we suggest a tentative classification for these
sources based on location and absorption.  Twenty-six sources are
lightly absorbed with $\log N_H < 22.0$ cm$^{-2}$; most of these are
probably new low-mass members of the ONC.  Several are concentrated
within $\sim 0.5$\arcmin\/ around $\theta^1$C Ori, while others are
distributed across the ACIS field.

The remaining 75 sources are deeply embedded or behind the cloud with
$\log N_H > 22.0$ cm$^{-2}$, some of which are probably protostars very
recently formed in dense molecular cores.  Ten lie on the OMC 1 (=
Orion KL) molecular core (Figure \ref{noid_fig}) and are likely new
members of the BN/KL young stellar cluster.  These are among the first
clearly identified low mass members of this cluster, as it is too
obscured for complete $JHK$ band study and mid-infrared observations to
date have been sensitive only to the $L \geq 1000$ L$_\odot$ high-mass
stars \citep{Gezari98}.  Five of the sources coincide with the OMC 1S
(= Orion S) molecular core.  Little is known about the young stellar
population of OMC 1S other than a luminous protostar FIR 4 and an
unknown protostar producing an unusually fast and young bipolar flow
\citep{Rodriguez99}.  Nine embedded sources lie along the dense
molecular filament running north from OMC 1 towards OMC 2/3.  A small
concentration of infrared-excess and photometrically variable young
stars has also been found in this region from observations with the
2MASS telescope \citep{Carpenter01}.  This molecular concentration is
thus likely a separate star forming region, and we classify these ACIS
sources `OMC 1N' (analogous to the OMC 1S designation) in Table
\ref{noid_table}.  No unidentified embedded X-ray sources are
associated with the Orion Bar (the NE-SW molecular structure south of
the OMC 1 and OMC 1S concentrations in Figure \ref{noid_fig}),
suggesting that it is not an active region of star formation.

Finally, we classify the 51 heavily absorbed ACIS sources which do not
coincide with dense molecular cloud cores as `Embd/Bk'.  Roughly 20 of
these are contaminants (see above) and the others are likely new PMS
stars, perhaps embedded low-mass ONC members or somewhat older members
of the star forming cores.

While the luminosity distribution for the unidentified ACIS sources is
similar to that of the identified sources, their spectral properties
differ: nearly 60\% have plasma components with fitted energies $>$10 keV
compared to only 10\% for identified sources.  This suggests that at
least 300 additional embedded X-ray emitting stars with lower plasma
temperatures exist in the field but are undetected due to the high
column densities.  Most of these will likely have counterparts among
the hundreds of heavily absorbed low-mass and VLM ONC stars
\citep{Hillenbrand00}.  Many of these would likely be detected with
longer ACIS exposures to increase sensitivity and repeated observations
to catch flares.

One unidentified source, \#881 or CXOONC 053524.6-052759, deserves
special note due to its extraordinarily high and constant flux.  It has
$\simeq$3700 total counts, $\log N_H = 22.2$ cm$^{-2}$, $kT > 10$ keV,
and $\log L_t = 31.3$ erg s$^{-1}$ (assuming $d = 450$ pc), placing it
among the brightest 2\% of sources in the field.  The spectrum is also
well-fit by a powerlaw model with photon index $\Gamma = 1.5$ over the
range $1-8$ keV, but an additional soft component may be present from
$0.5-1$ keV.  There is no evidence for spatial extent larger than
$\simeq 2$\arcsec; this limit is high because the source lies
5\arcmin\/ off-axis.  No flux variations above $\simeq 10$\% are present
within an observation, and $<7$\% ($3 \sigma$) flux difference is seen
between the two observations.  The source was detected with the {\it
ROSAT} HRI instrument at a level consistent with the $\log L_s = 30.1$
erg s$^{-1}$ found with ACIS in the soft band \citep{Gagne95}.

The properties of this source do not readily fit most categories of
X-ray sources.  It is: too constant and with a $\log L_t/L_{bol}$ ratio
too high for a typical ONC PMS star or protostar; too bright and hard
compared to typical extragalactic background sources; too constant for
a typical Galactic accretion X-ray binary system; too bright and too
hard for blackbody emission from an isolated young neutron star; and
too hard for Bondi-Hoyle accretion of molecular gas onto an isolated
neutron star.  Perhaps the most likely possibility is the hard powerlaw
component of a transient, low magnetic field, neutron star binary
system seen during quiescence.  Several examples of such systems are
known in the Galaxy including Cen X-4 and Aql X-1 \citep[][and
references therein]{Rutledge01}.  The soft spectral component seen in
CXOONC 053524.6-052759, with $L_x \simeq {\rm several} \times 10^{32}$
erg s$^{-1}$ ($0.1-1$ keV assuming a blackbody or thermal temperature
around 0.1 keV) after correcting for absorption, would then arise from
the neutron star surface or atmosphere.  We note that, if the transient
neutron star binary model is correct, then the system has not emerged
out of quiescence above $L_x \sim 10^{34}(d/{\rm kpc})^2$ erg s$^{-1}$
during the past $\sim 30$ years.

\subsection{Global X-ray properties \label{global_sec}}

Having established that 98\% of the CXOONC sources are young stars from
the ONC or nearby Orion star forming cores (91\% by precise spatial
coincidence with catalogued stars and 7\% by inference in \S
\ref{noid_sec}), we can treat the entire ACIS source population as a
unified sample of young Orion stars with considerable reliability.  We
consider here only univariate distributions of X-ray properties such as
flux, luminosity, variability and spectra.  Bivariate 
distributions comparing the X-ray properties with other stellar
properties are considered in \citet{Feigelson02b}.

The distribution of source fluxes (Figure \ref{xlum_hist_fig}, left
panel), where $F_t = L_t / 4 \pi (450~{\rm pc})^2$, is better described
as a lognormal rather than a power law as commonly seen in
extragalactic source populations. The mean and standard deviation are
$<\log F_t> = -12.9 \pm 0.7$ erg s$^{-1}$ cm$^{-2}$.  The hatched
region denotes the completeness limit, which ranges from $\log F_t =
-14.4$ to $-14.9$ erg s$^{-1}$ cm$^{-2}$ depending on location in the
field (\S \ref{limits_sec}).  We emphasize that the fall in source
counts in the $-15.5 < \log F_t < -14.5$ erg s$^{-1}$ cm$^{-2}$
interval (and the corresponding fall in luminosity counts in the $28.0
< \log L_t < 29.0$ erg s$^{-1}$ interval) is intrinsic to the source
population and is not caused by sensitivity limitations.

The luminosity distribution (middle panel) of course has a similar
shape, with $<\log L_t> = 29.4 \pm 0.7$ erg s$^{-1}$.  A link to
another stellar property is easily found:  the most luminous sources
are also the high- and intermediate-mass stars (hatched region, see \S
\ref{highmass_sec}-\ref{intermed_sec}).  Further analysis shows that
stellar mass accounts for more of the variance in X-ray luminosity than
any other stellar property \citep{Feigelson02b}.  The total luminosity
of all 1075 sources is $L_t = 3.2 \times 10^{33}$ erg s$^{-1}$ in the
total ($0.5-8$ keV) band and $L_h = 1.3 \times 10^{33}$ erg s$^{-1}$ in
the hard ($2-8$ keV) band.  The latter value compares very well to the
integrated luminosity of $1.3 \times 10^{33}$ erg s$^{-1}$ (adjusted
for a distance of 450 pc) in the $2-10$ keV band found with the
non-imaging Ginga satellite in a 0.2$^\circ$ region around the
Trapezium stars \citep{Yamauchi93}.  The dominant star of the
Trapezium, $\theta^1$C Ori, contributes $L_t = 2 \times 10^{33}$ erg
s$^{-1}$ or $\simeq 60$\% of the total band luminosity and $L_h = 5
\times 10^{32}$ erg s$^{-1}$ or $\simeq 40$\% of the hard band
luminosity.

Figure \ref{xlum_hist_fig} (middle panel) indicates that the presence
or absence of an infrared excess, an indicator of a circumstellar disk,
has no discernable effect on the distribution of X-ray luminosities.
Similarly, no effect is seen in $\log L_t/L_{bol}$.  There is thus
no evidence that a circumstellar disk, at least one sufficiently
massive and dusty to produce excess K-band emission, is required for
the elevated X-ray emission of PMS stars.  A similar result was found
in several $Einstein$ and $ROSAT$ studies of nearby T Tauri stellar
populations \citep{Feigelson99}, although \citet{Stelzer01} find that
weak-lined T Tauri stars are several-fold more X-ray luminous than
classical T Tauri stars in the Taurus-Auriga complex.

The distribution of the ratio of X-ray to stellar bolometric luminosity
has a mean and standard deviation of $\log L_t/L_{bol} = -3.9 \pm 0.7$
(Figure \ref{xlum_hist_fig}, right panel).  The non-Gaussian tail
around $-9 < \log L_t/L_{bol} < -6$ is due to mid-A to late-O type
stars which have high $L_{bol}$ but modest $L_t$ values (see Figure
\ref{highmass_ls_fig}, left panel). Several dozen low mass stars have
high values above the `saturation' level $\log L_t/L_{bol} \simeq -3.0$
that defines the maximum X-ray emission seen in magnetically active
main sequence stars \citep[e.g.][]{Vilhu87, Fleming95, Randich97}.
Some of these were observed during a flare, but others exhibit high but
relatively constant emission.

The variability class distribution is shown in Figure
\ref{xprop_hist_fig} (left panel).  The excess of sources with
`Constant' emission compared to the other classes is a selection
effect:  the `Constant' sources are dominated by sources with $C_{xtr}
< 50$ counts which are too weak to clearly show flaring activity
(hatched region).  If these weak sources are ignored, the distribution
among the four variability classes becomes roughly equal.  If we group
`Flare' and `Possible flare' sources together into a single category,
then 55\% of the stronger sources in the field exhibit some form of
intra-day variability.  It is difficult to convert this number into a
flare duty cycle because of the great range of flare durations seen in
the source lightcurves.

Figure \ref{xprop_hist_fig} (middle panel) show the distribution of
plasma energies for sources with $\geq 30$ extracted counts and
satisfactory one-temperature fits.  (Recall that there may be
systematic errors in $kT$ values; \S \ref{spec_sec}). The median plasma
energy $kT  = 2.6$ keV, and the distribution is asymmetrical with a
heavy tail to higher energies.  There is no apparent trend that flaring
sources exhibit harder spectra.

Two results emerge from these source temperatures.  First, nearly all
PMS stars have plasmas hotter than seen in the Sun, even during its
most powerful contemporary flares.  Integrated over its disk and viewed
with CCD spectral resolution, the Sun typically would be seen at a
plasma energy $\leq 0.2$ keV, rising to 0.6 keV during powerful flares
\citep{Peres00, Reale01}.  Note, however, that a soft solar-type
spectral component would often be undetectable in Orion stars due to
interstellar absorption.

Second, while $kT \simeq 10$ keV energies were found during an
extremely powerful T Tauri flare with the {\it ASCA} satellite
\citep{Tsuboi98}, we find that such high temperatures are commonly
present even at moderate X- ray luminosities and in stars not
exhibiting flaring lightcurves.  The plasma temperatures of sources
with intraday variability (`Flare' and `Possible flare' variability
classes) are nearly indistinguishable from those of non-flaring
sources.  This implies that the X-ray emission from pre-main sequence
stars, even those without apparent variations during an observation,
is predominantly flare emission with negligible contribution by a
softer `coronal' component.  This supports current ideas that the
`quiescent' emission in magnetically active stars arises from
microflares rather than coronal processes \citep[][and references
therein]{Drake00}.  The high ONC temperatures also indicate that stellar
flares during their formative years are considerably hotter than in the
later main sequence phase.  This extends a similar earlier finding
among main sequence stars \citep{Gudel97}.

The interstellar column densities derived from X-ray spectral fitting
(Figure \ref{xprop_hist_fig}, right panel) are not an intrinsic
property of PMS X-ray emission, but rather reflect the location of each
star in relation to the blister HII region and the bulk of molecular
cloud material behind the HII region.  The median $\log N_H = 21.7$
cm$^{-2}$ and most values lie in a lognormal distribution with a FWHM of
1.6 in $\log N_H$, but about one tenth of the sources suffer no
detectable absorption with $\log N_H < 20.0$ cm$^{-2}$.  Other sources
have absorptions equivalent to $A_V \sim 10-100$; these are likely to
include very young protostars recently emerged from the active star
forming molecular cores, many of which are previously unidentified (see
\S \ref{noid_sec}).

\section{X-ray emission along the Initial Mass Function
\label{IMF_sec}}

\subsection{High-mass stars \label{highmass_sec}}

It is well-accepted that X-ray emission from stars earlier than
B1.5$-$B2 arises from processes in their radiation-driven stellar
winds, in contrast to X-ray emission from lower mass T Tauri stars
which arises from magnetic reconnection activity (\S \ref{intro_sec}).
These models are supported by extensive data from the {\it Einstein}
and {\it ROSAT} satellites; for example, X-ray emission from O stars
have showed very little variability and their emission lines exhibit
Doppler broadening.  While the sample of OB stars in the ONC is small,
it is complete for low-obscuration regions.  We also have uniform
spectral and variability data with higher signal-to-noise ratios than
available from previous satellite observations.

We consider here and in \S \ref{intermed_sec} a sample of 53 ONC stars
with $M>1.5$ M$_\odot$ with $V<20$ lying in the ACIS field of view
\citep{Hillenbrand97}. These are listed in Table \ref{highmass_table},
ordered by decreasing mass.  Forty-eight are detected with ACIS and
appear in Tables \ref{src_list_table} and \ref{src_prop_table}, while
five are undetected: P 1772, JW 108, P 1892, JW 531 and JW
608\footnote{Two other undetected stars, JW 794 and JW 997, are omitted
from the sample due to low probability of ONC membership based on
proper motions \citep{Jones88}.}.  For the undetected sources,
full-band X-ray upper limits $L_{t,lim}$ were calculated as described
in \S \ref{limits_sec}\footnote{Parenago 1892, which lies in the PSF
wings of $\theta^1$C Ori, was treated manually and assigned an upper
limit of 40 source counts.} with values in the range $28.4 < \log
L_{t,lim} < 29.2$ erg s$^{-1}$. We adopt a soft-band upper limit 
\begin{equation} 
\log L_{s,lim} = \log L_{t,lim} - 0.3
\end{equation} 
for the undetected stars based on typical values seen in the detected
stars.  No significant differences are seen in the scatter plots made
using X-ray luminosities from the different bands, so we adopt the soft
band $L_s$ values to facilitate comparison with {\it Einstein} and {\it
ROSAT} studies.

Figure \ref{highmass_ls_fig} (left panel) shows the dependence of
$L_s/L_{bol}$ on mass superposed on the loci of stars reported in
previous studies.  The average of the six ONC stars with spectral types
earlier than B2 is $<\log(L_s/L_{bol})> \simeq -7.6$.  This is
understandably several fold lower than $<\log(L_s/L_{bol})> \simeq
-7.1$ found for a large sample of O stars by \citet{Berghofer97}, shown
as a dashed line in the diagram, as their value is based only on stars
detected in the shallow {\it ROSAT} All-Sky Survey and overestimates
the true mean of the underlying population.  Perhaps more important is
the wide scatter of 3 orders of magnitude about this mean for the ONC
stars.  In the standard theory of X-ray emission from many spatially
distributed shocks in the stellar wind, this scatter would be explained
by a wide range of shock filling factors \citep{Owocki99}.

However, our variability results cast doubt on the standard model for
some massive stars.  Figure \ref{highmass_var_fig} shows that most of
the 8 B2$-$O6 ONC stars which should be dominated by extended wind
emission exhibit variability within a 12-hour observation\footnote{The
fluctuations seen in the lightcurve of P 1891 = $\theta^1$C Ori may be
of instrumental origin, as these counts have been extracted from the
wings of a severely piled up ACIS source. P 1993 and P 1889 suffer mild
pileup such that the amplitudes, but not general characteristics, of
the variations may be affected.}. Indeed, the second-most massive star
in the cluster $-$ P 1993 = $\theta^2$A Ori, O9.5Vpe, V=5.1, with M=31
M$_\odot$ and time-averaged $\log L_t = 31.6$ erg s$^{-1}$ $-$ exhibits
the most dramatic X-ray variability ever recorded from an O star, with a
50\% drop in 10 hours superposed by multiple $10-20$\% flares with
$1-3$ hour durations.  The best previous case for rapid variations was
a $\Delta L_s \simeq 30$\% rise during 2 days in the $V=1.8$ O9.5Ib
supergiant $\zeta$ Ori \citep{Berghofer94b}.  Parenago 2031 (=
$\theta^2$B Ori, B1V, V=6.0, M=12 M$_\odot$, $\log L_t = 29.5$ erg
s$^{-1}$) shows a very high-amplitude but low luminosity flare similar
to many others seen from ONC T Tauri stars.  Other less dramatic cases
of intra-day variations, also at low luminosity levels, are seen in P
1889, P 2074, P 1863a and P 2085.  Except for $\theta^1$C Ori, all of
these stars have X-ray luminosities consistent with those of lower mass
cluster members (Figure \ref{highmass_ls_fig}, right panel). 

We consider three explanations for the rapid variable behavior seen in
these Trapezium B2$-$O6 stars.  \begin{enumerate}

\item Hydrodynamic calculations have shown that strong events as seen
in P 1993 can be produced in the occasional large shocks that may
propagate through a massive stellar wind \citep{Feldmeier97}.  However,
the characteristic temperature of the emitting regions is $\sim 10^6$ K
in these models, while the ACIS spectra of the Trapezium sources
require $1-2$ keV plasmas and three (P 1685, P 1993 and P 2074) show
hot components around $5-7$ keV.  These stars have rather modest winds
which may not be capable of producing sufficiently powerful shocks to
account for the X-ray flares.  For example, the wind of P 1993 has
$\log \dot{M} = -7.5$ M$_\odot$ yr$^{-1}$ and $v_\infty = 700$ km
s$^{-1}$ \citep{Howarth89}.

\item The X-ray variation and hard spectrum may be produced by a
stellar companion rather than by the massive star that dominates the
optical light.  For example, spectroscopy and speckle interferometry
have established that P 1993 is at least a triple system with a $\sim
10-15$ M$_\odot$ close secondary in an eccentric 21 day orbit and a
more distant $3-7$ M$_\odot$ companion \citep{Preibisch99}. Similarly,
P 1891 is at least a binary, P 1865 is at least a triple, P 2074 is at
least a triple, and P 1863 has at least 5 components.  Only P 1889 and
P 2031 do not have known companions among the Trapezium B2$-$O6 stars
\citep{Preibisch99}. The companion model is attractive for most of
these systems where the X-ray luminosity is $\log L_t \simeq 29-30$ erg
s$^{-1}$, similar to hundreds of other lower-mass T Tauri stars in the
ONC (Figure \ref{highmass_ls_fig}, right panel).  However, this model has
difficulty explaining the flare of P 1993 where, with time-averaged
$\log L_t = 31.6$ erg s$^{-1}$, it would be in the top $\simeq 0.2$\%
of the lower-mass ONC X-ray luminosity function.

\item The X-ray flares may arise from magnetic reconnection events near
the stellar surface of the OB stars themselves.  While OB X-ray phenomenology
is generally attributed to thermal wind rather than magnetic processes,
there is some evidence for solar-type magnetic activity on such stars.
This includes: optical spectroscopic and X-ray variability evidence for
magnetically confined plasma on the B0.5 IVe star $\gamma$ Cas
\citep{Smith99}; X-ray spectroscopic evidence for very high-density
plasma in the O9.7 Ib supergiant $\zeta$ Ori \citep{Waldron01}; and
variable nonthermal radio continuum emission from 25\% of OB stars
\citep{Bieging89}. If a sufficiently strong dipole field is present, OB
winds may be guided into an equatorial disk structure with shocks
heating the gas to X-ray temperatures \citep{Babel97}.  We note that
the column density of the wind of P 1993 should be relatively
transparent to X-ray emission near the stellar surface, with
\begin{equation}
N_H = \frac{\dot{M}}{4 \pi \mu m_p v_\infty R_*} \simeq 2 \times 10^{21}
~~{\rm cm}^{-2},
\end{equation}
assuming $\log \dot{M} = -7.5$ M$_\odot$ yr$^{-1}$, $v_\infty = 700$ km
s$^{-1}$, $\mu=1.3$, $R_* = 8$ M$_\odot$, unity filling factor and an
isotropic geometry.

\end{enumerate}

We tentatively reach the following conclusions.  The three O stars
exhibit X-ray properties consistent with the strong and constant
emission expected from distributed shocks in line-driven stellar
winds.  During one of the two observations, however, the O9.5 star P
1993 exhibited a remarkable rapid flaring behavior.  From the
discussion above, perhaps the most reasonable explanation is that the
lower constant level seen in the October 1999 exposure represents the
underlying emission from the O star wind, while the April 2000 flare
arises from a magnetic process (either reconnection event or shock from
magnetically funneled wind material) near the base of the P 1993 wind.
The emission from early B stars, despite previous reports that they lie
on a $L_x/L_{bol} \simeq 10^{-7}$ locus associated with wind emission,
generally exhibits rapid variability and lower X-ray luminosities
similar to that commonly seen in ONC T Tauri stars.  Their X-ray
emission thus likely arises from lower mass companions. The wind
emission from B0$-$B2 stars themselves thus probably has been
undetected and lies below $\log L_s < 29$ erg s$^{-1}$ and their
$L_x/L_{bol} < -8$ or even $< -9$.

\subsection{Intermediate-mass stars \label{intermed_sec}}

The source of X-rays from late B and A type stars, which have neither
strong winds nor outer convective zones conducive to a magnetic dynamo,
has been the subject of some concern (\S \ref{intro_sec}).  While some
researchers have argued that the emission arises from late-type
companions, others call this model into question.  The hypothesis is
more readily testable in a PMS population like the ONC than in the
field main sequence stars that are usually examined, as the T Tauri
emission is elevated and more easily studied in very young stars.

Figure \ref{highmass_ls_fig} (left panel) compares the distribution of
$L_s/L_{bol}$ for ONC BA stars with the loci of stars from past
studies: the regression line for B stars detected in the {\it ROSAT}
All-Sky Survey \citep[dashed line;][]{Berghofer97}, a pointed {\it
ROSAT} survey of mid-B stars \citep[lower open region;][]{Cohen97}, and
several pointed {\it Einstein} and {\it ROSAT} surveys of late-B and A
stars \citep[upper open region;][]{Caillault89, Berghofer94a,
Zinnecker94, Simon95}.

While the $L_s/L_{bol}$ diagram appears to show a huge rise in X-ray
emissivity as one considers stars of decreasing mass, this effect is
entirely due to changes in the bolometric luminosity rather than the
X-ray luminosity.  This is clearly seen in Figure \ref{highmass_ls_fig}
(right panel) which plots $L_s$ against mass.  Here we see that the
distribution of X-ray luminosities is virually unchanged from spectral
types F5 ($M \simeq 1.5$ M$_\odot$) through B0 ($M \simeq 20$
M$_\odot$) with a mean $<\log L_s> \simeq 30.4$ erg s$^{-1}$, and
remains at a similar level for the $0.7 < M < 1.4$ M$_\odot$ mass range
where the emission clearly arises from magnetic flaring \citep[\S
\ref{lowmass_sec} and ][]{Feigelson02a}.

Although our findings do not conclusively exclude intrinsic X-ray
emission from late-B and A stars, the ONC intermediate mass star
properties can be fully attributed to G and F companions\footnote{A
{\it Chandra} ACIS study of the central region of the Pleiades cluster
finds that sources associated with B6$-$F4 stars have high fluxes,
non-variable light curves, and soft hardness ratios which point to
intrinsic emission by the intermediate mass stars not by low mass
companions \citep{Krishnamurthi01}.  However, their result is based on
only four stars in this mass range and is considered tentative.}.  Note
that the level of X-ray emission we see in the late-B and A stars
cannot be explained by lower mass K and M companions.  This implies
either that each intermediate-mass star is preferentially formed with a
star with higher than average mass, or is accompanied by several
companions, one of which is likely to be a G or F star.  The presence of
F5$-$A0 ONC stars with somewhat stronger X-ray emission ($\log
L_s/L_{bol} \simeq -4$) than reported in {\it ROSAT} studies is likely
due to the higher X-ray emission in ONC PMS compared to the main
sequence companion stars that dominate the earlier samples.  We cannot
address here whether Herbig Ae/Be stars produce extra X-rays than
ordinary young A/B stars, as there is no well-established subsample of
Herbig Ae/Be stars in the ONC.

\subsection{Low mass stars \label{lowmass_sec}}

Figure \ref{lowmass_ls_fig} shows the distribution of soft X-ray flux
as a function of mass for $M<1.5$ M$_\odot$ stars in the $V<20$ sample
of \citet{Hillenbrand97}\footnote{Versions of Figures
\ref{highmass_ls_fig} and \ref{lowmass_ls_fig} based on {\it ROSAT}
observations of the ONC and its vicinity are given by \citet{Gagne95}.
They show some of the effects discussed here, with a larger sample of
high luminosity sources due to a wider field of study, but with a
factor $\simeq 100$ lower sensitivity to low luminosity sources than
achieved here.}.  Whereas for higher mass stars $L_s/L_{bol}$ varies
with mass and $L_s$ was invariant, the opposite pattern is seen here.
Over the mass range $0.1 < M < 1$ M$_\odot$, the fraction of bolometric
energy emerging in the X-ray band is invariant with mean and standard
deviation $< L_s/L_{bol} > = -4.2 \pm 0.6$, while the X-ray luminosity
rises steeply with mass.  The behavior of these relations for higher
mass stars was explained by the inappropriate use of the easily
measured $L_{bol}$ value of the massive companion rather than the
unavailable $L_{bol}$ value of a lower mass companion (\S
\ref{highmass_sec}).  For low mass stars, it is likely that the star
that dominates the optical luminosity $L_{bol}$ also dominates the
X-ray luminosity $L_s$, so that the constancy of $L_s/L_{bol}$ for a
wide range of low mass stars should be astrophysically meaningful.  A
steep $L_x-$mass relation was seen in $ROSAT$ study of the Chamaeleon
I cloud \citep{Feigelson93}.

A constant value of $L_s/L_{bol}$ is usually interpreted as a constant
X-ray surface flux\footnote{The quantities $L_s/L_{bol}$ and $F_s$ are
related to each other according to $F_s = \sigma T_{eff}^4
(L_s/L_{bol})$ where $\sigma$ is the Stefan-Boltzmann constant and
$T_{eff}$ is the effective temperature given in Table
\ref{src_list_table}.  $\log F_s$ and $\log L_s/L_{bol}$ do not
differ by more than $\pm 0.3$ for ONC stars in the $0.1 < M < 1$
M$_\odot$ mass range, and correlation plots of low mass ONC stars using
the two variables look very similar.} so that X-ray luminosity $L_s$
scales with the surface areas of different stars.  Although this
accounts for the general behavior of stars in Figure
\ref{lowmass_ls_fig}, we recognize that the average ONC low mass star
has a $L_s/L_{bol}$ value an order of magnitude below the `saturation'
value around $L_s/L_{bol} \simeq -3.0$ seen in samples of main sequence
G$-$M stars \citep[e.g.][]{Vilhu87, Fleming95, Randich97}.  This result
is not unique to the ONC:  sufficiently sensitive {\it ROSAT} studies
of nearby young stellar clusters showed a similar effect.  We conclude
that low mass T Tauri X-ray emission appears to scale with stellar
surface area but, if the mechanism is similar to that in main sequence
stars, in most T Tauri stars the magnetic activity may saturate at a
level $\simeq 10$ times lower than in main sequence stars.  This issue
is discussed further in \citet{Feigelson02b}.

The X-ray spectral characteristics of low mass ONC stars also confirm
results obtained in earlier work, though with some additional
insights.  Both the Sun and late-type stars exhibit a scaling between
plasma temperature and X-ray emission, roughly $L_s \propto T^{3 \pm
1}$, that emerges from simple models of plasma heated in magnetic loops
\citep{Rosner78}.  Figure \ref{lowmass_kt_fig} shows such an
association as a rise in the lower envelope of the $kT$ distribution
with increasing X-ray luminosity, which agrees with the locus found
with {\it ROSAT} for magnetically active late-type stars \citep[][heavy
dashed line]{Preibisch97}.  The effect is also present, though less
clearly, in a $L_s/L_{bol}$ $vs.$ $L_t$ diagram. While the majority of
sources follow the standard $L_x-T$ correlation, 10\% of the ONC
sources have fitted plasma energies $kT > 10$ keV and another $\sim
10$\% have energies considerably higher than expected from the standard
$L_x-T$ relation.  These temperatures are too high to have been
measured with {\it ROSAT}.  Such ultra-hot plasmas have been found in
{\it ASCA} studies during powerful T Tauri and protostar flares
\citep[e.g.][]{Koyama96, Tsuboi98} but have not been previously
reported for T Tauri stars with relatively constant lightcurves and
ordinary luminosities around $\log L_t \simeq 28-30$ erg s$^{-1}$.

We find no clear pattern in the properties of these ultra-hot ONC
stars.  While a few are attributable to unusually violent flares, most
of these stars are deeply embedded with average X-ray luminosities.
Many may also have soft components that we can not observe, similar to
the ONC stars with two-temperature spectral fits which include a hot
component above 5 keV.  There is no evidence for the simple solar-type
model of a hotter, high-luminosity, high-variability `flare' component
superposed on a cooler, low-luminosity, low-variability `coronal'
component.  We conclude that T Tauri stars of all types can produce
ultrahot plasmas, even at modest X-ray luminosities.

\subsection{Very low mass objects (brown dwarfs) \label{bd_sec}}

The ONC is perhaps the best laboratory available to study the magnetic
activity of PMS brown dwarfs (BDs) as over 100 such objects have been
found in recent deep near-infrared imaging of the cluster
\citep{Hillenbrand00, Luhman00, Lucas00}.  As noted in Paper I,
relatively few of these young (proto) BDs appeared in the first ACIS-I
image.  Here we consider the merged ACIS dataset and discuss in detail
the frequency and properties of X-ray detected BDs (see \S
\ref{intro_sec}).

Table \ref{bd_table} lists, in right ascension order, the 30 ACIS
sources associated with very low mass (VLM) ONC objects.  By a
considerable factor, this is the largest sample of X-ray detected PMS
VLM objects yet obtained; previously samples are reported by
\citet{Neuhauser99}, \citet{Imanishi01} and \citet{Preibisch01}.  An
asterisk in column 1 indicates that the source lies in the central
$5\arcmin \times 5\arcmin$ region with deep $JHK$ coverage by
\citet{Hillenbrand00}.  Columns $3-6$ give $K$ and $H-K$ apparent
magnitudes from \citet{Hillenbrand00} and
\citet{Carpenter01}\footnote{Magnitudes for 2MASS sources not
identified as infrared variables were not published in the tables of
\citet{Carpenter01} but can be found at
\url{http://astro.caltech.edu/\~jmc/papers/variables\_orion}},
photometrically dereddened $M_K$ absolute magnitudes, and corresponding
masses from Figure 8 of \citet{Hillenbrand00}, assuming no $K$-band
excess from a disk and ages between 0.1-1 Myr.  If the age is older,
the mass would be larger than the listed value, while if an excess is
present the mass would be smaller than the listed value.  For
comparison, masses estimated from the optical spectroscopy of
\citet{Hillenbrand97} are given in table notes.

We emphasize the difficulty in establishing the masses of young
pre-main sequence stars when only near-infrared photometry is
available.  For example, ACIS sources associated with PSH 116 and H
5096 \citep[found by][]{Garmire00} lie considerably above the stellar
boundary in the K $vs.$ $H-K$ diagram but have spectroscopic
temperatures corresponding to $M \simeq 0.05$ M$_\odot$ PMS BDs.  Such
misleading infrared magnitudes and colors may be attributed to
circumstellar disks.  In contrast, an optical or near-infrared spectrum
with type $\sim$M6 or later places objects securely on a sub-stellar BD
mass track for ages $<$1 Myr \citep{Burrows97}, even accounting for
uncertainty in both the empirical measurements (surface temperatures
from spectroscopy and bolometric luminosities from reddening-corrected
photometry and a bolometric correction) and the theoretical tracks.  We
acknowledge these uncertainties by adopting the neutral label `very low
mass objects' (VLM objects) rather than `candidate BDs' and `BDs'.
Despite these cautions, the preponderance of evidence indicates that
most of the objects listed in Table \ref{bd_table} will never undergo
hydrogen ignition and thus are {\it bona fide} PMS BDs.

Columns $7-11$ of Table \ref{bd_table} reproduce X-ray properties given
in Table \ref{src_prop_table}.  The bolometric luminosity  values used
in the final column are estimated from
\begin{equation}
\log L_{bol} \simeq 0.4 (M_{\odot,bol}- M_K - BC_K) + \log L_{\odot,bol} 
\hspace{0.2in}{\rm erg~s}^{-1},
\end{equation}
where $BC_K \simeq 2.9$ for a dwarf star with spectral type M7 
\citep{Leggett92, Leggett01}.  If a disk $K$-band excess is present, the 
true value of $L_t/L_{bol}$ is larger than the listed value.  

The distinctive X-ray characteristic of the 30 X-ray detected PMS VLM
objects in the ONC is their faintness:  only 7\% (2/30) have total band
time-averaged luminosities $\log L_t \geq 29.5$ erg s$^{-1}$ compared
to 37\% (136/369) of well-characterized $-1.0 < \log M < -0.5$
M$_\odot$ ONC stars.  However, when considered in terms of X-ray
luminosity per unit bolometric luminosity (or, nearly equivalently,
X-ray flux per unit surface area), the VLM objects are X-ray luminous:
53\% (16/30) have $\log L_t/L_{bol} \geq -3.5$ compared to 34\%
(124/369) for the low-mass PMS stars.  Several of the VLM objects lie
above the $\log L_t/L_{bol} \simeq -3.0$ `saturation' level for
late-type stars.  These strongest emitters are typically caught during
a flare, shown in Figure \ref{bd_var_fig}.  VLM flaring is comparable
in frequency and morphology to flares from similarly weak low-mass ONC
stars.  Most have several hours of elevated emission and are often
truncated by the limited duration of the observation.  Spectral
characteristics are also similar to the general ONC low mass
population: the VLM objects are about equally divided between light and
heavy absorption; and the plasma temperatures range from $<$1 to $>$10
keV.

Comparing the $\log L_t/L_{bol}$ values in Table \ref{bd_table} to the
(somewhat uncertain) underlying VLM population, we find that roughly
1/4 of ONC VLM objects have X-ray emission within an order of magnitude
of the saturation level.  This is comparable to the fraction near
saturation for the lowest mass PMS stars which will evolve into late-M
main sequence stars.  It thus appears that the future turn-on of
hydrogen fusion in the core has no effect on the magnetic dynamo or
other processes leading to surface activity in PMS objects.

The situation is dramatically different in older (roughly gigayear) L-
and T-type field BDs, where surface magnetic activity traced by
H$\alpha$ emission nearly always lies $10^2$ or more below saturation
levels \citep{Gizis00}.  This decline in magnetic activity might be a
consequence of the drop of ionization fraction in the outer layers of
$M < 0.08$ M$_\odot$ objects as they descend their Hayashi tracks and
cool.  While a magnetic dynamo may still be present in the ionized
interiors of older BDs, the eruption and reconnection of surface
magnetic structures is impeded by their thick neutral atmospheres
(G.\ Basri, private communication).

\section{Summary \label{sum_sec}}

The Orion Nebula Cluster is the closest and most spatially concentrated
rich young star cluster.  It is thus the best available laboratory for
studying large samples of stars in a single CCD field, providing
impressive subsamples of stars covering all phases of PMS evolution,
masses from $<$0.05 to nearly 50 M$_\odot$,  and ages from $10^5$ to
$10^7$  yr.  The value of the present {\it Chandra} study is greatly
amplified by extensive studies of the stellar population in the optical
and infrared bands. With 1075 CXOONC sources detecting nearly all
low-obscuration $V<20$ stars, and a considerable number of embedded
stars, we present here the largest and most homogeneous sample of PMS
stars yet studied in the X-ray band.  The ACIS detector provides
variability and spectral properties as well as precise positions and
broad-band X-ray luminosities.

\begin{figure}
\centering
  \begin{minipage}[t]{1.0\textwidth}
  \centering
  \includegraphics[width=4.5in]{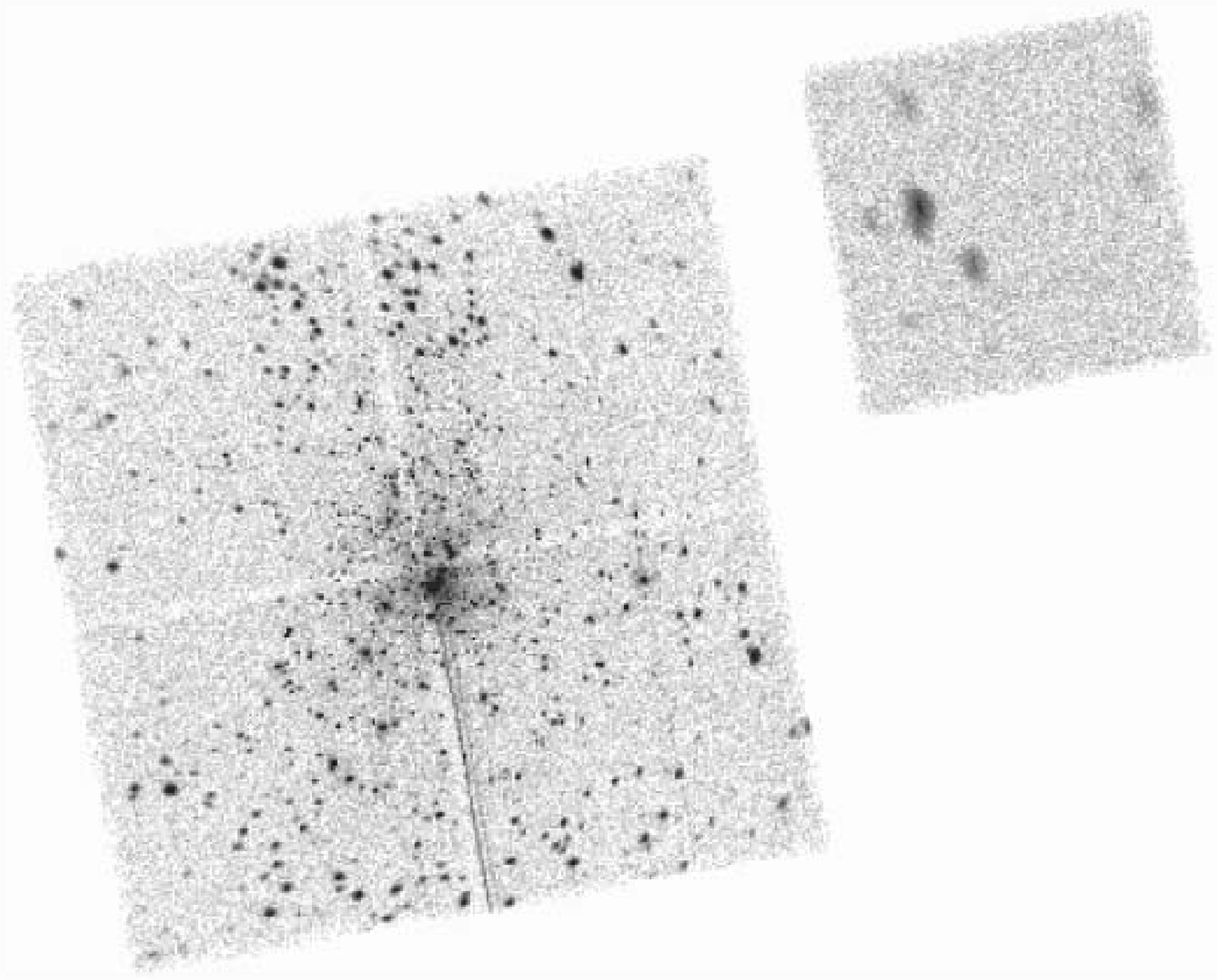}
  \end{minipage} \\ [0.3in]
  \begin{minipage}[t]{1.0\textwidth}
  \centering
  \includegraphics[width=4.5in]{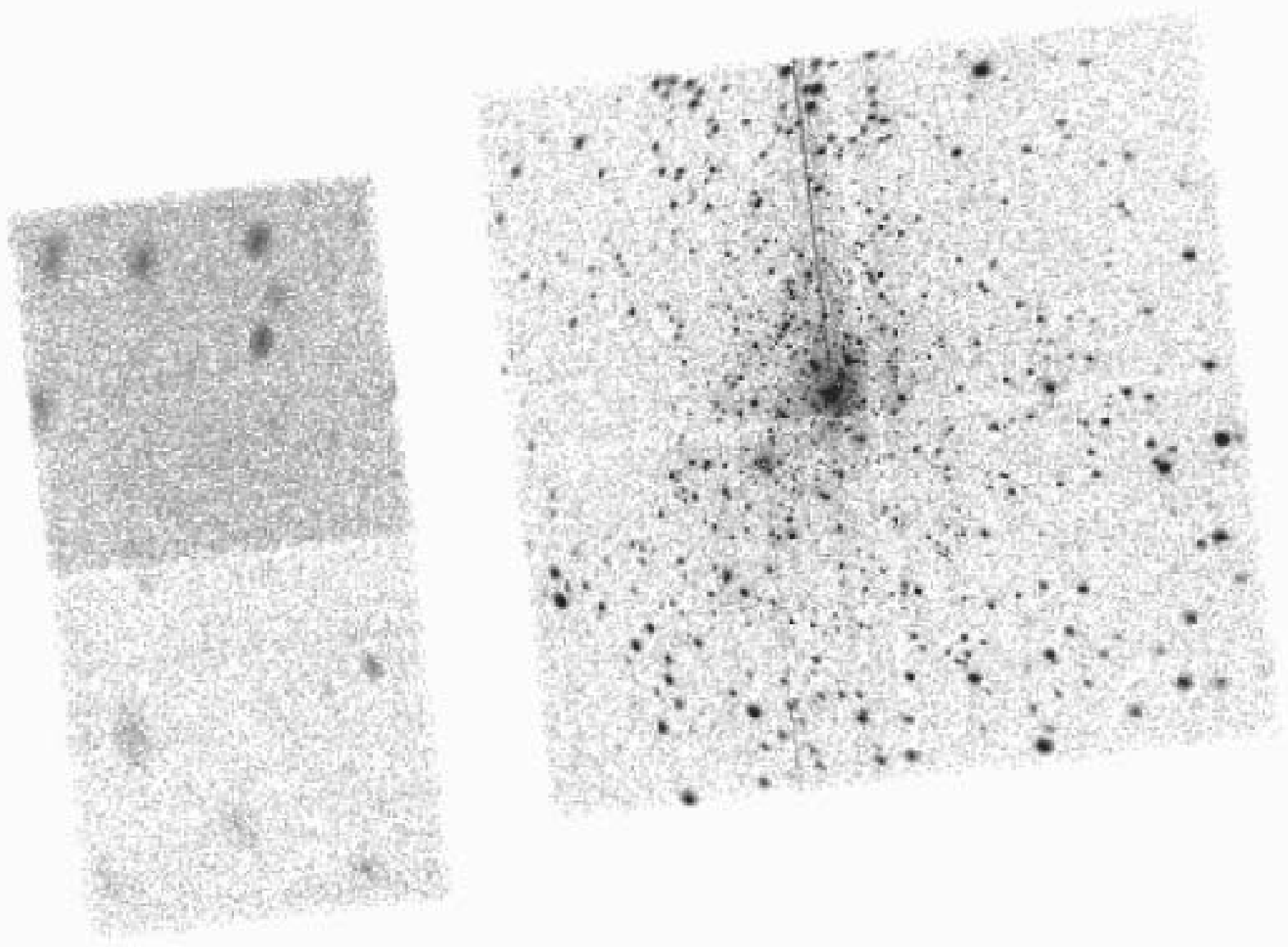}
\caption{ Low resolution views of the (top) October 1999 and (bottom)
April 2000 ACIS-I observations of the Orion Nebula Cluster after data
selection.  North is up and East is to the left. Gray hues are scaled
to the log of the counts in each 4\arcsec\/ element. Results from the
spectroscopic array chips outside of the square ACIS-I array are not
discussed in this or the accompanying studies.
\label{fullfields_fig}}
  \end{minipage}
\end{figure}

\begin{figure}
\centering
  \includegraphics[angle=-90.,scale=0.45]{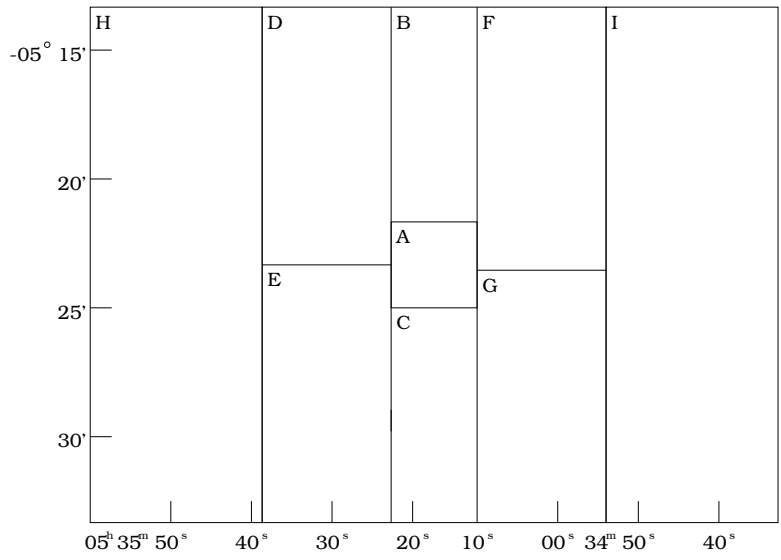}
\caption{Guide to Figure \ref{image_fig}.}
\label{guide_fig}
\end{figure}

\begin{figure}
\centering
  \includegraphics[scale=0.8]{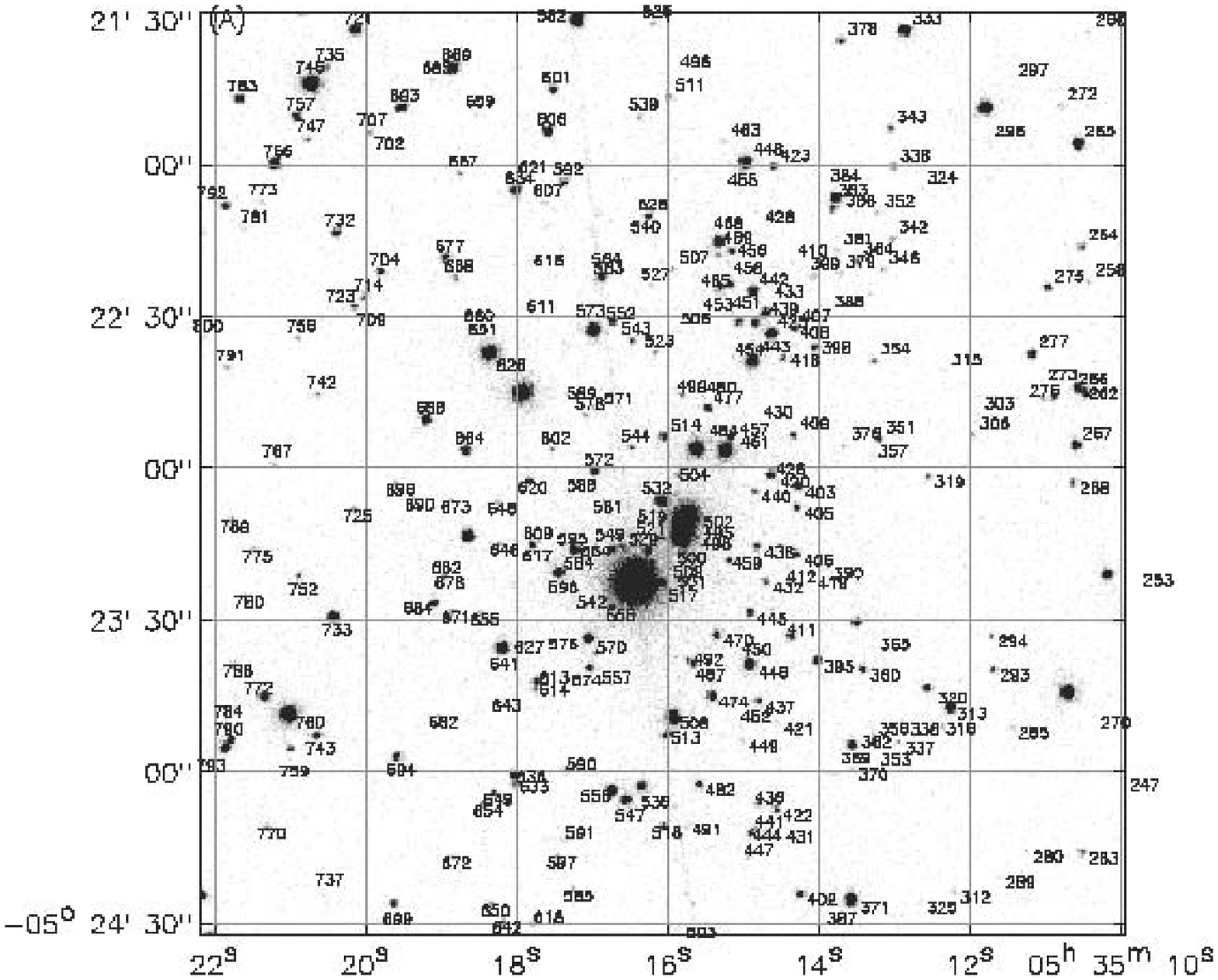}
\caption{(a)-(i) Expanded view of the merged ACIS image with sources 
indicated.}
\label{image_fig}
\end{figure}

In addition to an X-ray atlas of the region (Figures
\ref{fullfields_fig}-\ref{image_fig}), detailed description of our data
analysis (\S \ref{obs_sec}, Table \ref{obs_table}, Figures
\ref{offsets_fig}-\ref{avnh_fig}), and a comprehensive database of X-ray
sources and properties (Tables \ref{src_list_table} and
\ref{src_prop_table}, available in full in the electronic edition), we
present some of the many results that will emerge from these
observations.  Other results will appear in forthcoming papers
\citep[e.g.][]{Feigelson02a, Feigelson02b}.  \begin{enumerate}

\item We detect 1075 X-ray sources with sub-arcsecond on-axis
(arcsecond off-axis) precision in absolute celestial positions (\S
\ref{detect_sec}).  The limiting sensitivity is 9 counts on-axis
($15-30$ counts off-axis) corresponding to a limiting luminosity of
$\log L_t = 28.0$ erg s$^{-1}$ on-axis for a lightly absorbed star in
the total $0.5- 8$ keV band (\S \ref{lum_sec}-\ref{limits_sec}).
Relatively few sources are seen near the detection limit: only 12\%
have $<20$ extracted counts compared to 47\% with $>100$ counts and 9\%
with $>1000$ counts.  The X-ray catalog is estimated to be 99\% reliable
(\S \ref{spurious_sec}). 

\item Ninety-one percent of the ACIS sources are confidently associated
with young Orion stars catalogued in optical and near-infrared surveys
(\S \ref{ids_sec}, \S \ref{noid_sec}, Tables \ref{src_list_table} and
Figure \ref{offsets_fig}). While most are members of the ONC that
ionizes the Orion Nebula, both massive and low-mass members of deeply
embedded populations around the OMC 1, OMC 1S and OMC 1N molecular
cloud cores (but not the Orion Bar) are seen.  Some of these were
previously catalogued but others are discovered in the ACIS image (Table
\ref{noid_table} and Figure \ref{noid_fig}). Hundreds of additional
sources, mainly deeply embedded and VLM stars, should emerge in deeper
{\it Chandra} exposures of the region.  One of the unidentified
sources, CXOONC 053524.6-052759, has an unusual combination of high and
constant flux, hard spectrum, and no stellar counterpart.  It may be a
background transient neutron star binary system in quiescence.

\item The X-ray luminosity function of the full sample is approximately
lognormal with mean and standard deviation $<\log L_t> = 29.4 \pm 0.7$
erg s$^{-1}$ and X-ray to bolometric ratio $<\log L_t/L_{bol}> = -3.9 \pm
0.7$ (\S \ref{global_sec} and Figure \ref{xlum_hist_fig}). The
corresponding values for the full underlying population of ONC and
Orion cloud stars is uncertain, as the means will drop with inclusion
of undetected stars but increase with compensation for low energy
absorption. Half of the observed X-ray emission is produced by the
luminous O6 star $\theta^1$C Ori, and half by the remaining 1074
sources.

\item More than half of PMS stars with $>50$ counts show intraday flux
variations, often exhibiting dramatic flaring on timescale of 2 to
$>12$ hours, in our 13 and 10 hour observations (\S \ref{global_sec}
and Figure \ref{xprop_hist_fig}).  Half of the remaining sources showed
different flux levels in the observations separated by 6 months.

\item There is no indication that the presence or absence of a
circumstellar disk significantly affects the X-ray luminosities of PMS
stars (Figure \ref{xlum_hist_fig}, middle panel).

\item The X-ray emission from 5 of the 7 B2$-$O6 stars constituting the
Orion Trapezium shows surprising variability (\S \ref{highmass_sec},
Table \ref{highmass_table} and Figures \ref{highmass_ls_fig} and
\ref{highmass_var_fig}).  The $M \simeq 31$ M$_\odot$ O9.5 star
$\theta^2$A Ori (P 1993), in particular, exhibited multiple flares on
timescales of hours during one observation.  This result either
requires an extraordinary flare from a lower mass stellar companion, or
revision of the conventional model of OB stellar X-ray production in a
myriad small-scale wind shocks.  The X-ray emission from B0$-$B2 stars
can be attributed to T Tauri companions, and their intrinsic emission
is much weaker than predicted by the long-standing $\log L_s/L_{bol} = -7$
relation for wind-dominated OB stars.

\item The X-ray emission from intermediate-mass stars with spectral
types from mid-B through A is consistent with emission from lower mass
companions, although it requires that most stars have companions of a
solar mass or greater(\S \ref{intermed_sec}, Table
\ref{highmass_table}, Figure \ref{highmass_ls_fig}).  This supports
well-accepted views that mid-B to A stars themselves are X-ray quiet,
as they are insufficiently luminous to radiatively accelerate massive
winds and lack outer convection zones that generate magnetic activity
via a dynamo.

\item The average low mass G$-$M PMS star exhibits a moderate level of
X-ray emission with $<\log L_s/L_{bol}> = -4.2$, an order of magnitude
below the `saturation' level seen in magnetically active main sequence
stars (\S \ref{lowmass_sec}, Figures \ref{lowmass_ls_fig}$-
$\ref{lowmass_kt_fig}).  In contrast, plasma energies of PMS stars are
often remarkably high, with $kT \simeq 5$ to $>$10 keV ($T \simeq 60$
to $>$120 MK) components often dominating the spectrum of even
low-luminosity T Tauri stars.  These high temperatures appear to
violate the standard $L_x-T$ relation seen in the Sun and magnetically
active stars.  Abundance anomalies may also be present in many of the
brighter sources, consistent with recent high-resolution spectroscopic
studies of nearby older magnetically active stars.  Implications for
the astrophysical origins of low mass PMS X-ray emission based on the
absence of a statistical X-ray/rotation relation are discussed in
\citet{Feigelson02b}.

\item We present the largest sample to date of X-ray detected very low
mass pre-main sequence objects, most of which will probably evolve into
brown dwarfs rather than stars (\S \ref{bd_sec}, Table \ref{bd_table},
Figure \ref{bd_var_fig}).  Though typically having low X-ray
luminosities near our detection limit, the detected objects have X-ray
surface fluxes near the saturation level $\log L_t/L_{bol} \simeq -3$ and
exhibit flaring.  The underlying VLM population appears to have X-ray
properties similar to the lower mass PMS stars, indicating that the
processes giving rise to magnetic activity in the PMS phase are
independent of whether hydrogen burning will eventually turn-on in the
stellar core.  Magnetic activity appears to decline as the VLM objects
evolve into older brown dwarfs, which is attributable to the drop in
ionization fraction in their cooling atmospheres.

\end{enumerate}

\acknowledgments We thank: George Chartas and Konstantin Getman (Penn
State) for expert assistance with data analysis; William Herbst
(Wesleyan) for access to rotation data prior to publication;
participants of the 12th Cool Star Workshop (including Gibor Basri of
Berkeley, James Liebert of Arizona and Hans Zinnecker of Potsdam) for
thoughtful comments; advice from Deepto Chakrabarty (MIT), David Cohen
(Swarthmore) and George Pavlov (Penn State) concerning unusual sources;
and Thierry Montmerle (Saclay) and the anonymous referee for helpful
critiques of the entire manuscript.  This work was principally
supported by NASA grant NAS 8-38252 (Garmire, PI).  Additional support
was obtained from the Japan Society for the Promotion of Science (YT)
and the Jet Propulsion Laboratory, California Institute of Technology,
under contract with the National Aeronautics and Space Administration
(SHP).  The study benefited from on-line databases including SIMBAD,
Astronomical Data Center and the Astrophysics Data System.

\newpage

\bibliography{aj-jour} 

\clearpage
\newpage

\clearpage
\newpage

\begin{figure}
\centering
\includegraphics[scale=0.7]{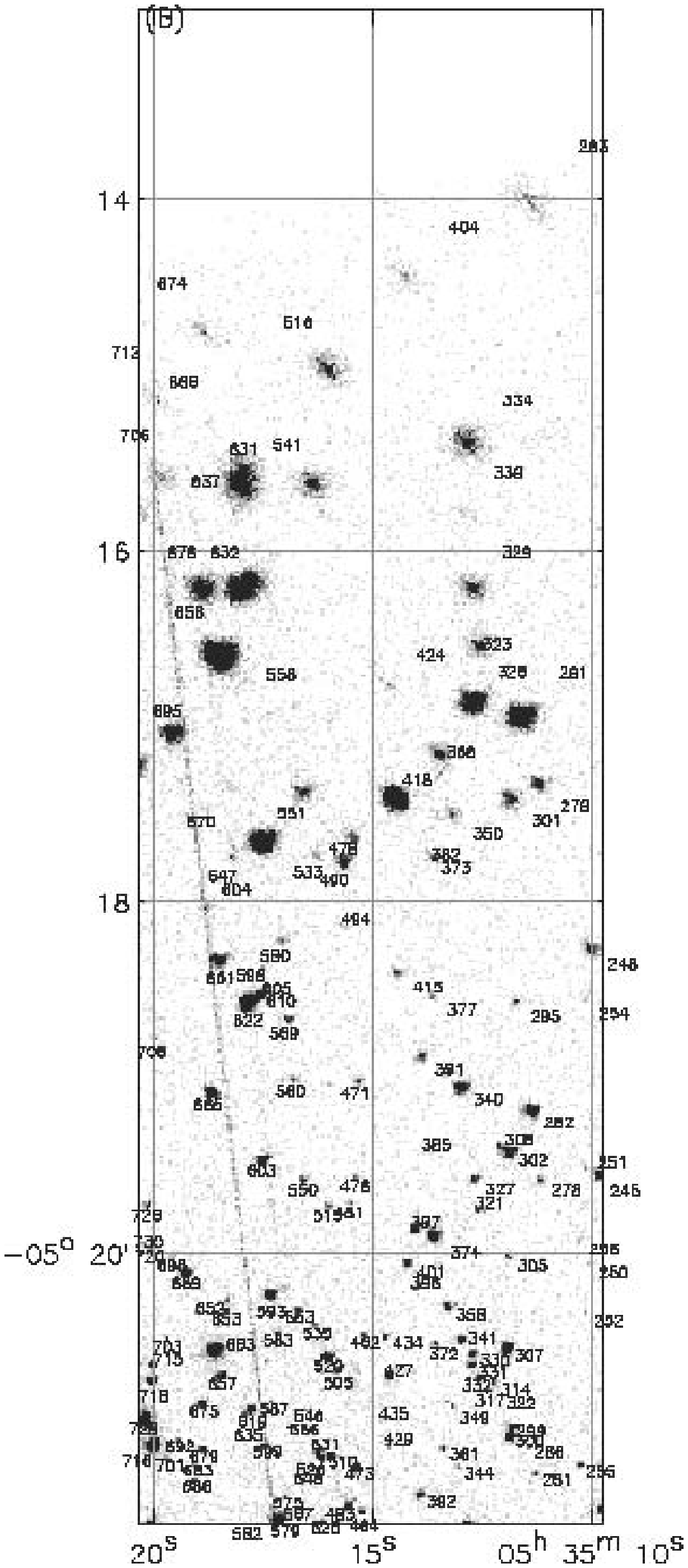}
\end{figure}

\clearpage
\newpage

\begin{figure}
\centering
\includegraphics[scale=0.7]{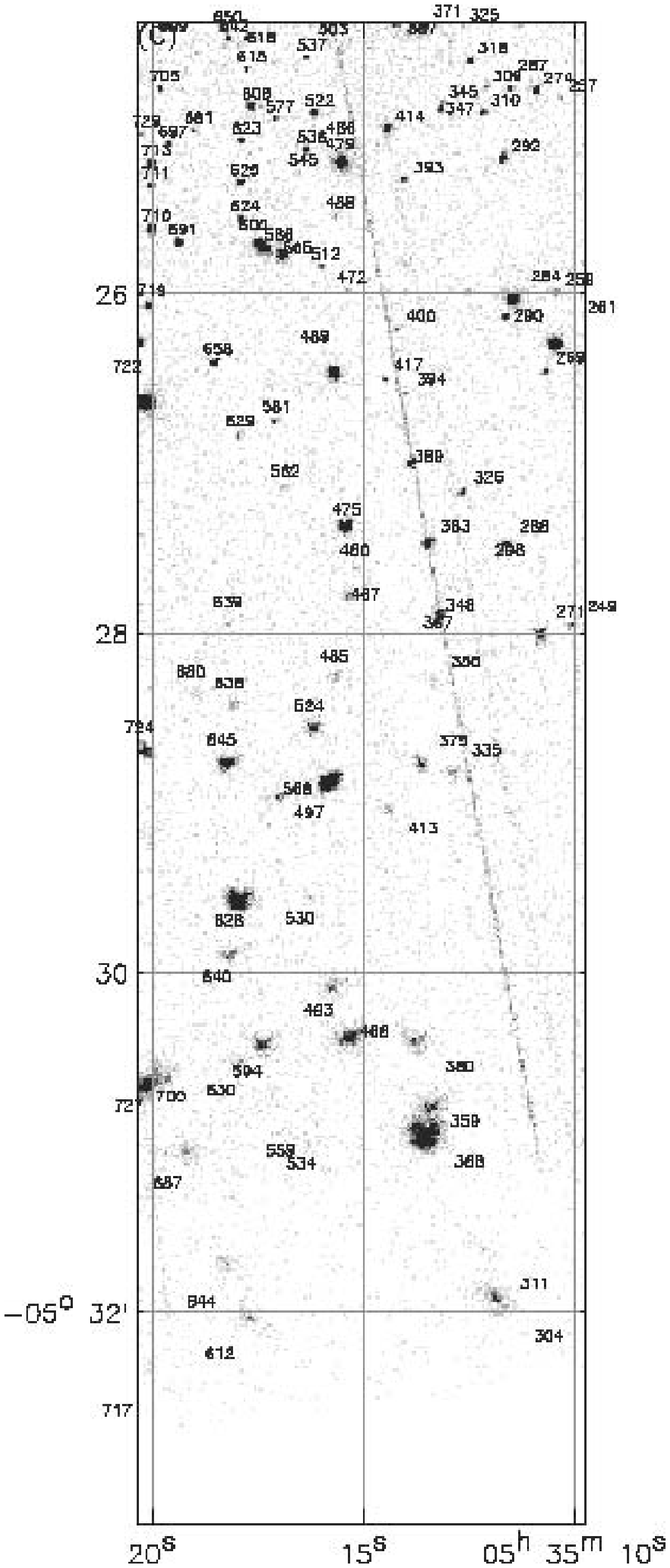}
\end{figure}

\newpage
\begin{figure}
\centering
\includegraphics[scale=0.6]{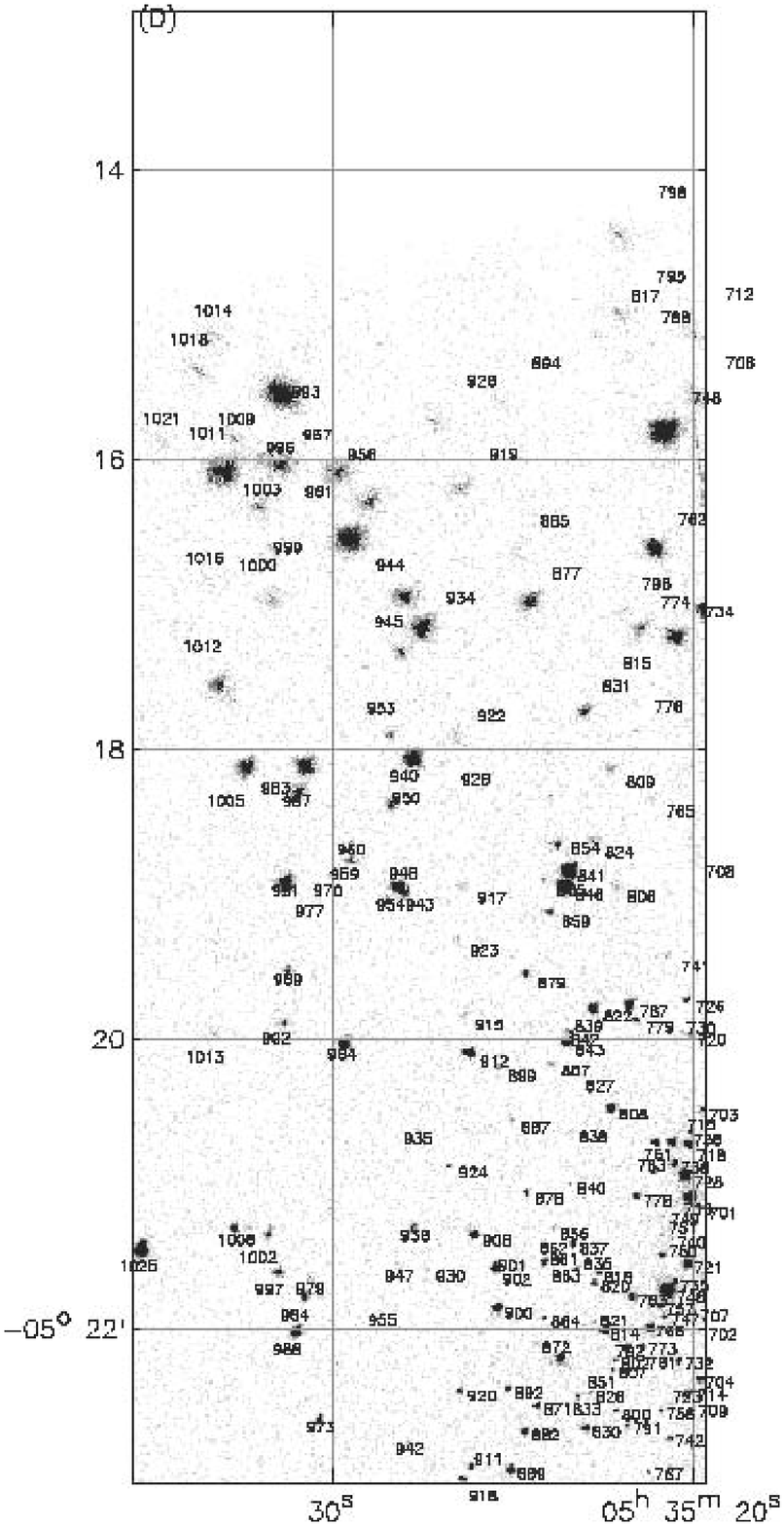}
\end{figure}

\clearpage
\newpage

\begin{figure}
\centering
\includegraphics[scale=0.6]{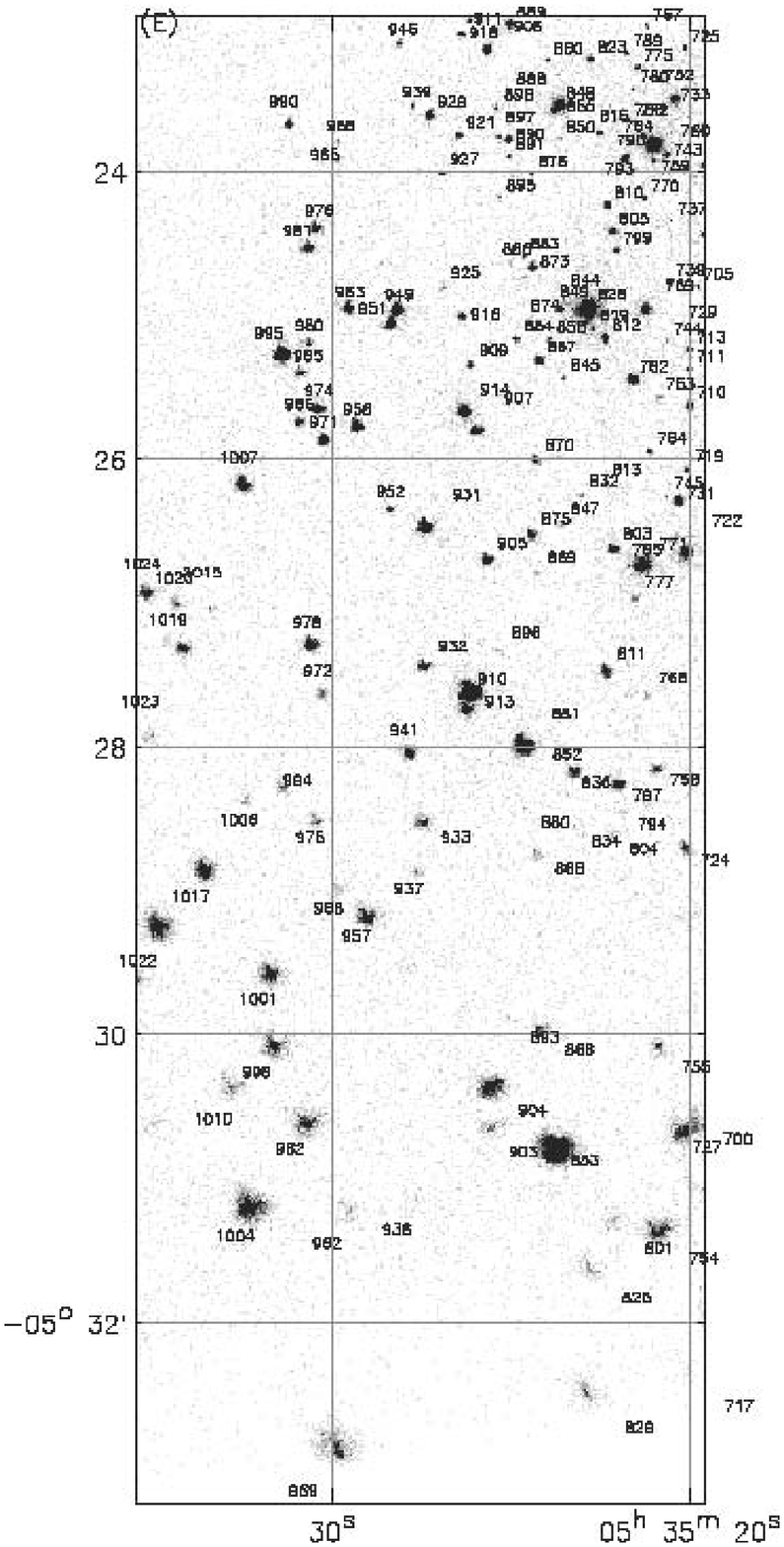}
\end{figure}

\newpage
\begin{figure}
\centering
\includegraphics[scale=0.6]{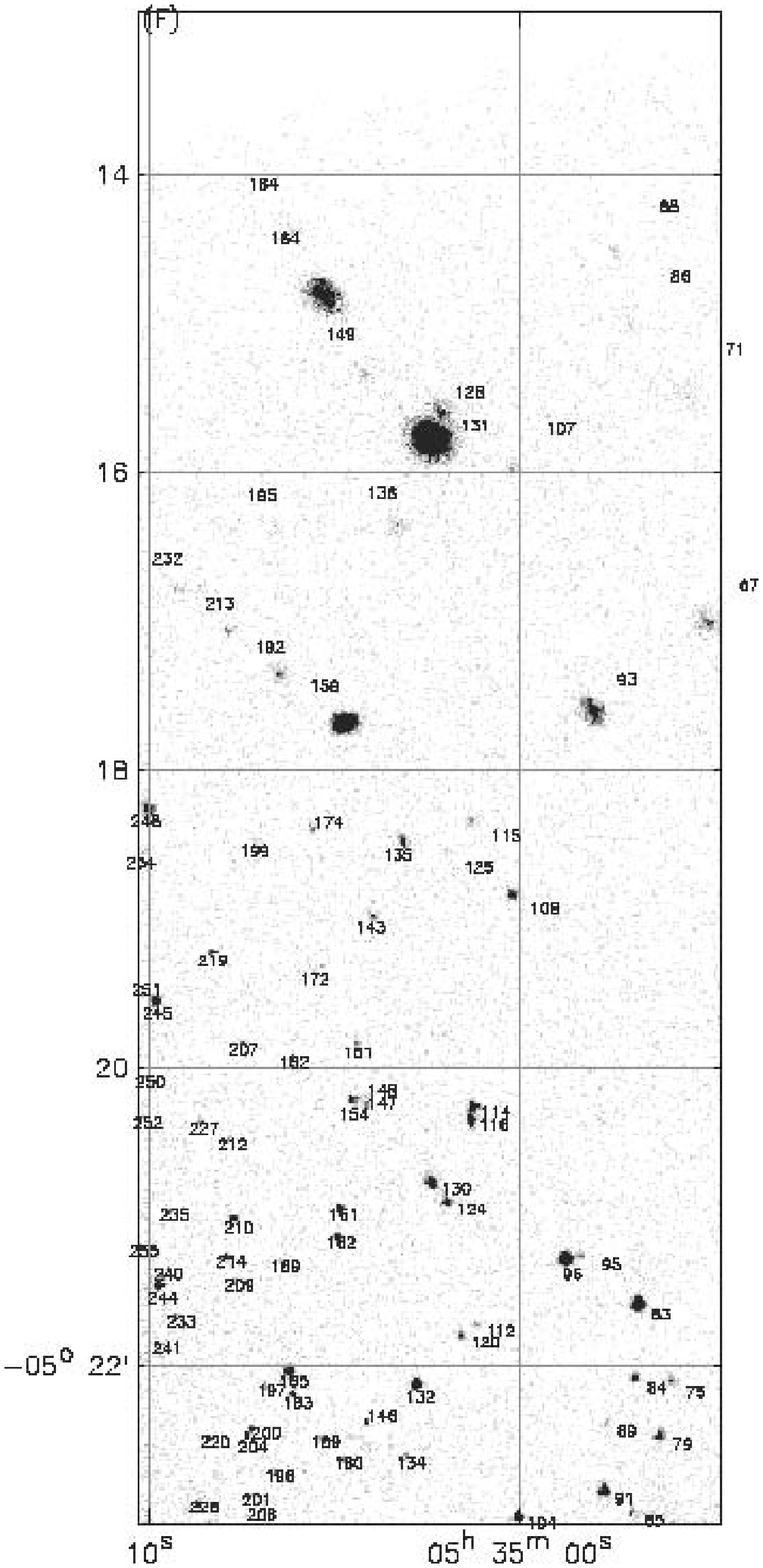}
\end{figure}

\clearpage
\newpage

\begin{figure}
\centering
\includegraphics[scale=0.6]{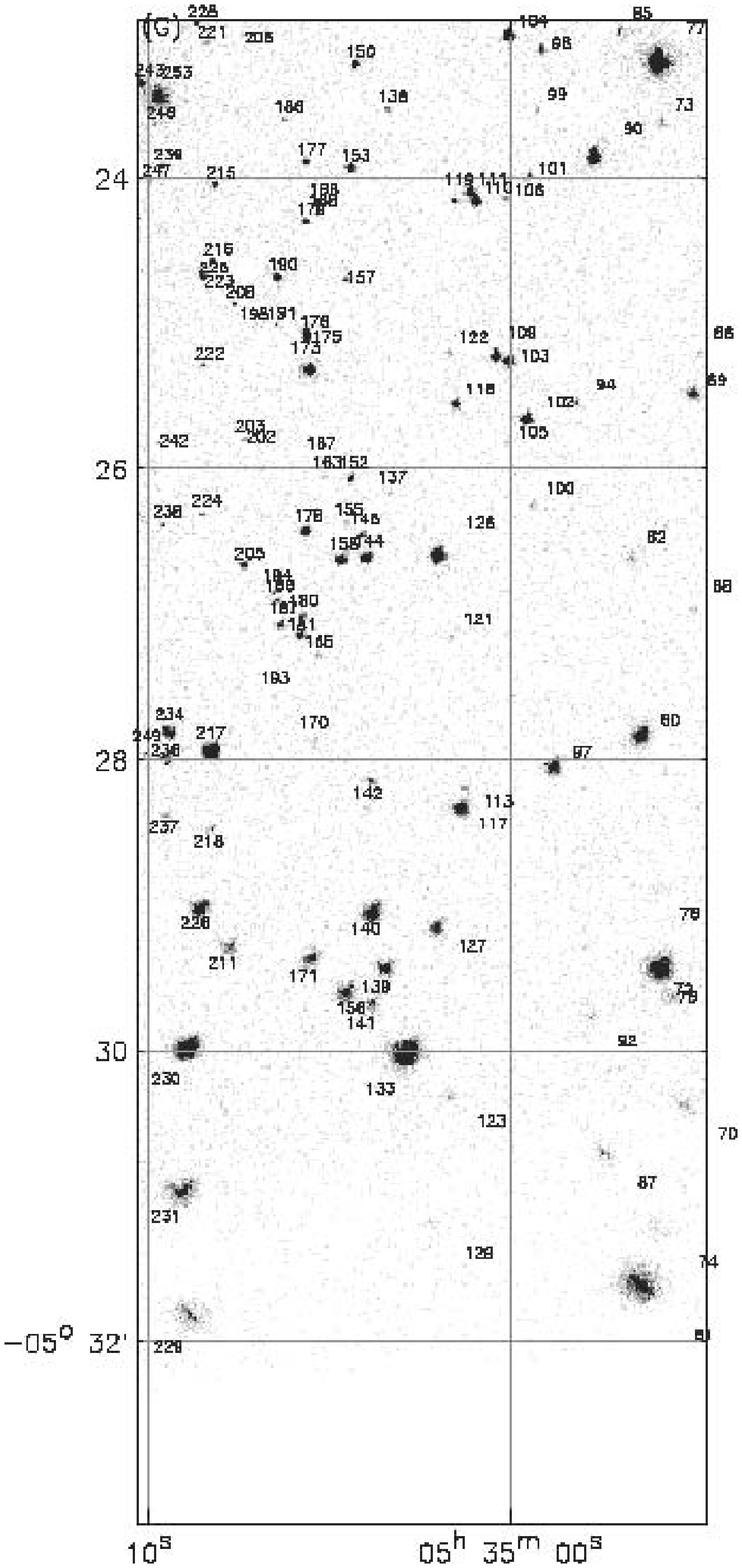}
\end{figure}

\newpage
\begin{figure}
\centering
\includegraphics[scale=0.75]{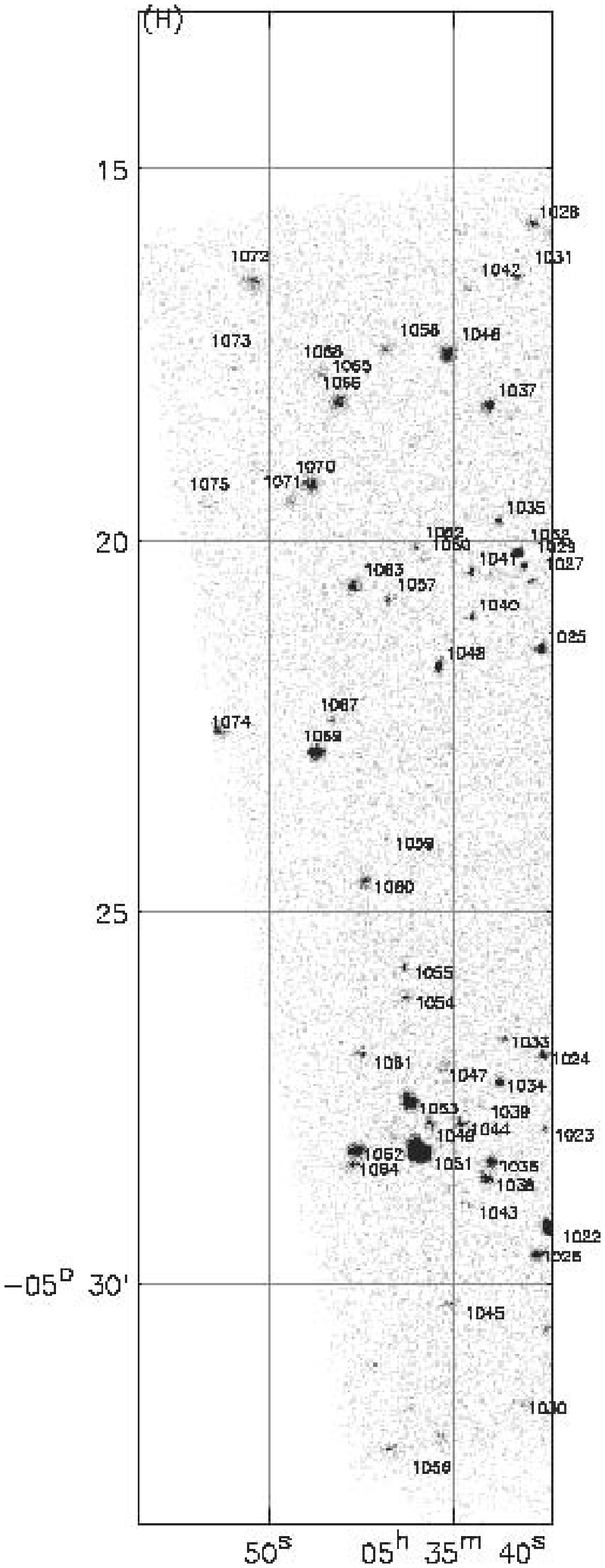}
\end{figure}

\clearpage
\newpage

\begin{figure}
\centering
\includegraphics[scale=0.75]{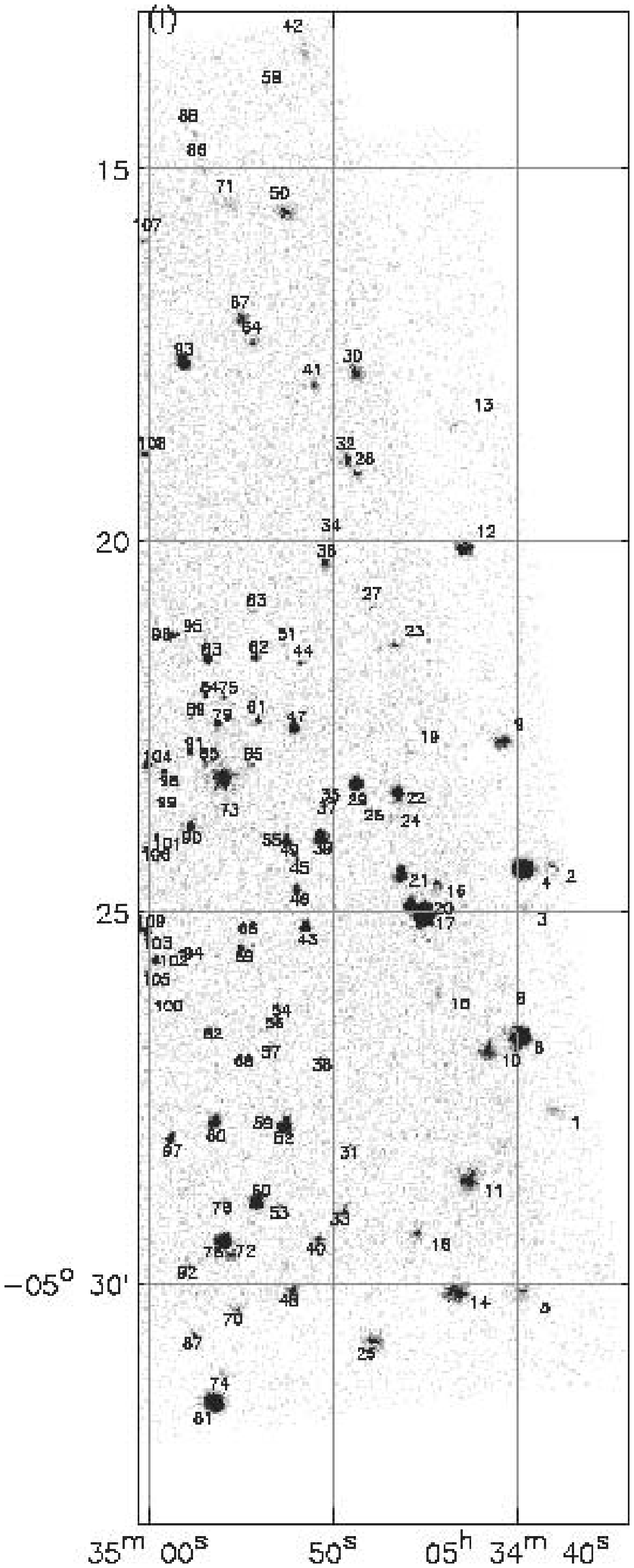}
\end{figure}

\clearpage
\newpage

\begin{figure}
\centering
\includegraphics[scale=0.8]{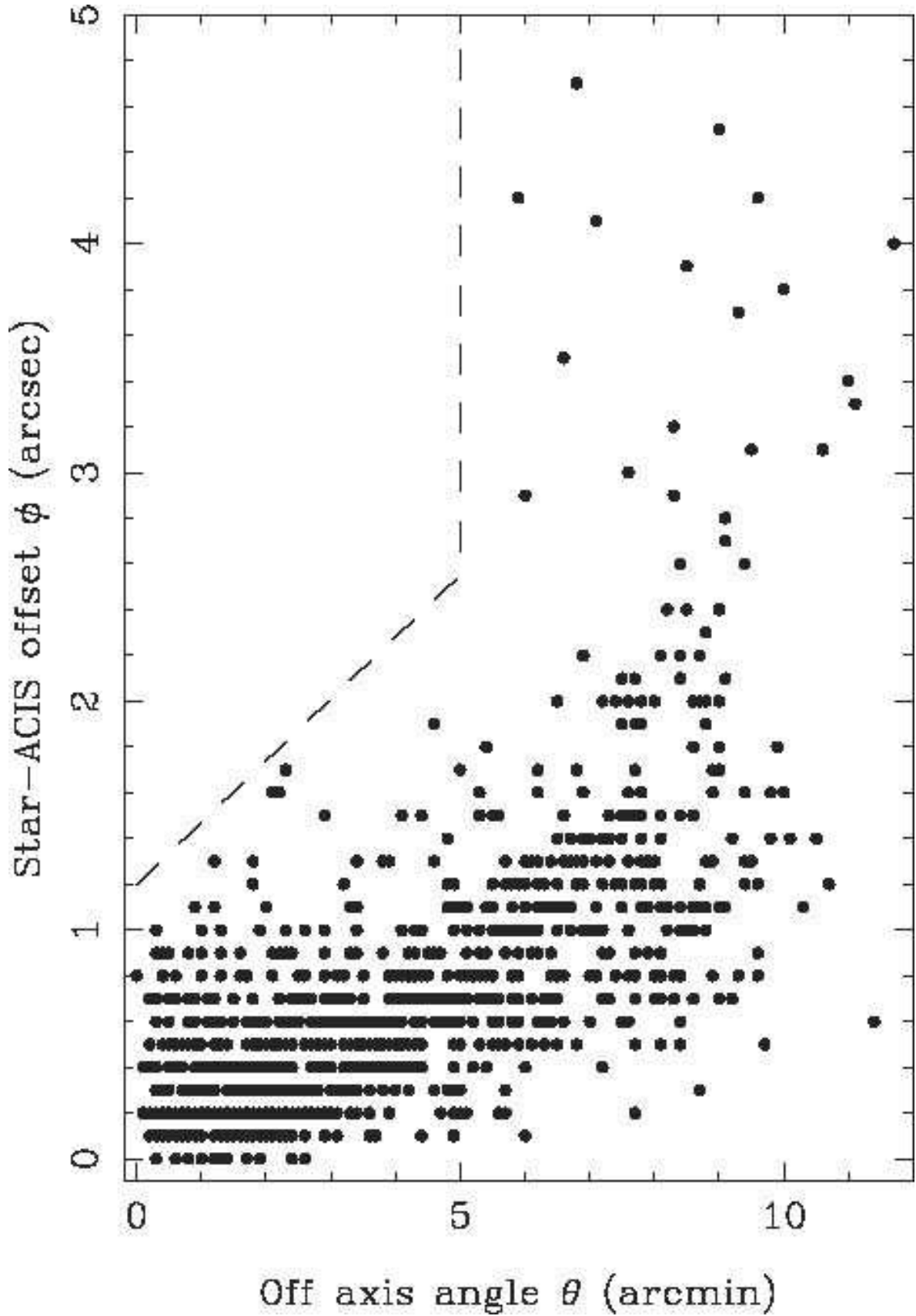}
\caption{Positional offsets $\phi$ (in arcsec) between ACIS sources and
proposed stellar counterparts plotted against the off-axis angle
$\theta$ (in arcmin).  Sources falling to the left of the dashed line were
rejected as false matches.
\label{offsets_fig}}
\end{figure}

\clearpage
\newpage

\begin{figure}
\centering
\includegraphics[scale=0.8]{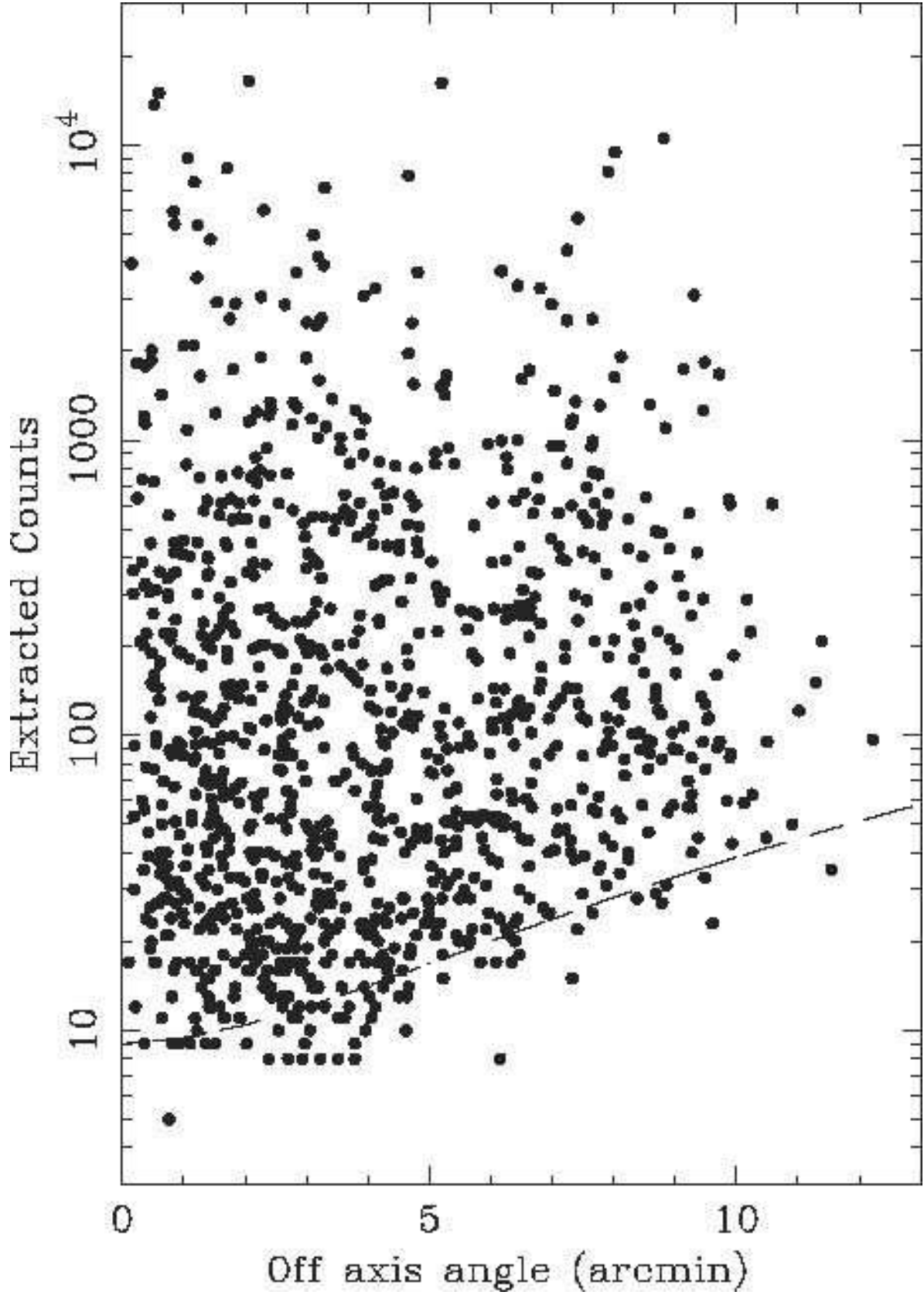}
\caption{Distribution of extracted counts for ONC sources as a function
of off-axis angle in the ACIS-I detector.  The dashed curve shows the
estimated completeness limit.
\label{cts_theta_fig}}
\end{figure}

\clearpage
\newpage

\begin{figure}
\centering
  \begin{minipage}[t]{1.0\textwidth}
  \centering
  \includegraphics[scale=0.45]{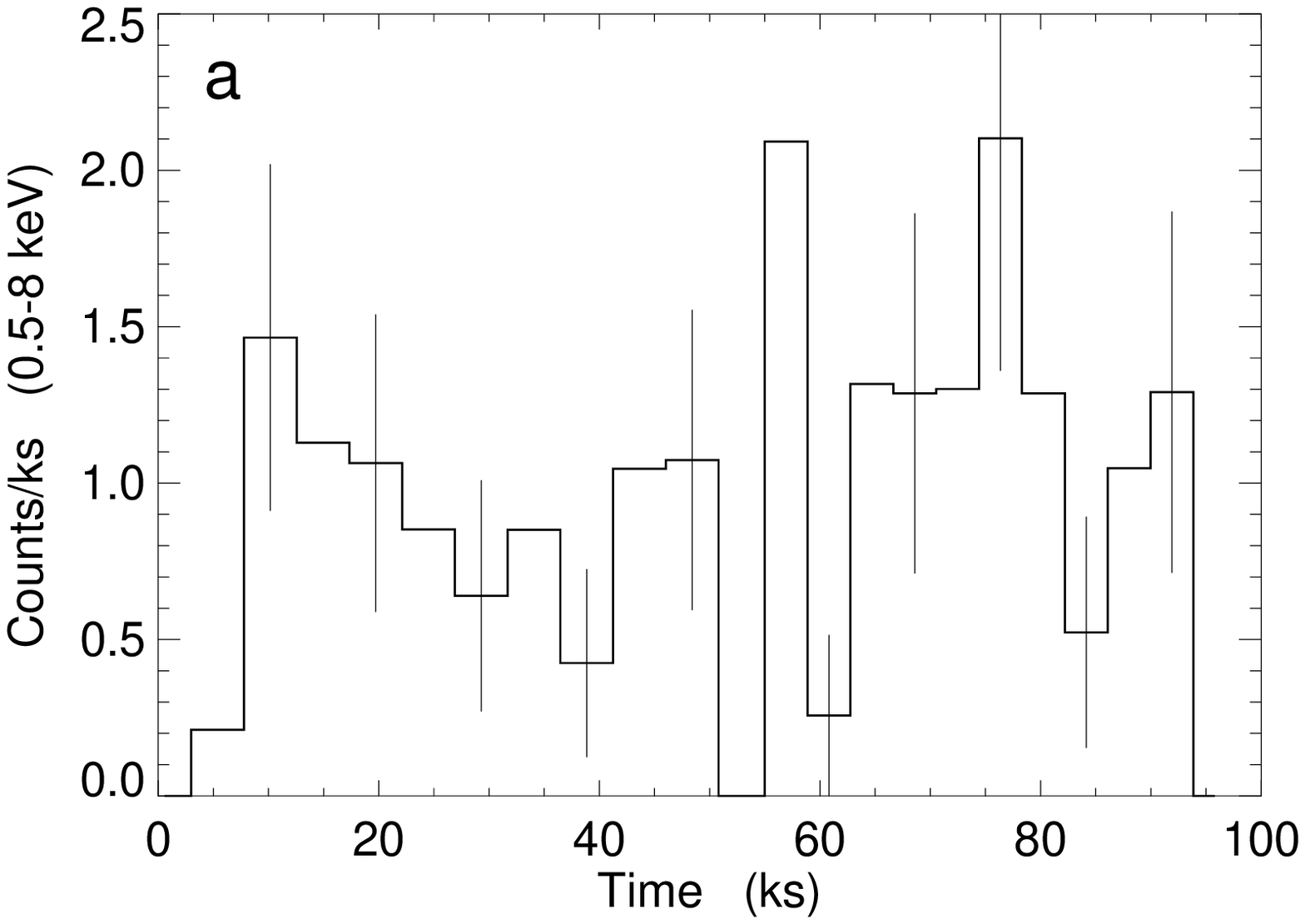}  
  \includegraphics[scale=0.45]{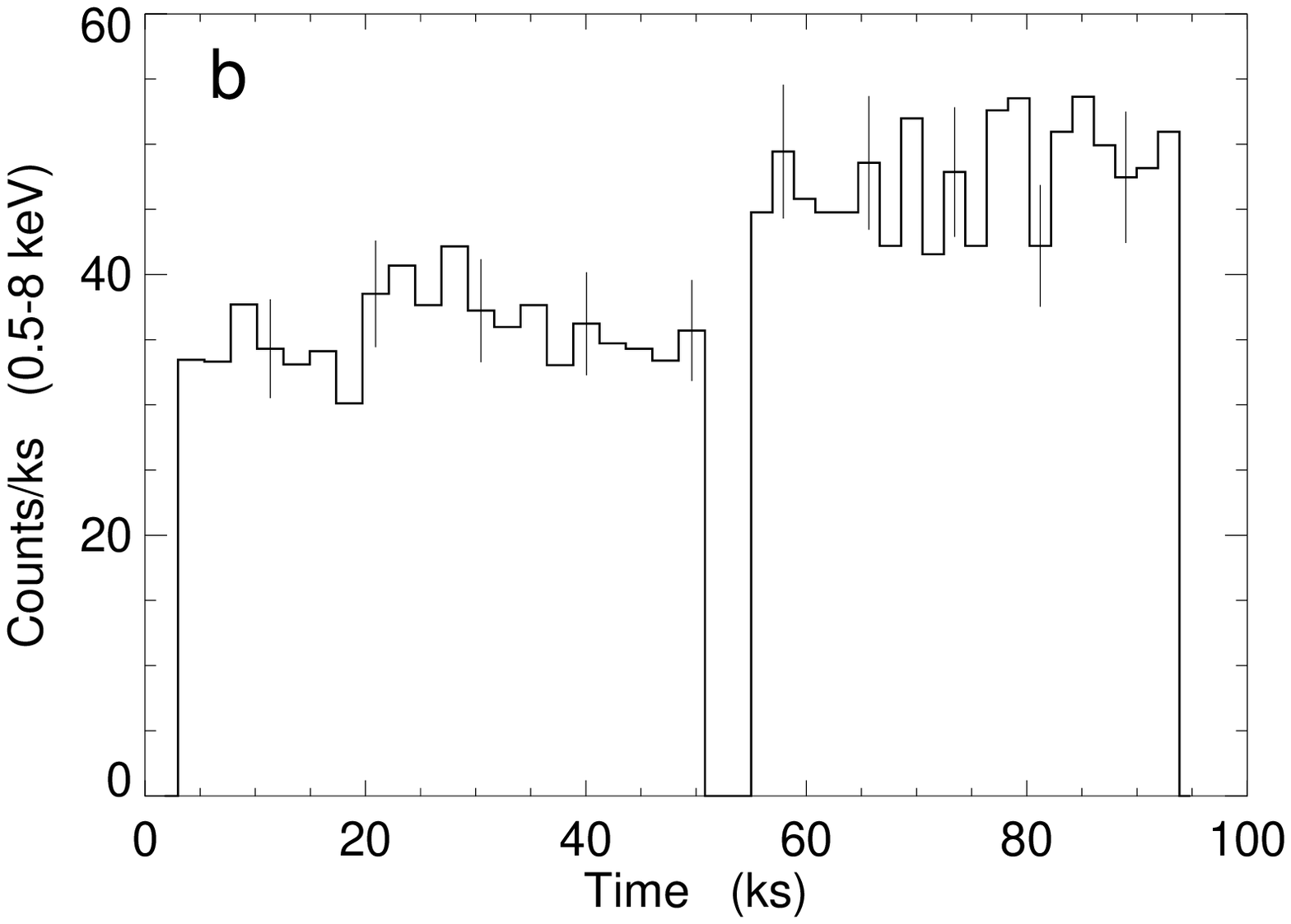}
  \end{minipage} \\ [0.3in]
  \begin{minipage}[t]{1.0\textwidth}
  \centering
  \includegraphics[scale=0.45]{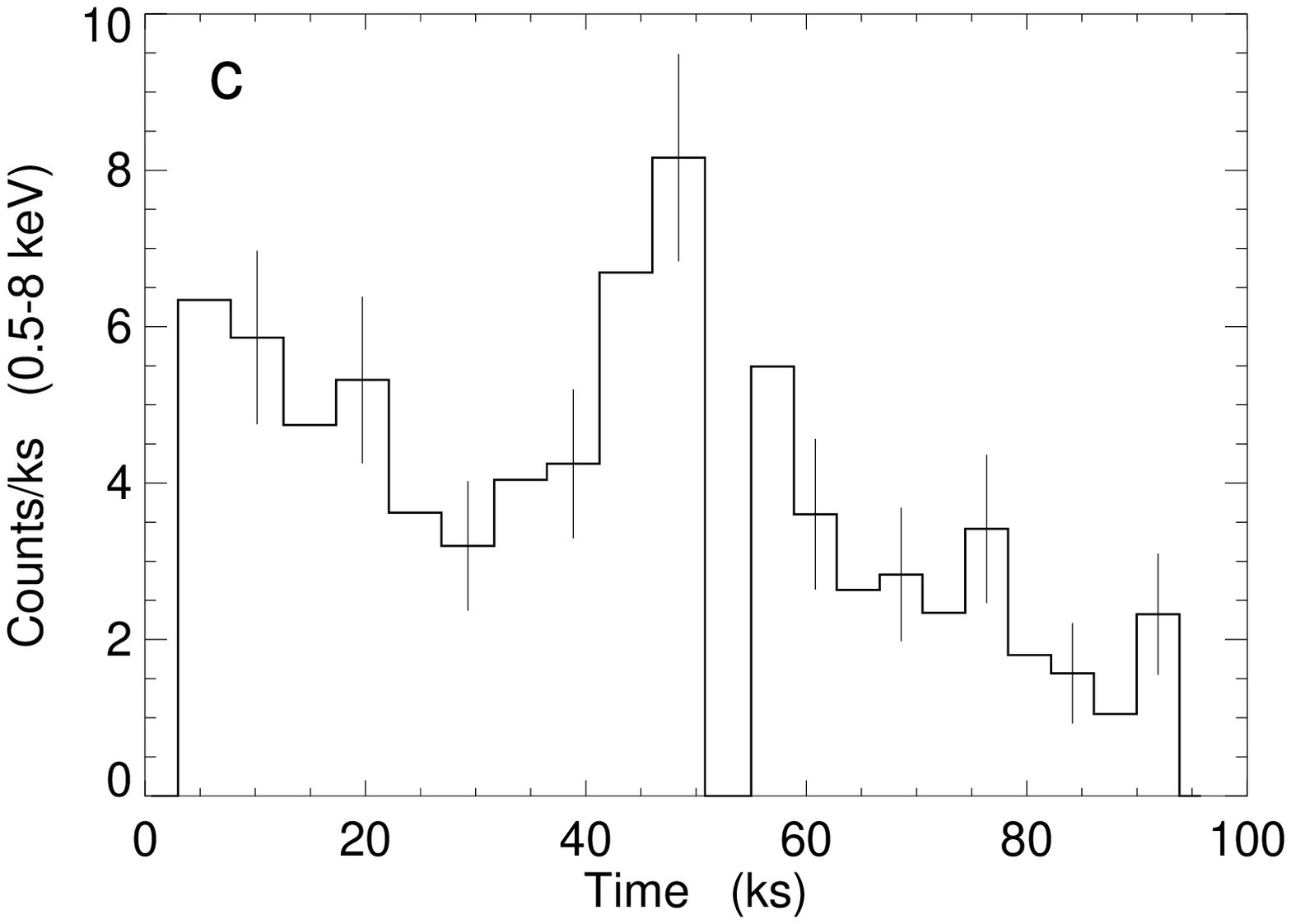}
  \includegraphics[scale=0.45]{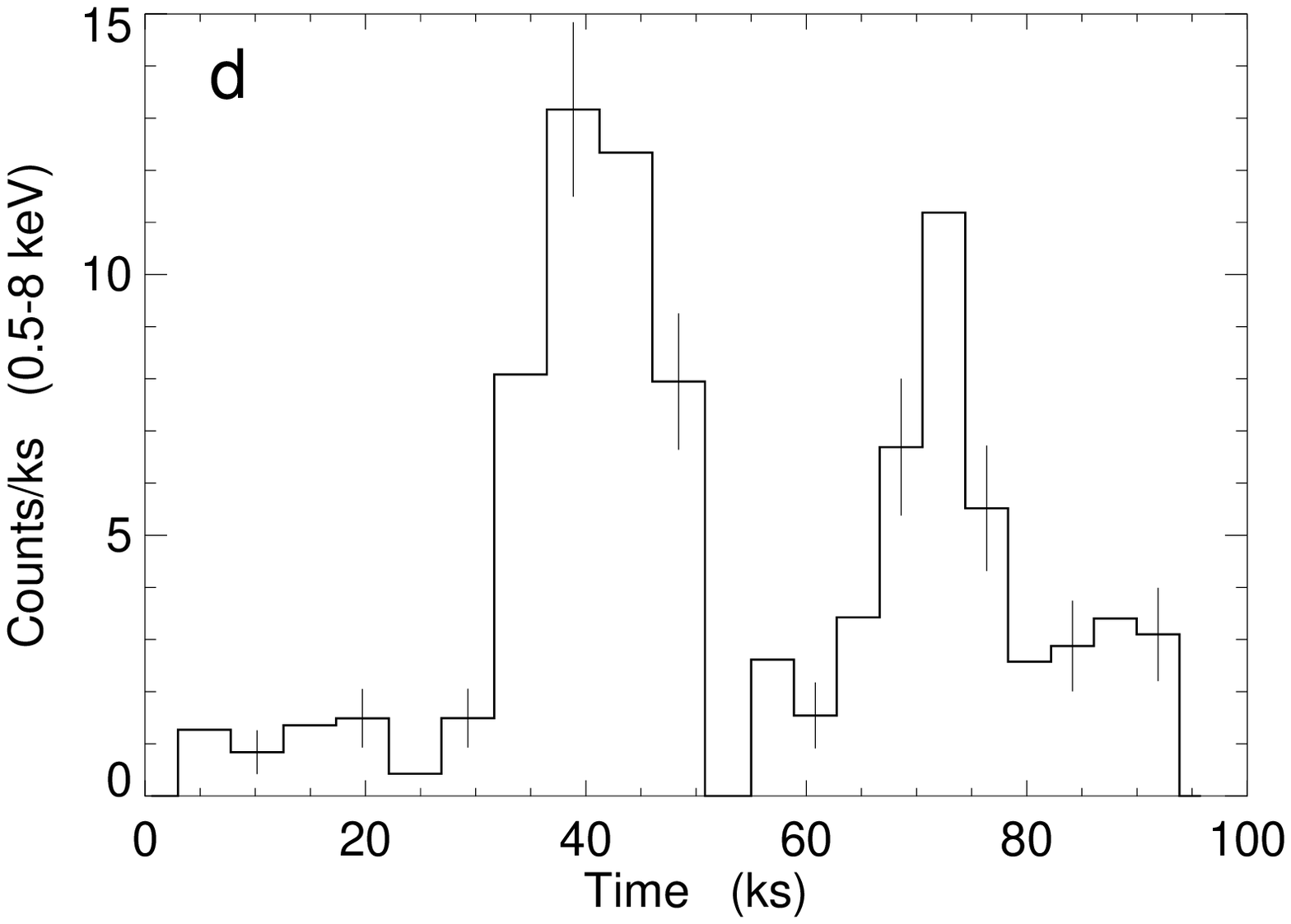}
\caption{Examples of ACIS ONC variability classes: (a) Constant (source
\#149 = JW 240, 0.4 M$_\odot$); (b) Long Term Variability (\#663 = JW
595, 3 M$_\odot$); (c) Possible Flare (\# 466 = CHS 8664, uncharacterized
2$\mu$m source); and (d) Flare (\#657 = JW 594, M=0.2 M$_\odot$).  The
ordinate gives counts ks$^{-1}$ in the total $0.5-8$ keV band. Error
bars show typical $\sqrt{N}$ uncertainties.  The abscissa gives time in
ks, with 10 (20) bins per observation for weaker (stronger) sources.
For graphical convenience, the two observations are plotted
consecutively separated by 5 ks, though in fact they are separated by
$\simeq 6$ months.  \label{var_class_fig}}
  \end{minipage} 
\end{figure}

\clearpage
\newpage

\begin{figure}
\centering
  \begin{minipage}[t]{1.0\textwidth}
  \centering
  \includegraphics[angle=-90.,scale=0.3]{f7a.eps}
  \end{minipage} \\ [0.3in]
  \begin{minipage}[t]{1.0\textwidth}
  \centering
  \includegraphics[angle=-90.,scale=0.3]{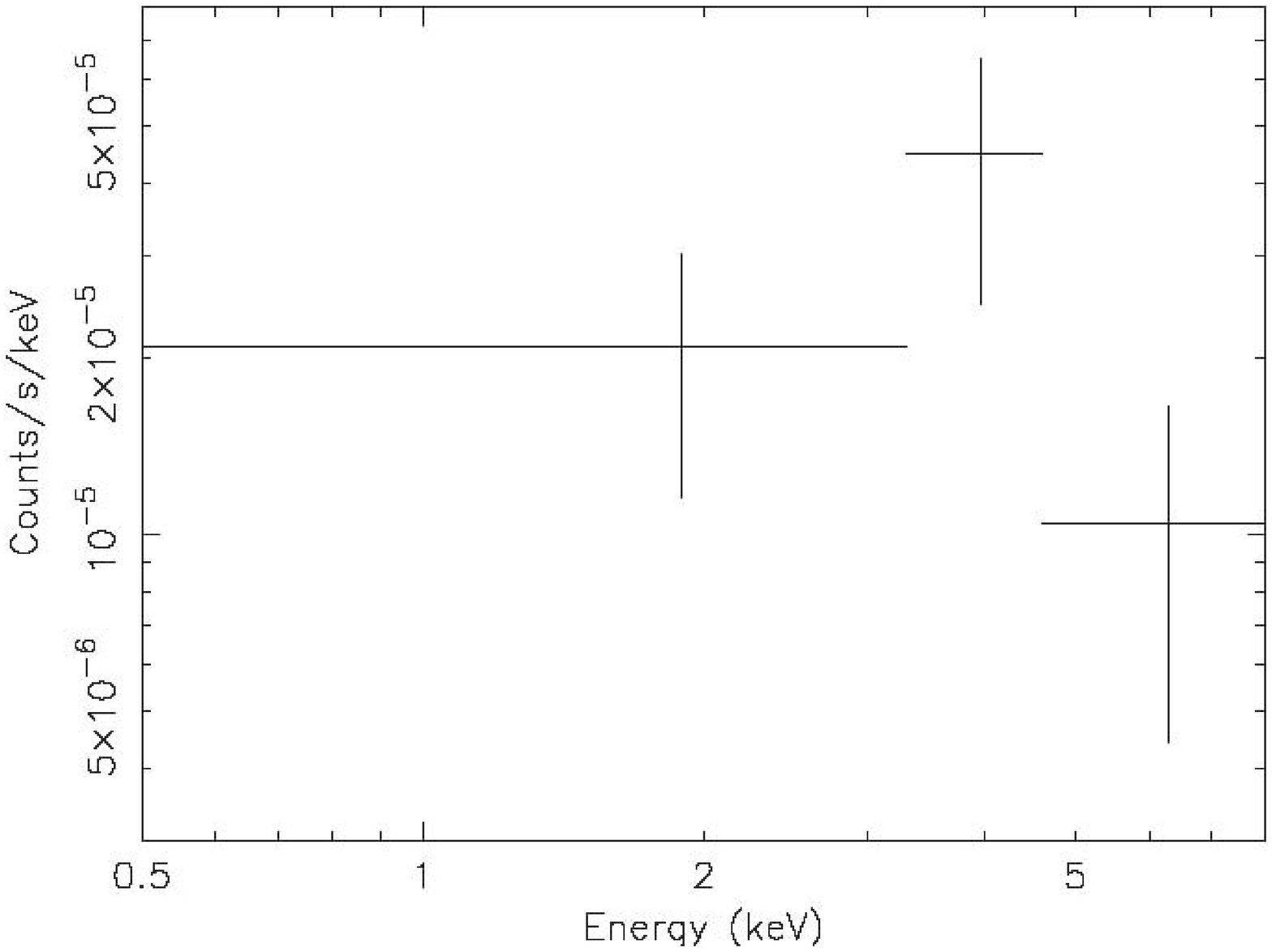}
  \end{minipage} \\ [0.3in]
  \begin{minipage}[t]{1.0\textwidth}
  \centering
  \includegraphics[angle=-90.,scale=0.3]{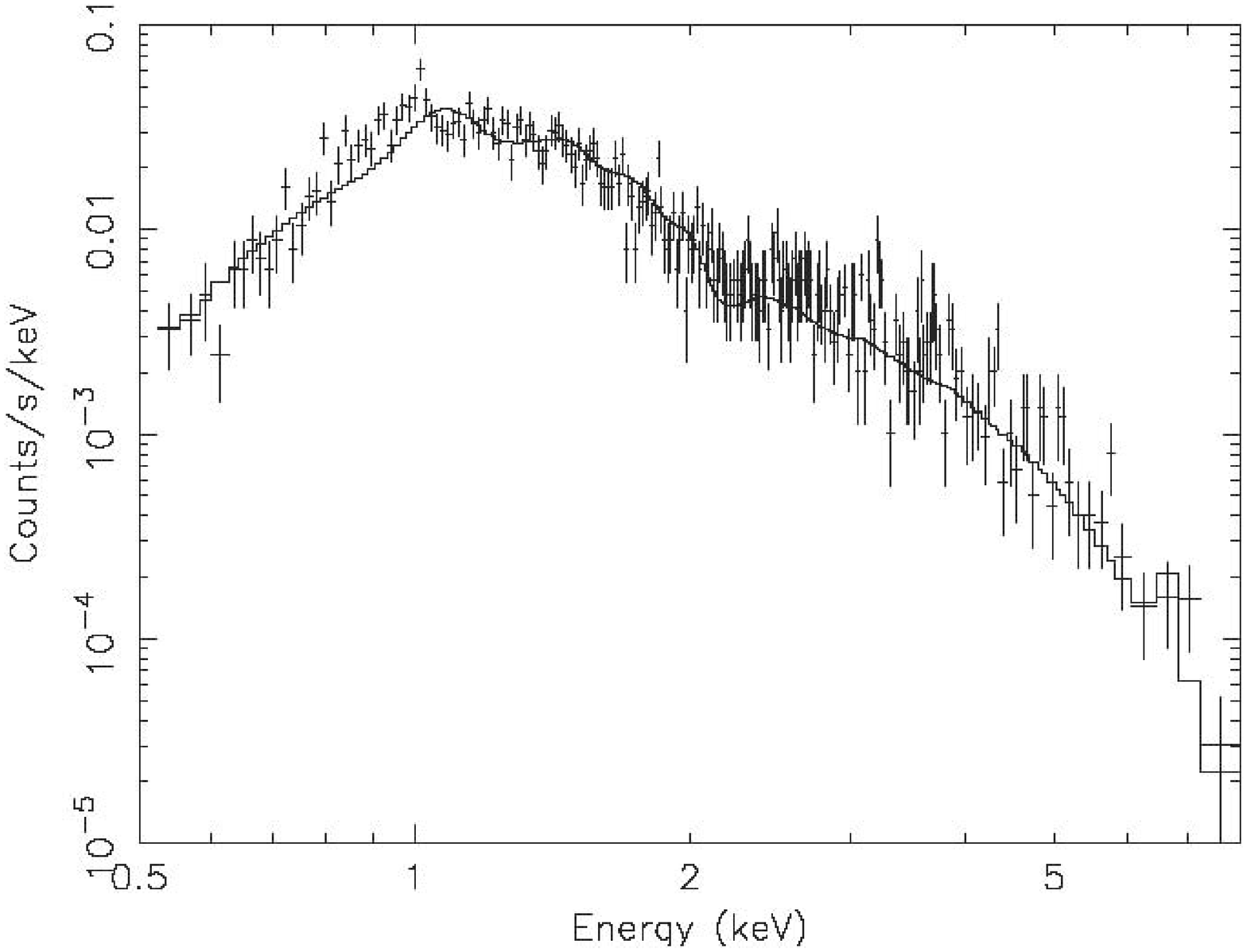}
\caption{Examples of ACIS ONC X-ray spectra: (top) source \#180;
(middle) source \# 388; and (bottom) source \#573.  See \S
\ref{spec_sec} for description.
\label{spec_fig}}
  \end{minipage} 
\end{figure}

\clearpage
\newpage

\begin{figure}
\centering
\includegraphics[scale=0.8]{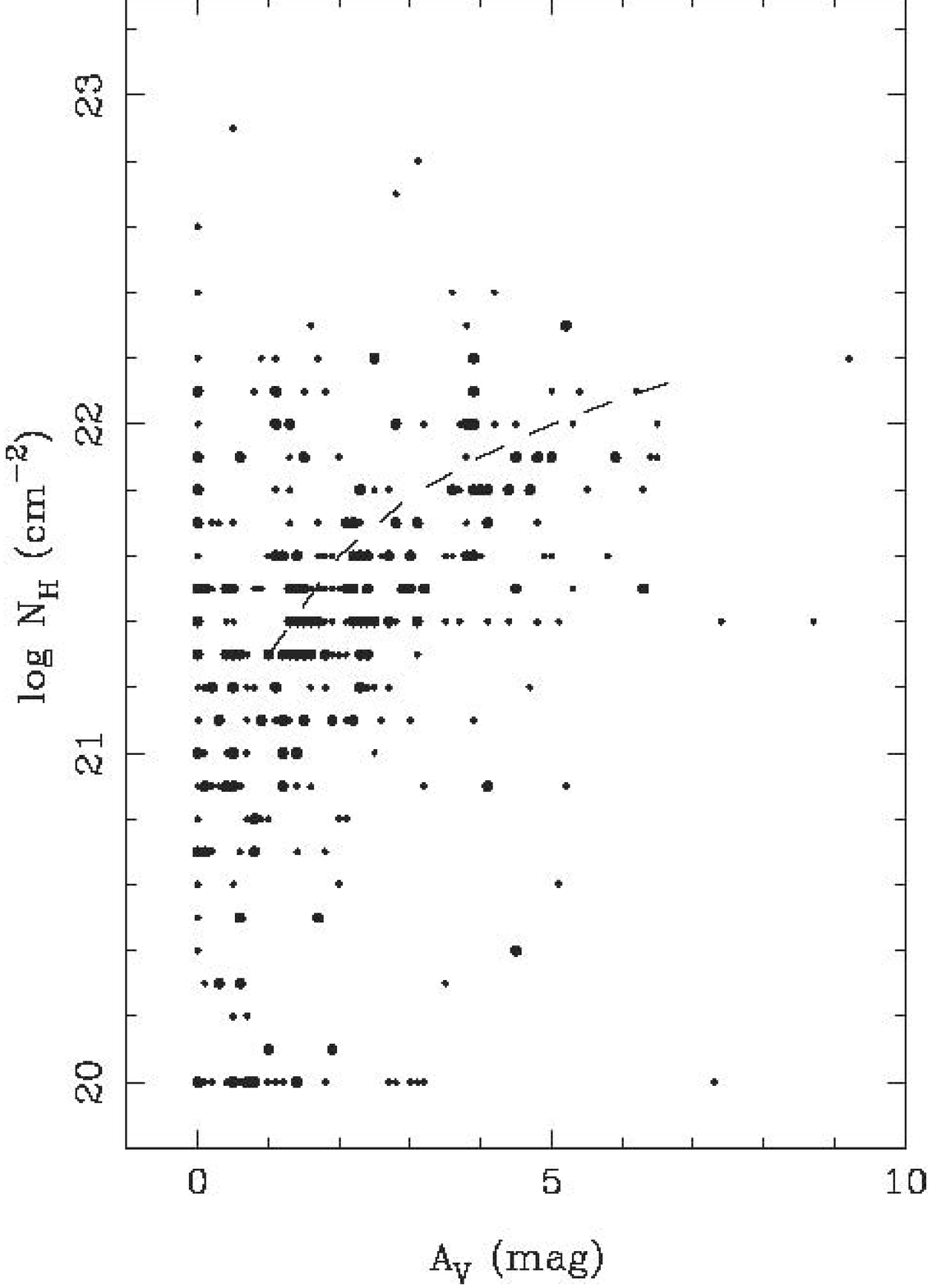}
\caption{Plot of intervening column density derived from X-ray spectral
analysis $vs.$ visual absorption derived from optical studies for
sources with both quantities known.  Large circles are bright sources
with $> 500$ counts while small circles are sources with $30-500$
counts.  Sources along the $\log N_H = 20.0$ cm$^{-2}$ locus have upper
limits to the X-ray absorption.  The dashed curve is the relation $N_H
= 2 \times 10^{21} ~ A_V$.
\label{avnh_fig}}
\end{figure}

\clearpage
\newpage

\begin{figure}
\centering
\includegraphics[scale=0.8]{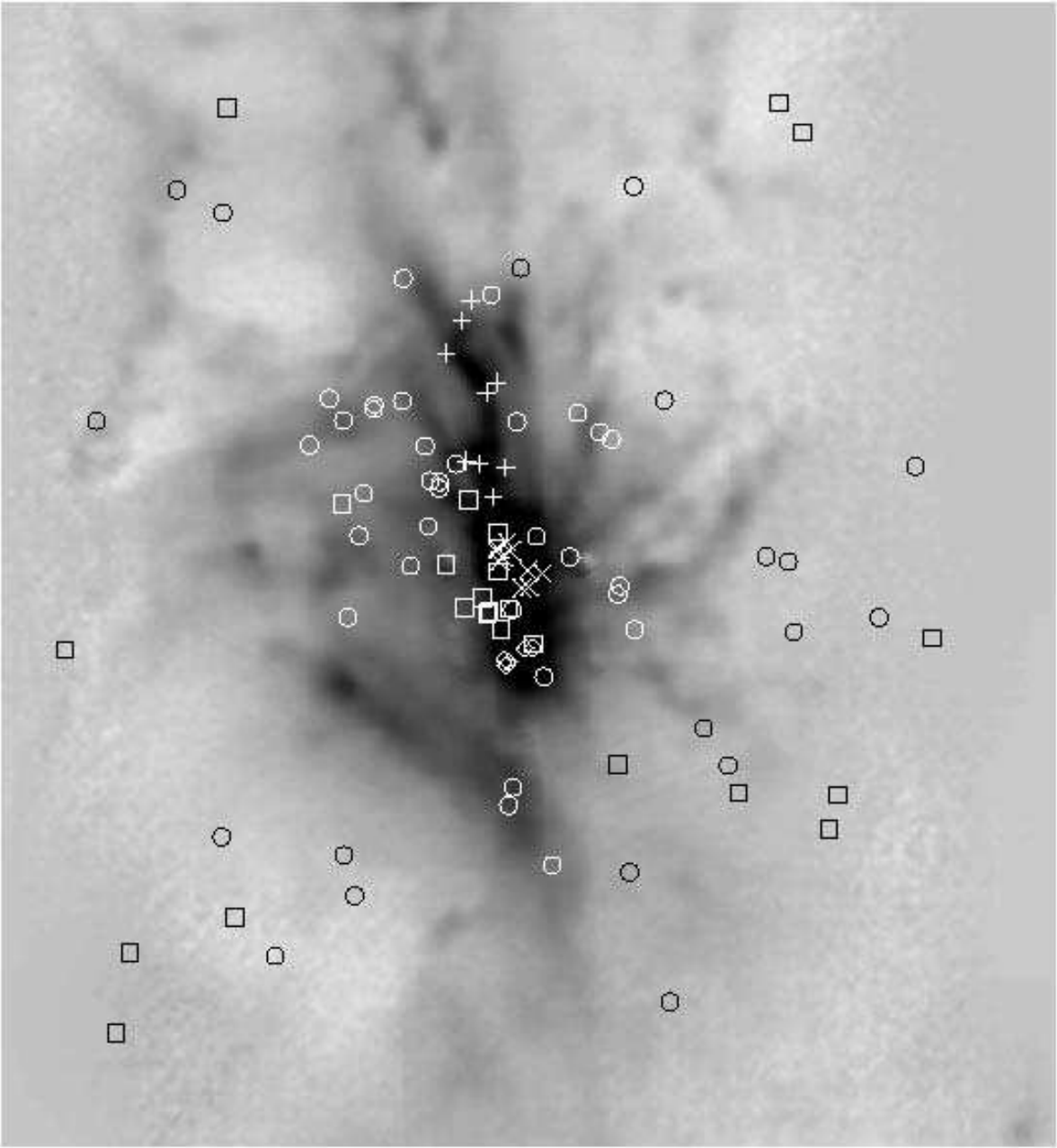}
\caption{ACIS sources without stellar counterparts plotted on a
greyscale SCUBA submillimeter map of the Orion Nebula showing the
distribution of dense molecular material over a $17.6\arcmin \times
16.3\arcmin$ region \citep{Johnstone99}.  The ACIS sources are coded by
our suggested classification: lightly absorbed members of the ONC or
other Orion OB association (squares); embedded stars associated with
the OMC 1 = Orion KL core (crosses), OMC 1S = Orion S core (diamonds),
and OMC 1N core (plusses); and dispersed absorbed sources with both
embedded stars and Galactic and/or extragalactic background sources
(circles).
\label{noid_fig}}
\end{figure}

\clearpage
\newpage

\begin{figure}
\centering
\includegraphics[scale=0.30]{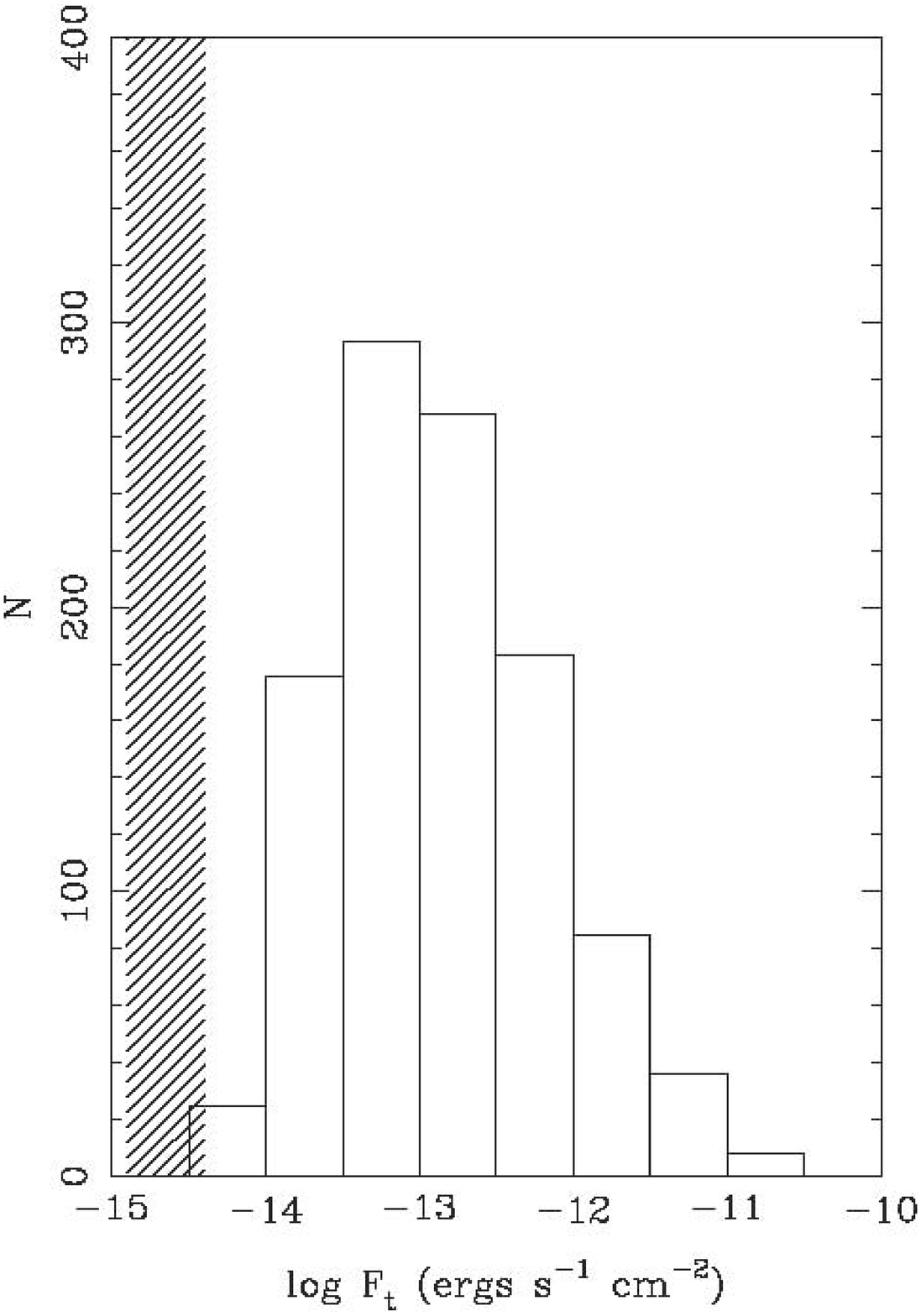} \hspace{0.1in}
\includegraphics[scale=0.30]{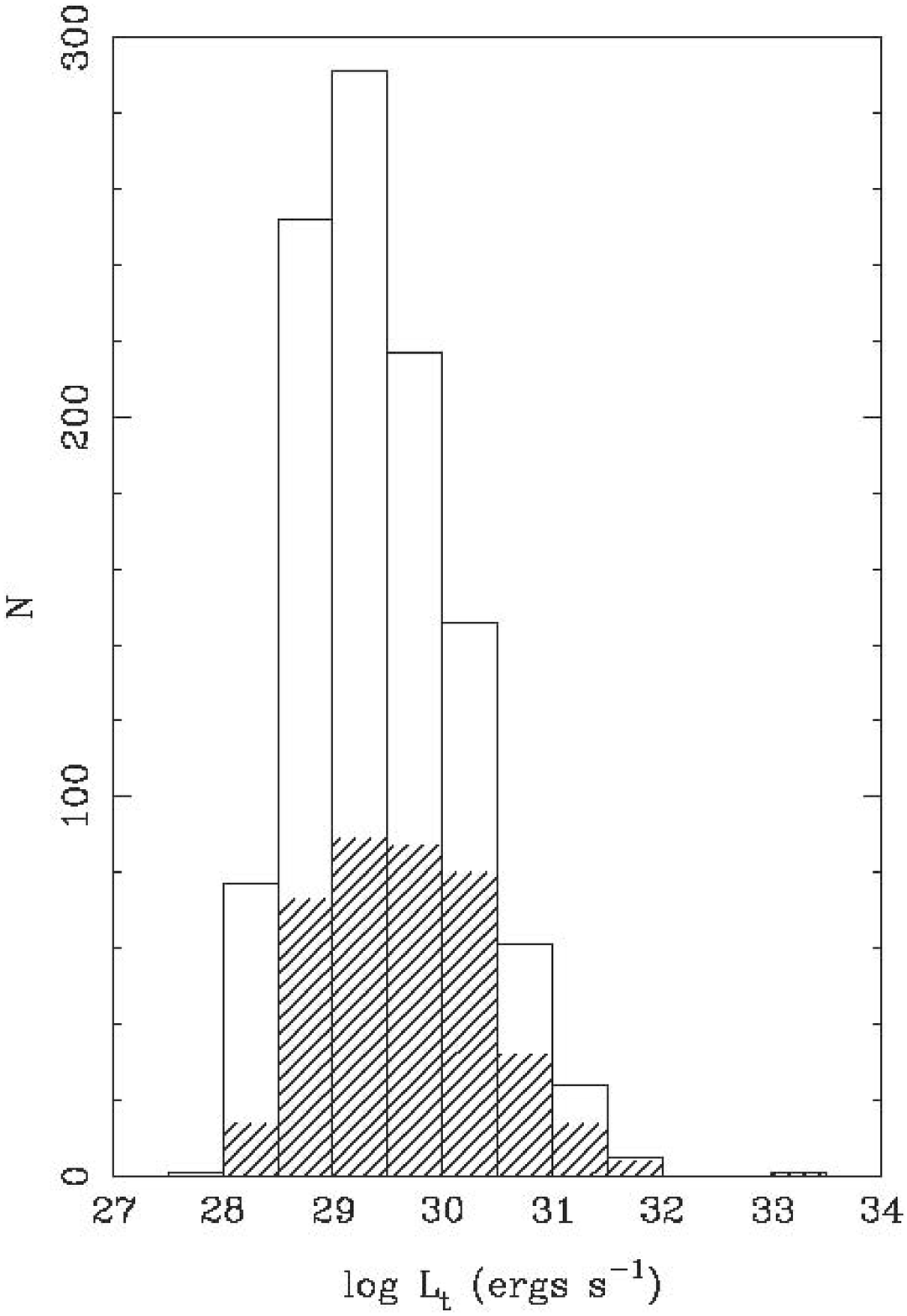} \hspace{0.1in}
\includegraphics[scale=0.30]{f10c.eps}
\caption{Distributions of X-ray emission ($0.5-8$ keV band) for the
ACIS population: (left) flux with hatching indicating the completeness
limit; (middle) luminosity with hatching indicating stars with $K$-band
excess disks; and (right) X-ray to stellar
bolometric luminosity ratio for well-characterized stars
\citep{Hillenbrand97}.
\label{xlum_hist_fig}}
\end{figure}

\clearpage
\newpage

\begin{figure}
\centering
  \includegraphics[scale=0.30]{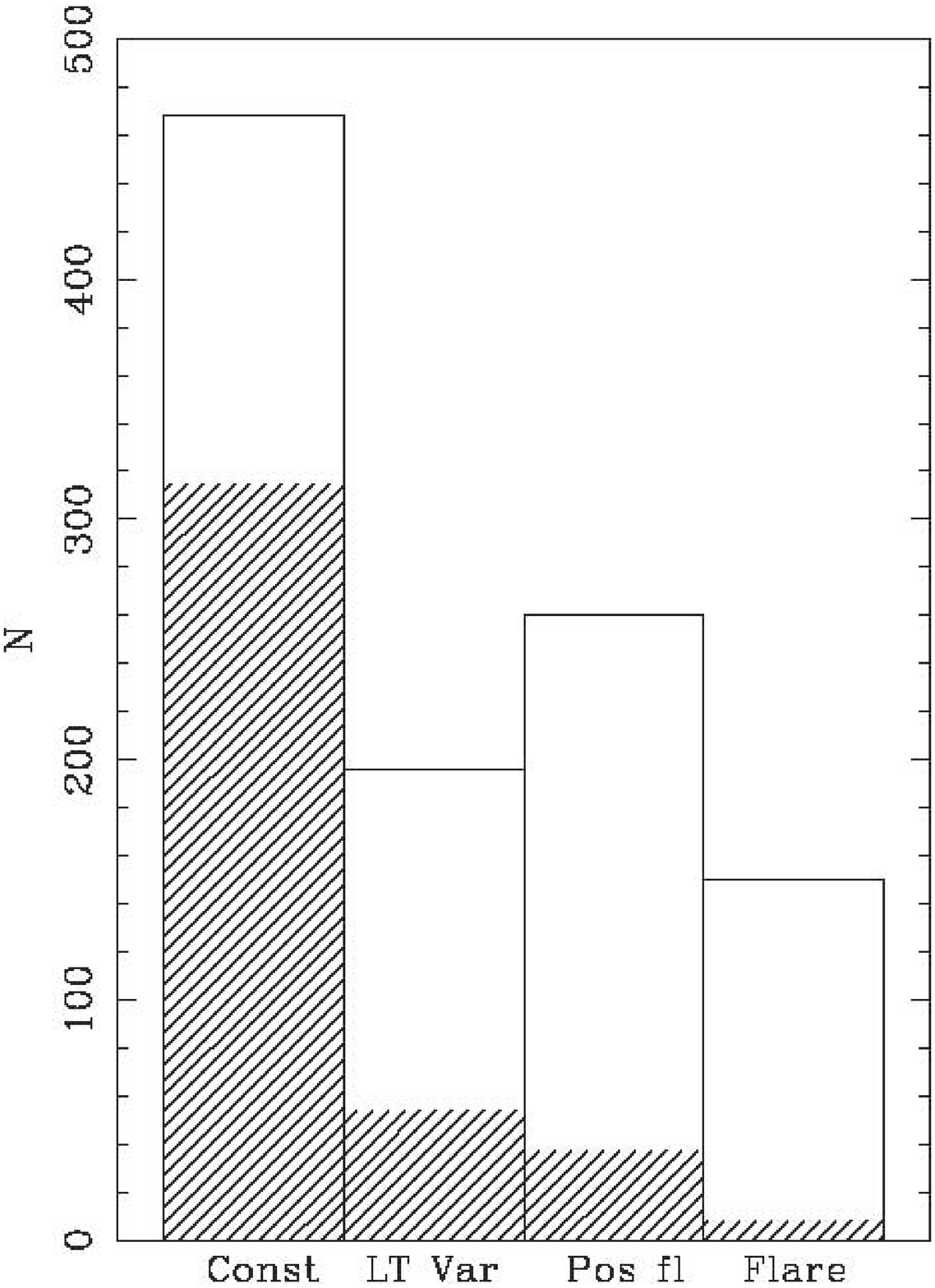} \hspace{0.1in}
  \includegraphics[scale=0.30]{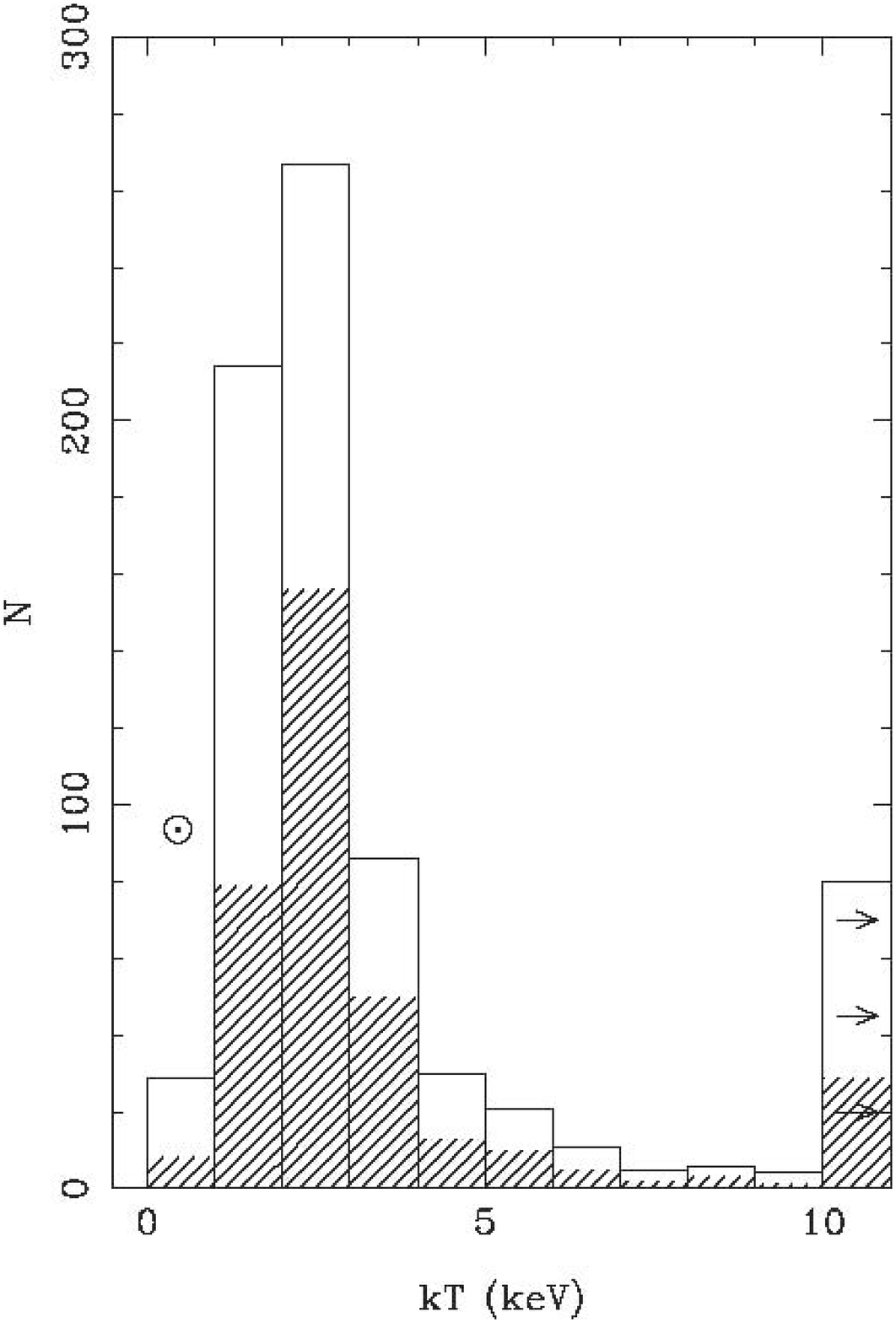} \hspace{0.1in}
  \includegraphics[scale=0.30]{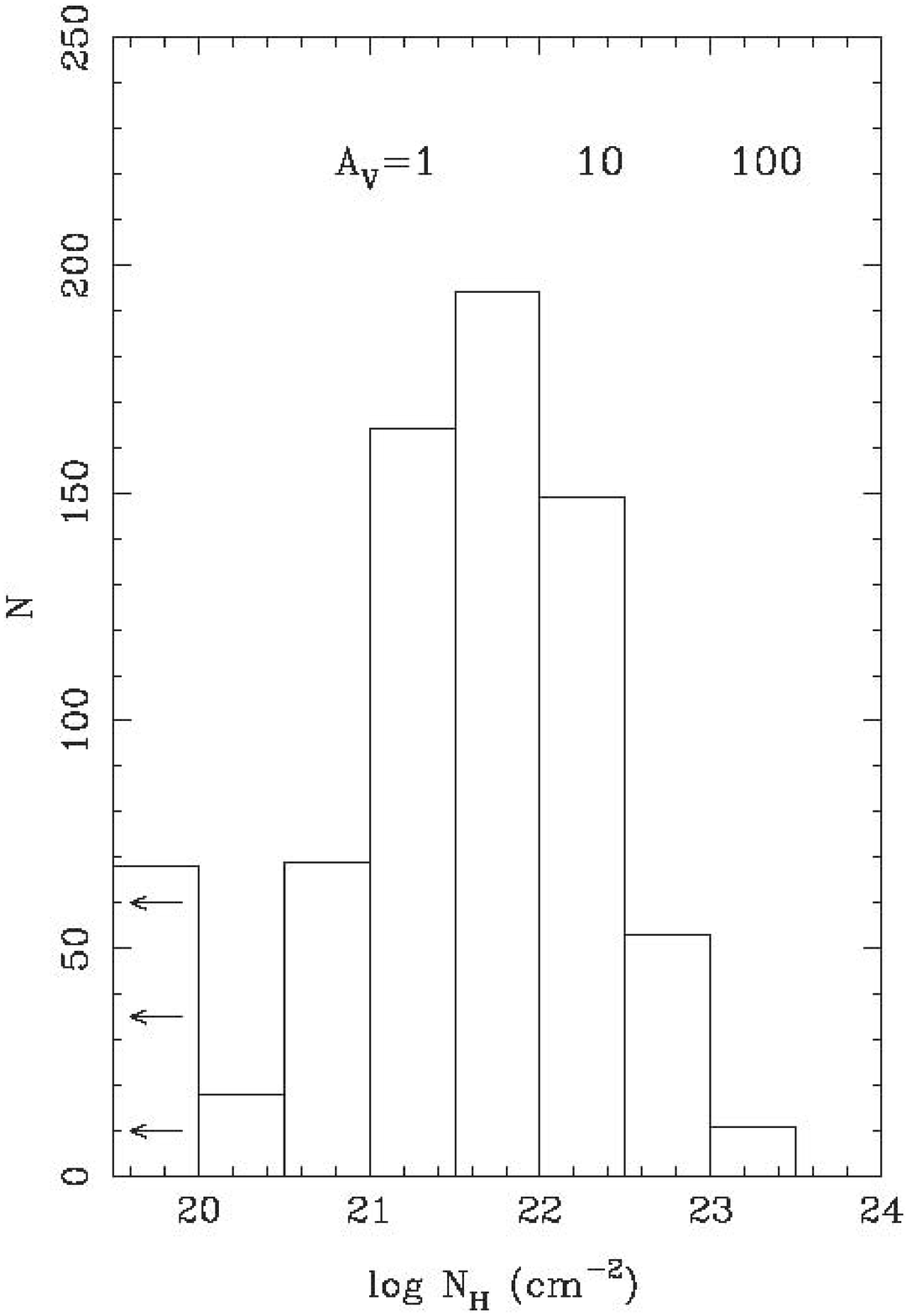}
\caption{Distributions of X-ray variability and spectral properties of
the ACIS sources:  (left) variability classes with hatching indicating
sources with $<50$ extracted counts; (middle) plasma energies with
hatching indicating sources with intraday variability, where $\odot$
indicates the characteristic X-ray temperature of the contemporary
flaring Sun; and (right) absorbing column densities with corresponding
visual absorptions.  Only sources with $\geq 30$ counts are included
for the spectral parameters.  Bins with arrows indicate sources with
very high plasma energies or very low column densities.  
\label{xprop_hist_fig}}
\end{figure}

\clearpage
\newpage

\begin{figure}
\centering
  \includegraphics[scale=0.45]{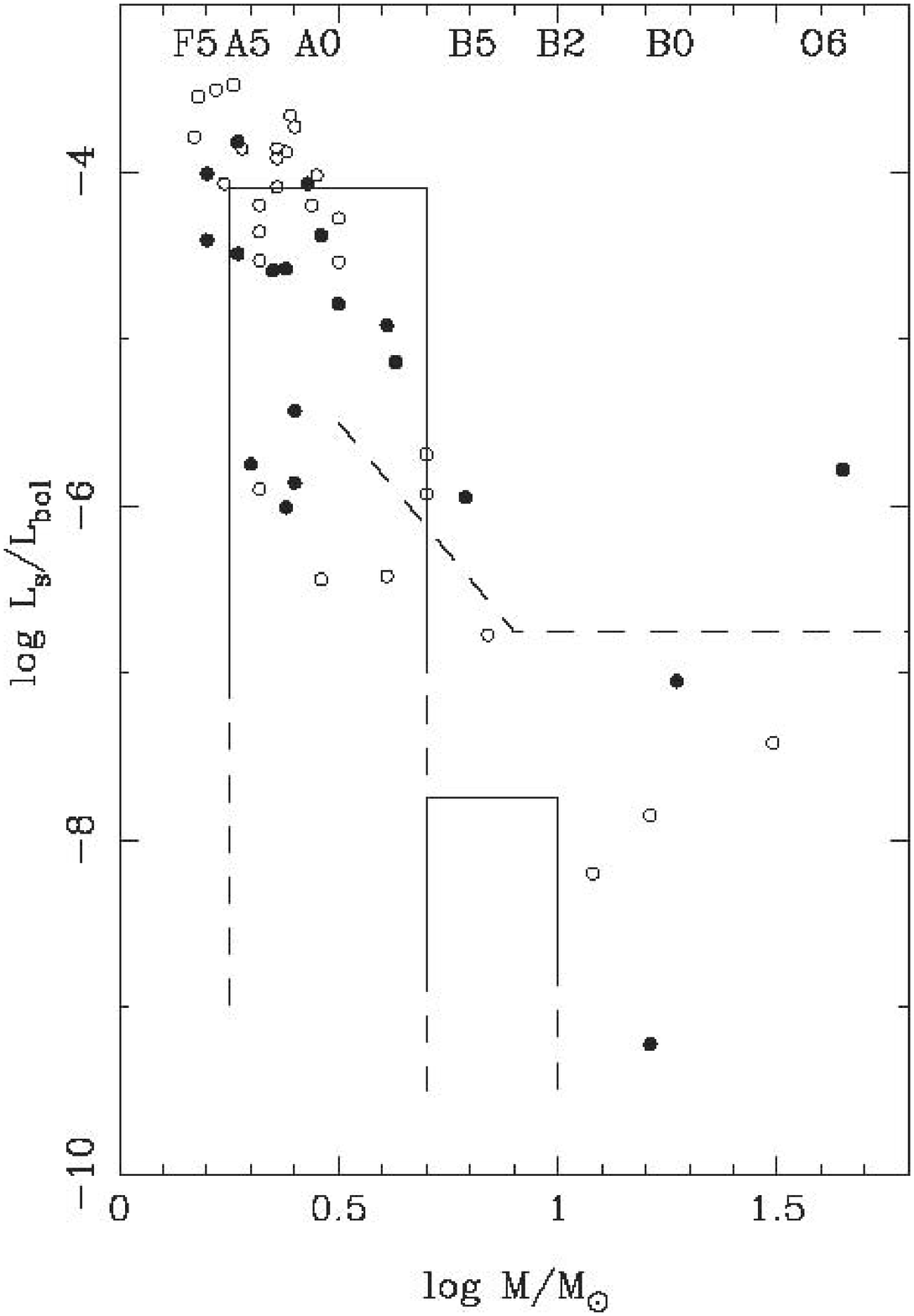}
  \hspace{0.3in}
  \includegraphics[scale=0.45]{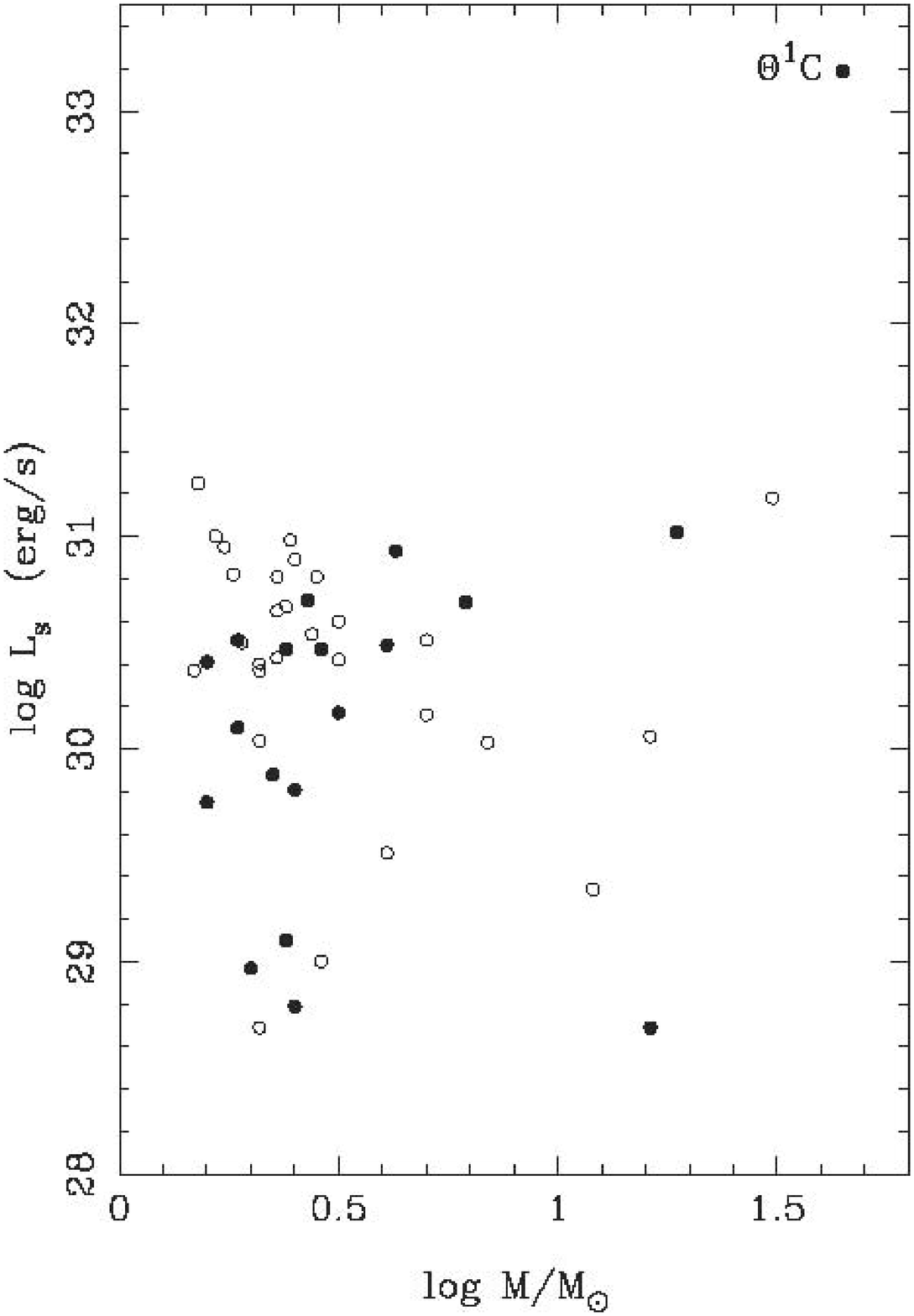}
\caption{X-ray emission of high- and intermediate-mass ONC stars:
(left) soft band $\log L_s/L_{bol}$ $vs.$ spectral type and mass (see
\S \ref{highmass_sec}-\ref{intermed_sec} for explanation of marked
regions and line); and (right) $\log L_s$ $vs.$ mass.  Open circles
denote stars exhibiting intra-day variability (Variability Class
`Flare' or `Possible flare').
\label{highmass_ls_fig}}
\end{figure}

\clearpage
\newpage

\begin{figure}
\centering
  \begin{minipage}[t]{1.0\textwidth}
  \centering
  \includegraphics[scale=0.4]{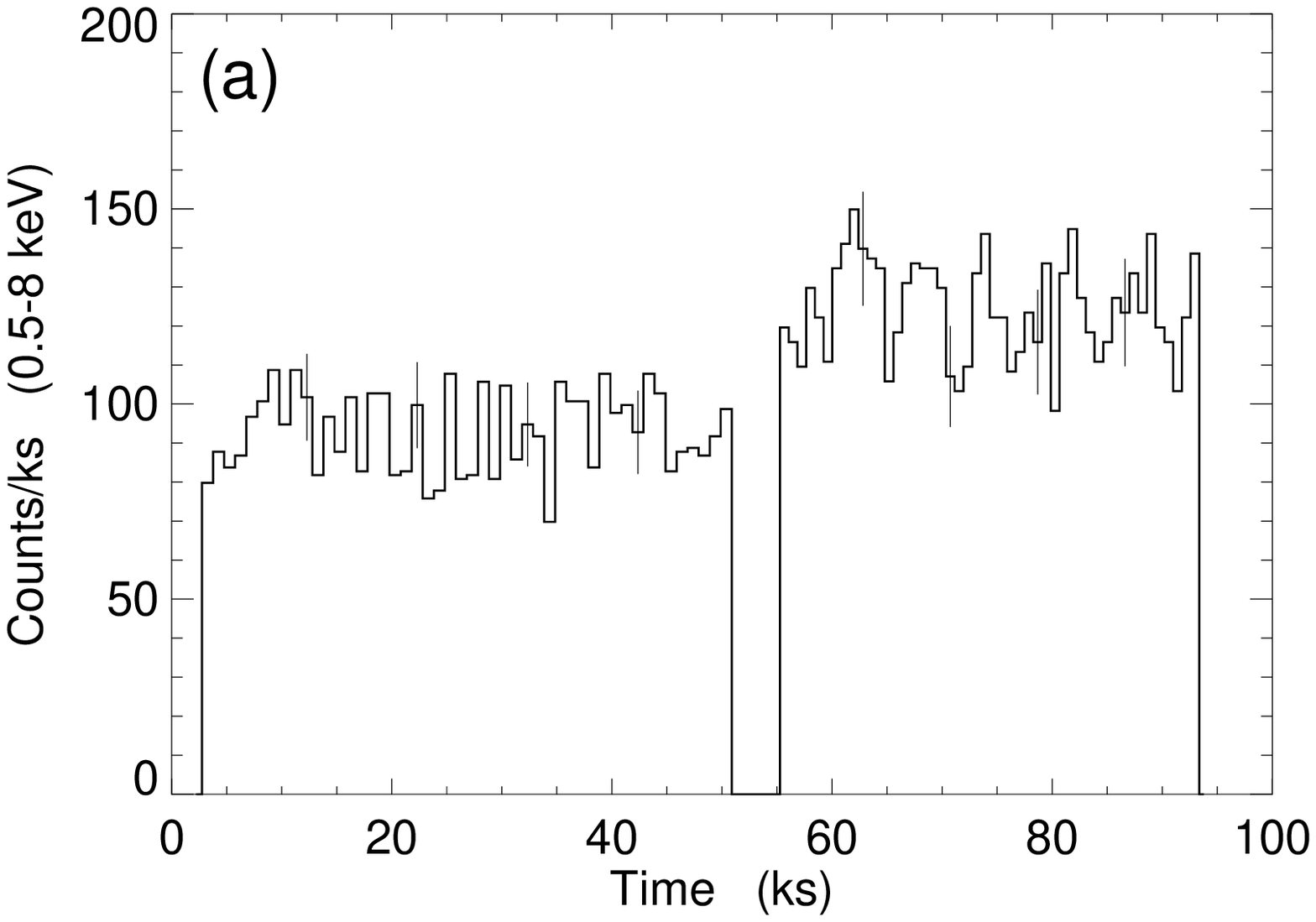}  
  \includegraphics[scale=0.4]{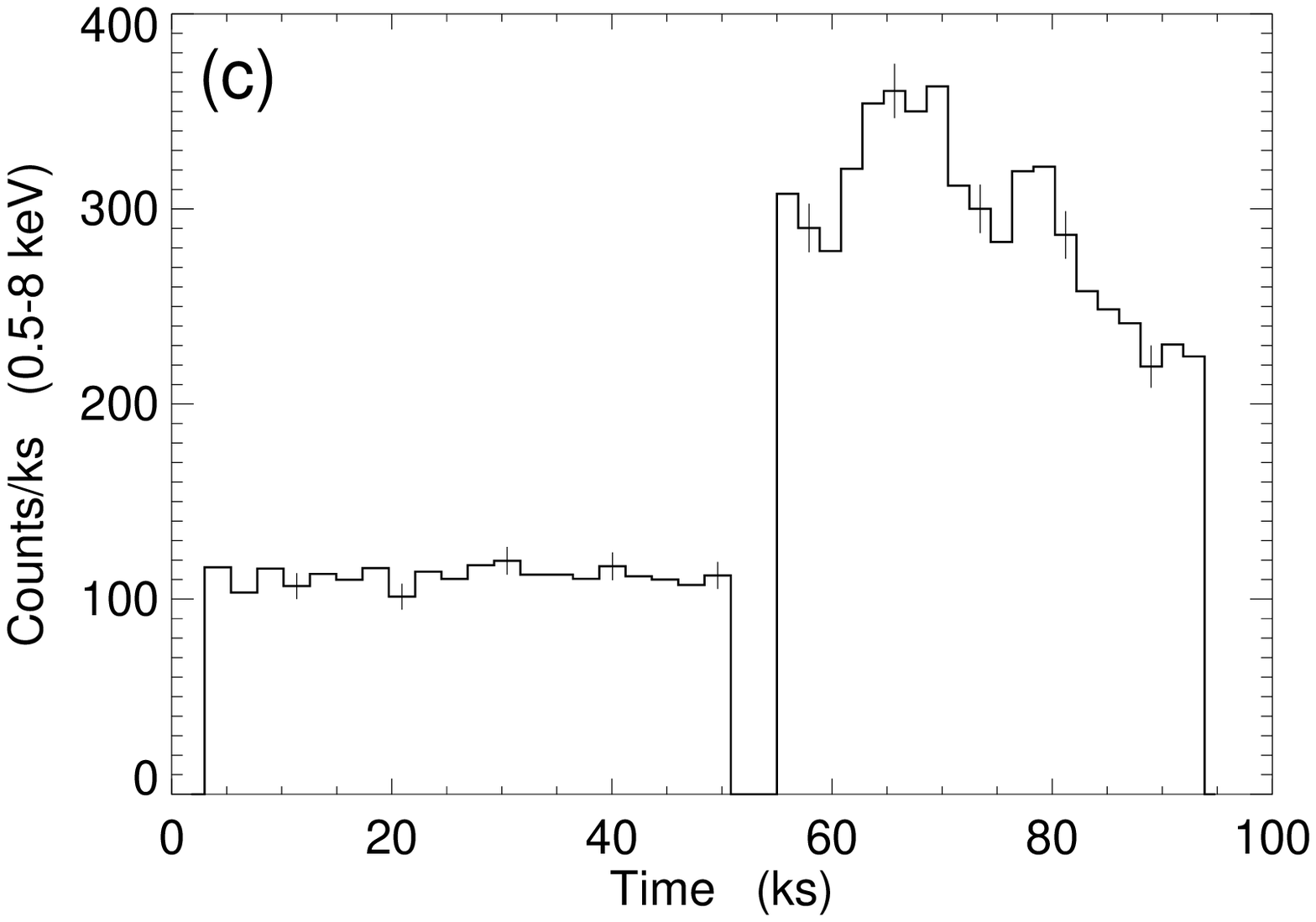}
  \end{minipage} \\ [-0.15in]
  \begin{minipage}[t]{1.0\textwidth}
  \centering
  \includegraphics[scale=0.4]{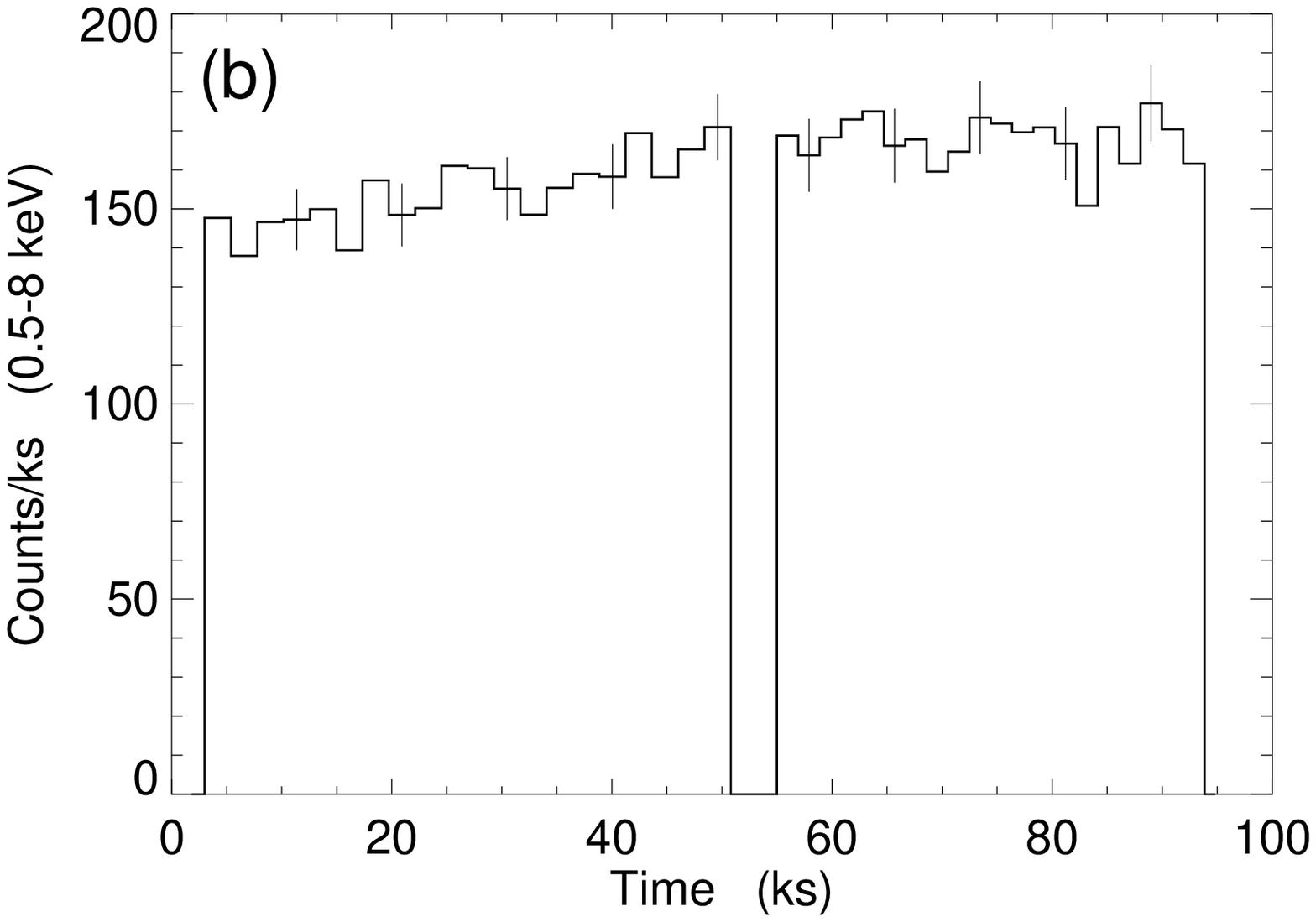}
  \includegraphics[scale=0.4]{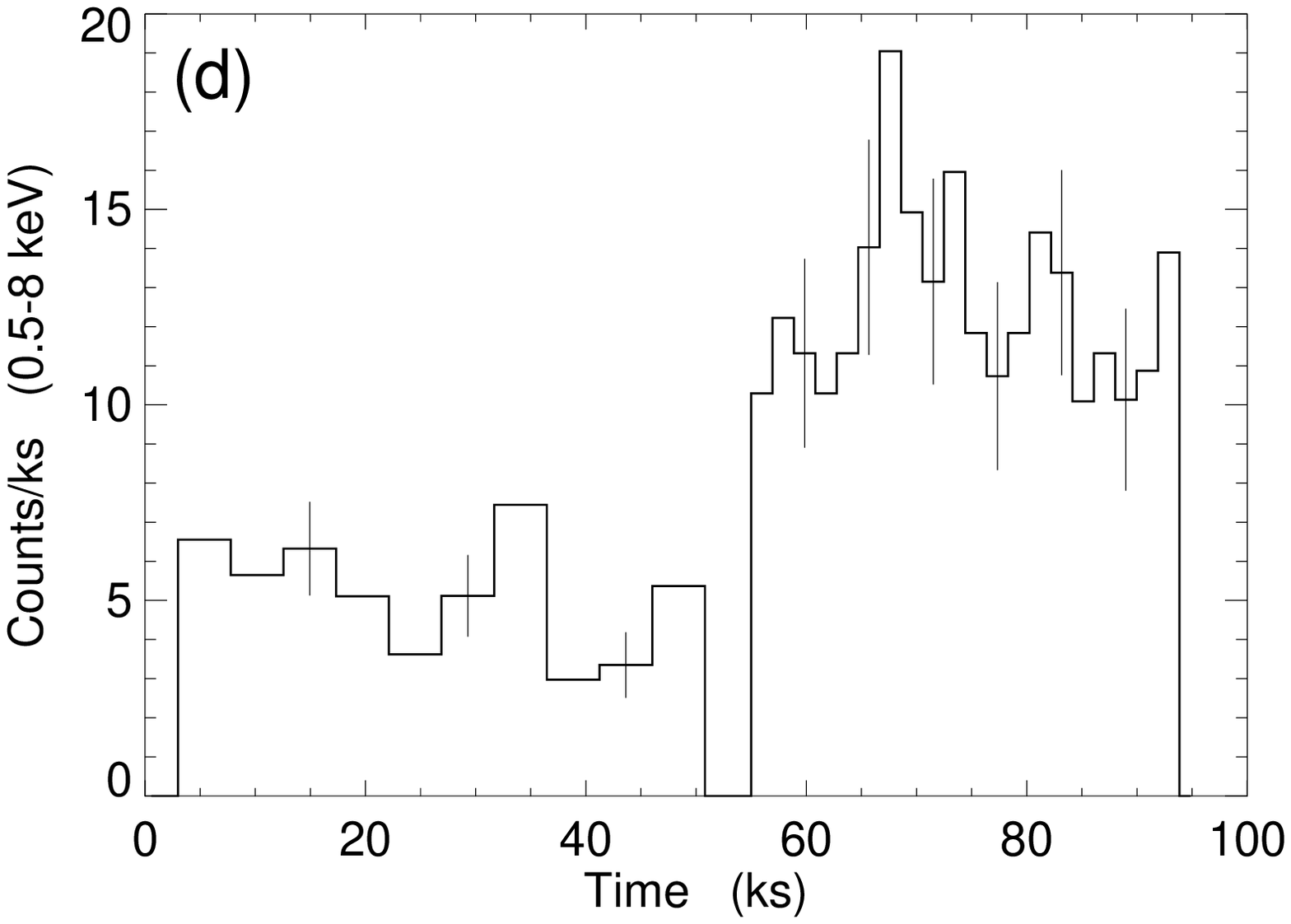}
  \end{minipage} \\ [-0.15in]  
  \begin{minipage}[t]{1.0\textwidth}
  \centering
  \includegraphics[scale=0.4]{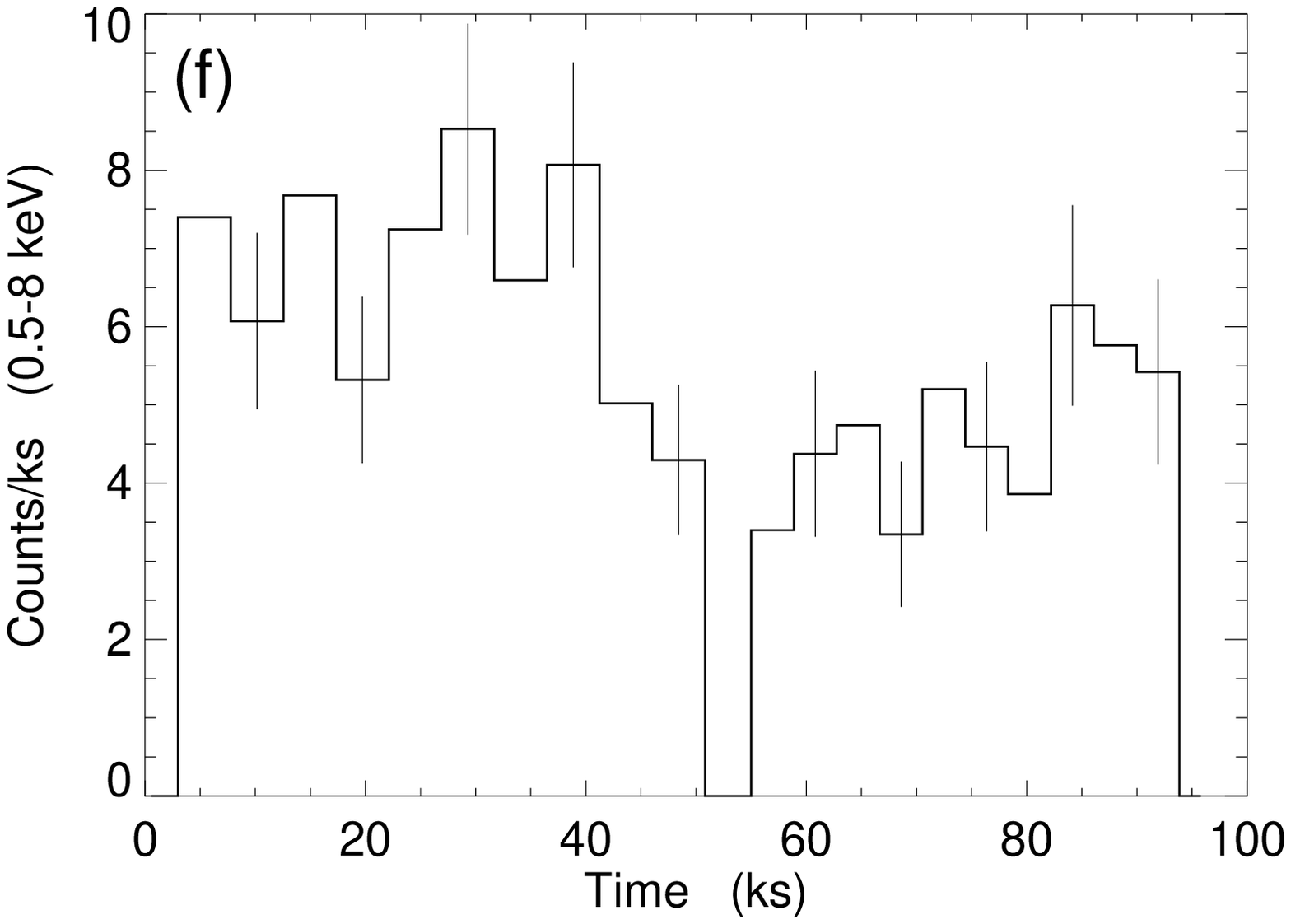}
  \includegraphics[scale=0.4]{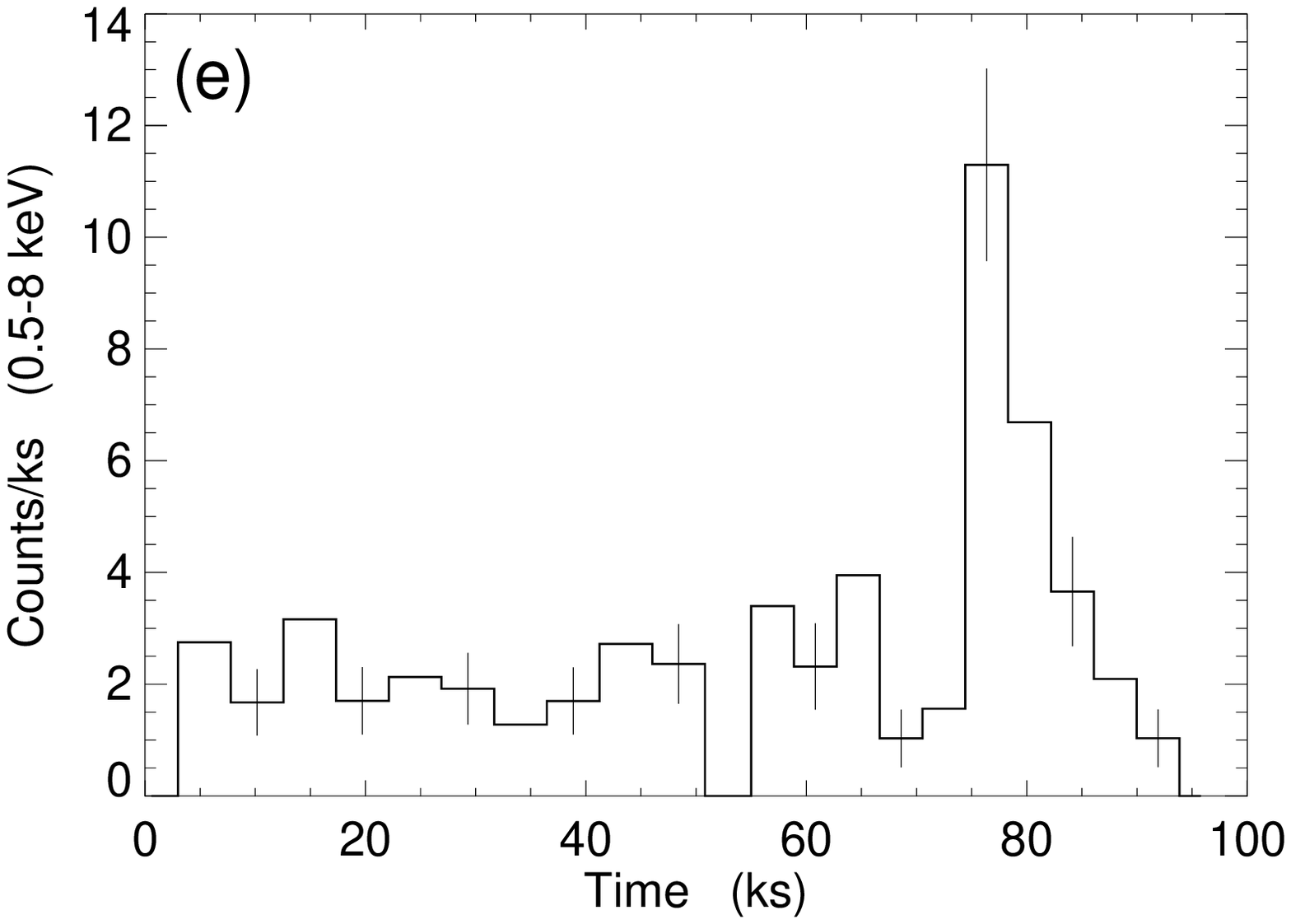}
  \end{minipage} \\ [-0.15in]
  \begin{minipage}[t]{1.0\textwidth}
  \centering
  \includegraphics[scale=0.4]{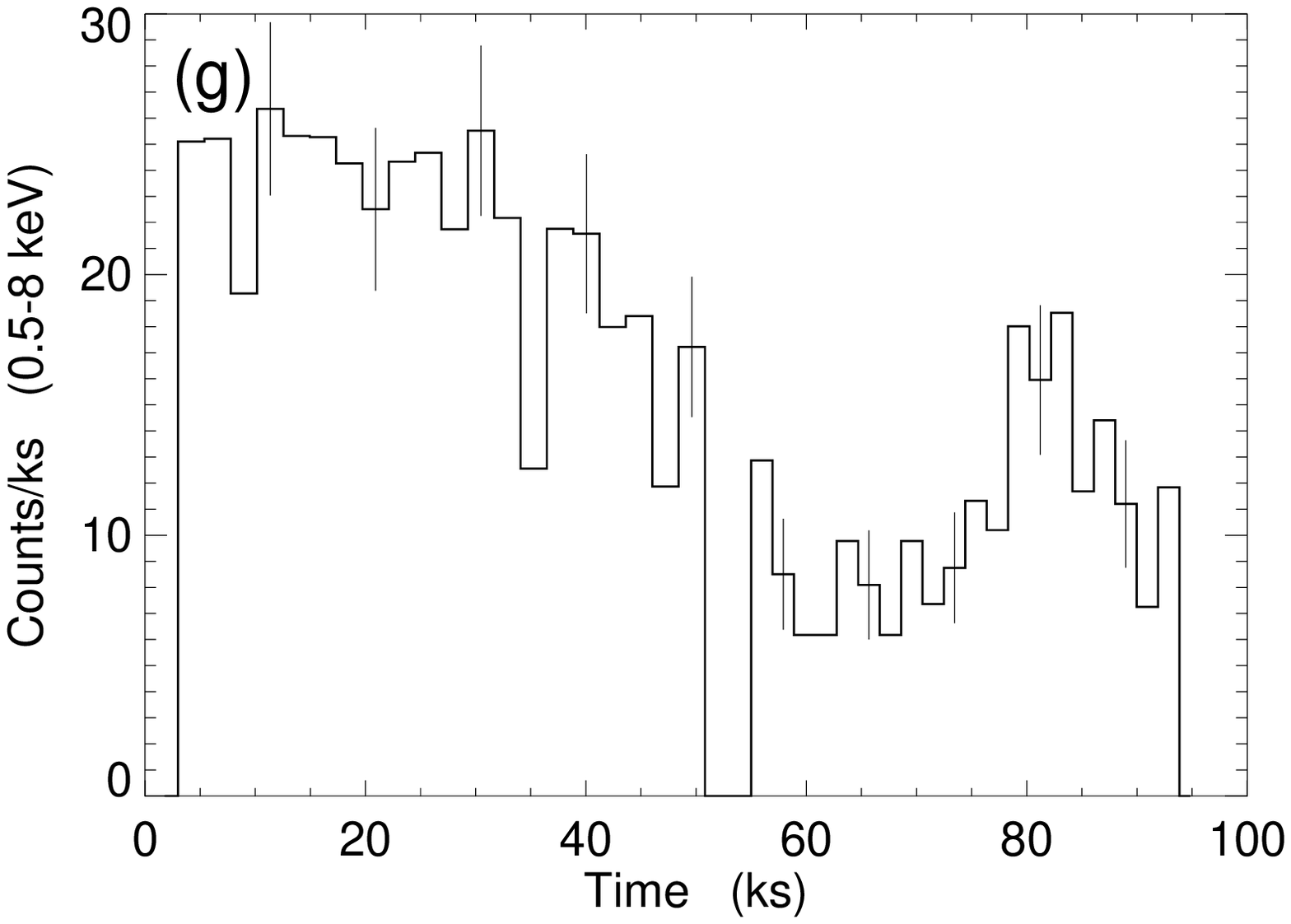}
  \includegraphics[scale=0.4]{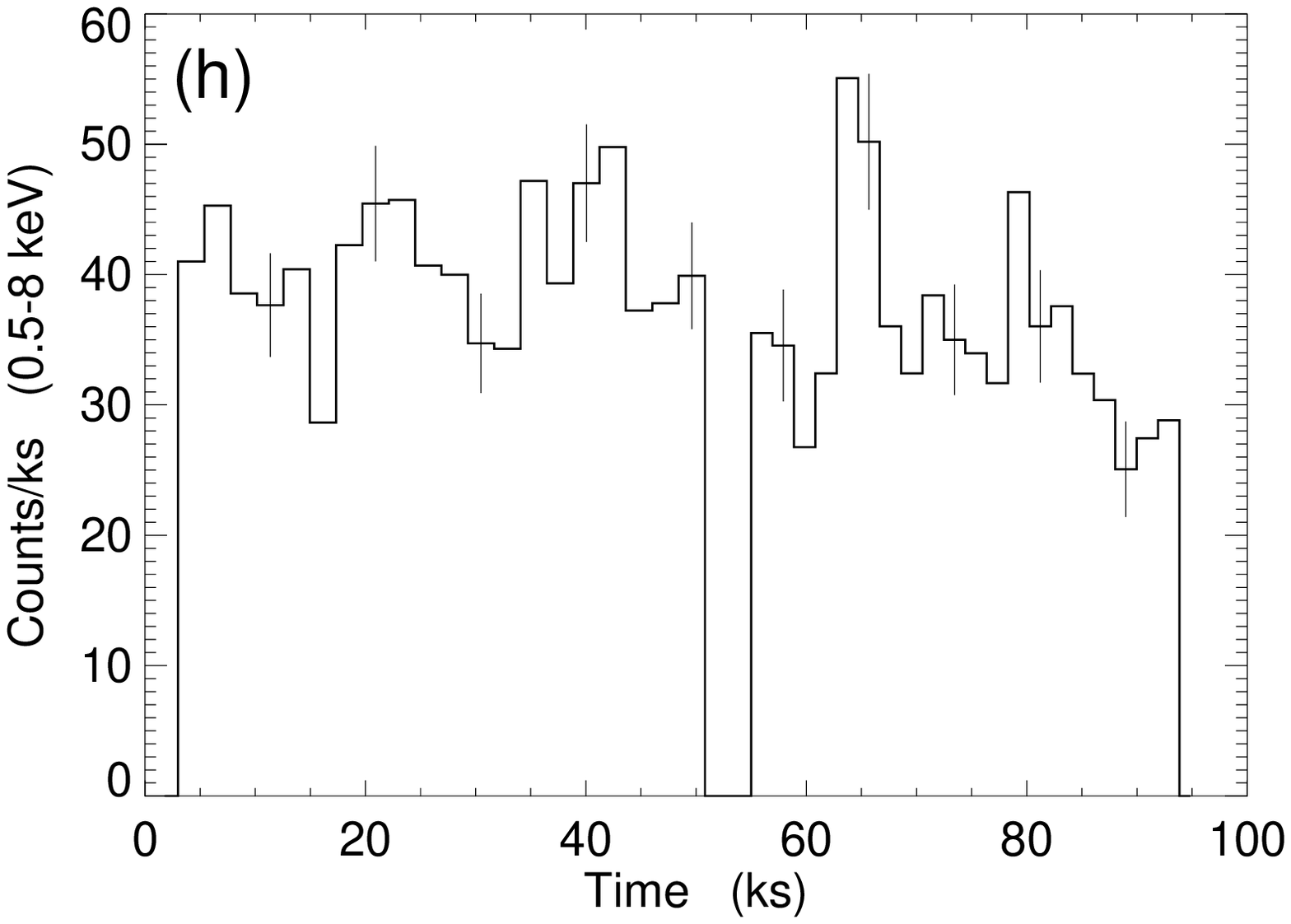}
\caption{Lightcurves of massive Trapezium stars in order of decreasing
mass: (a) P 1891 = $\theta^1$C Ori (O6, V=5.1); (b) P 1993 =
$\theta^2$A Ori (O9.5, V=5.1); (c) P 1865 = $\theta^1$A Ori (O7,
V=6.7); (d) P 1889 = $\theta^1$D Ori (B0.5, V=6.7); (e) P 2074 = NU Ori
(B1, V=6.9); (f) P 2031 = $\theta^2$B Ori (B1, V=6.0);  (g) P 1863a =
$\theta^1$B Ori (B0, V=8.0); and (h) P 2085 = $\theta^2$C Ori (B4,
V=8.2). Spectral types and magnitudes from SIMBAD.  For graphical
convenience, the two observations are plotted consecutively separated
by 5 ks, though in fact they are separated by $\simeq 6$ months.
\label{highmass_var_fig}}
  \end{minipage} 
\end{figure}

\clearpage
\newpage

\begin{figure}
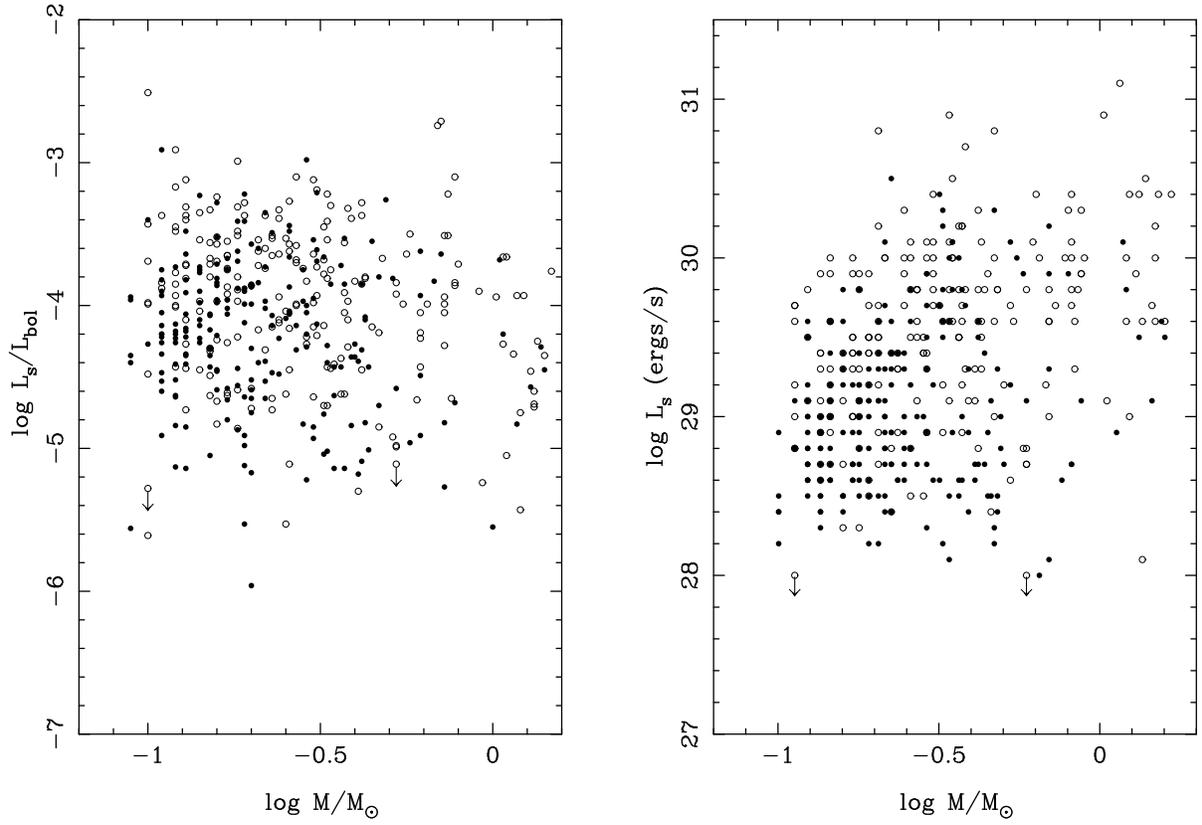

\centering
  \includegraphics[scale=0.45]{f14a.eps}  
  \hspace{0.3in}
  \includegraphics[scale=0.45]{f14b.eps}
\caption{X-ray emission of low mass ONC stars:  (left) soft band $\log
L_s/L_{bol}$ $vs.$ spectral type and mass; and (right) soft band $\log
L_s$ $vs.$ mass.  Open circles denote stars exhibiting intra-day
variability. 
\label{lowmass_ls_fig}}
\end{figure}

\clearpage
\newpage

\begin{figure}
\centering
\includegraphics[scale=0.8]{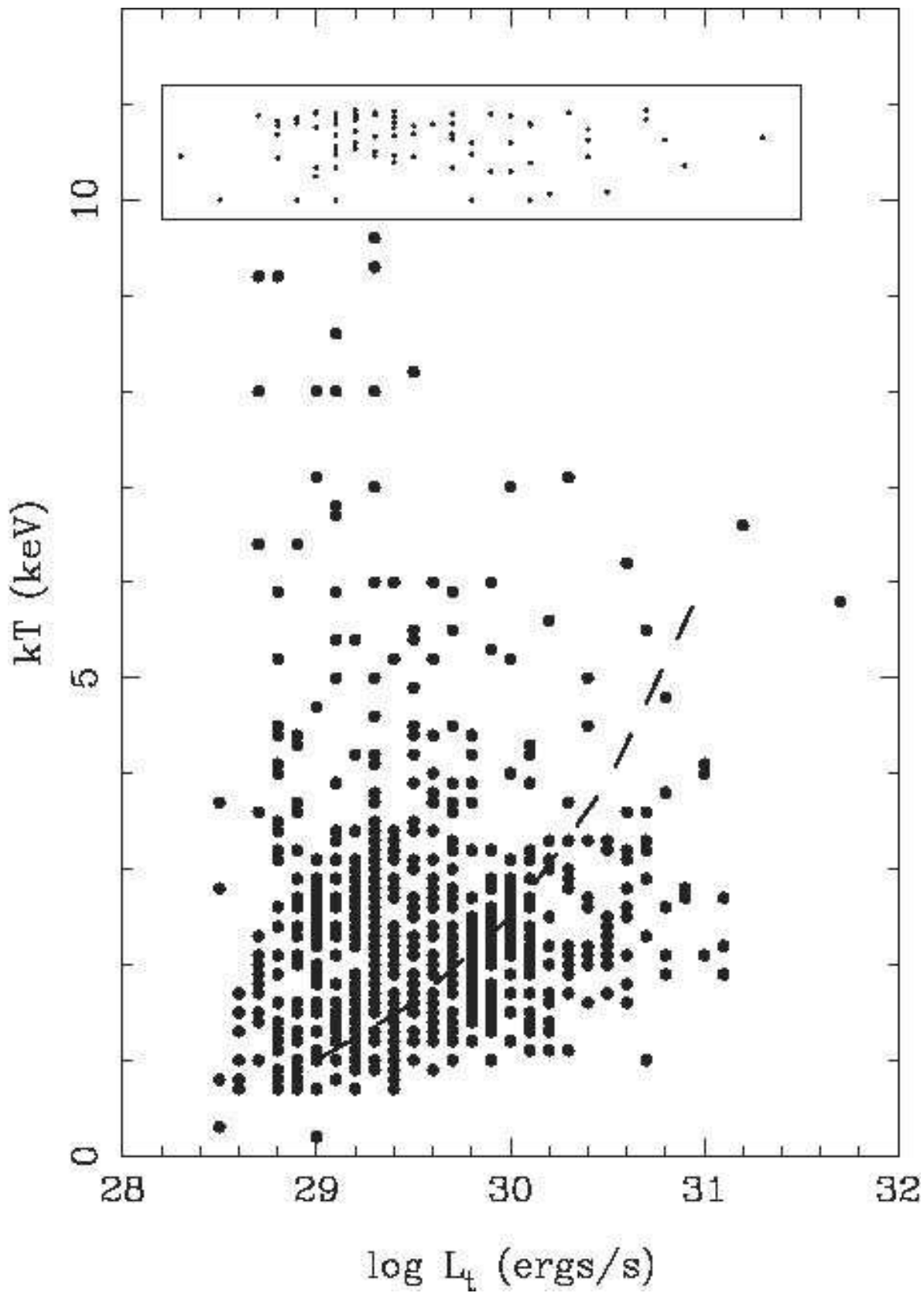}
\caption{Plasma energies $vs.$ full band X-ray luminosities for ONC
stars.  Only sources with $\geq 30$ counts, masses $\leq 2$ M$_\odot$
and successful 1-temperature spectral fits are included here.  The
inset box shows sources with fitted plasma energies $>10$ keV.  The
dashed curve shows the approximate locus for young stars found by
\citet{Preibisch97} from {\it ROSAT} data.
\label{lowmass_kt_fig}}
\end{figure}

\clearpage
\newpage

\begin{figure}
\centering
  \begin{minipage}[t]{1.0\textwidth}
  \centering
  \includegraphics[scale=0.4]{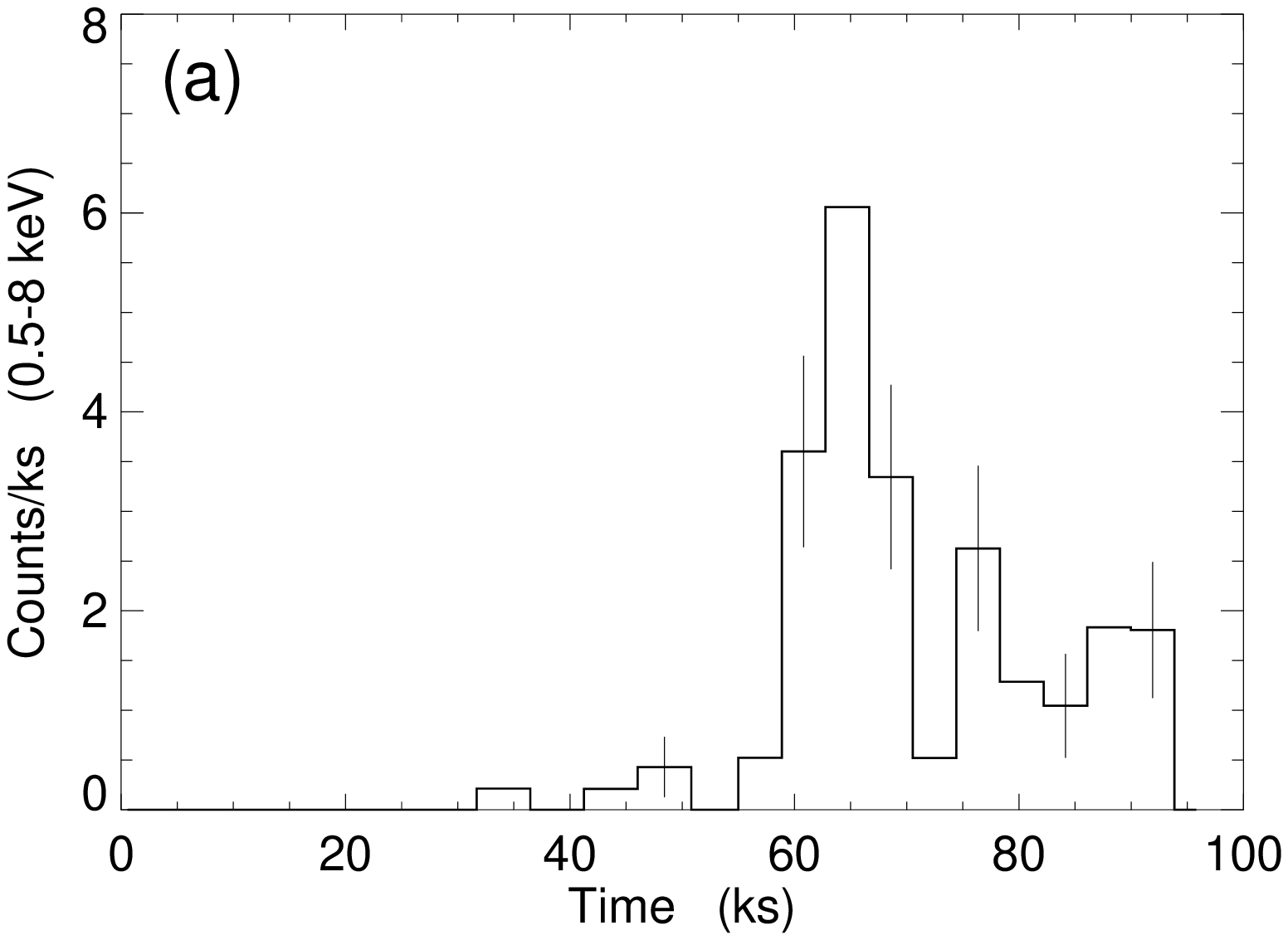}  
  \includegraphics[scale=0.4]{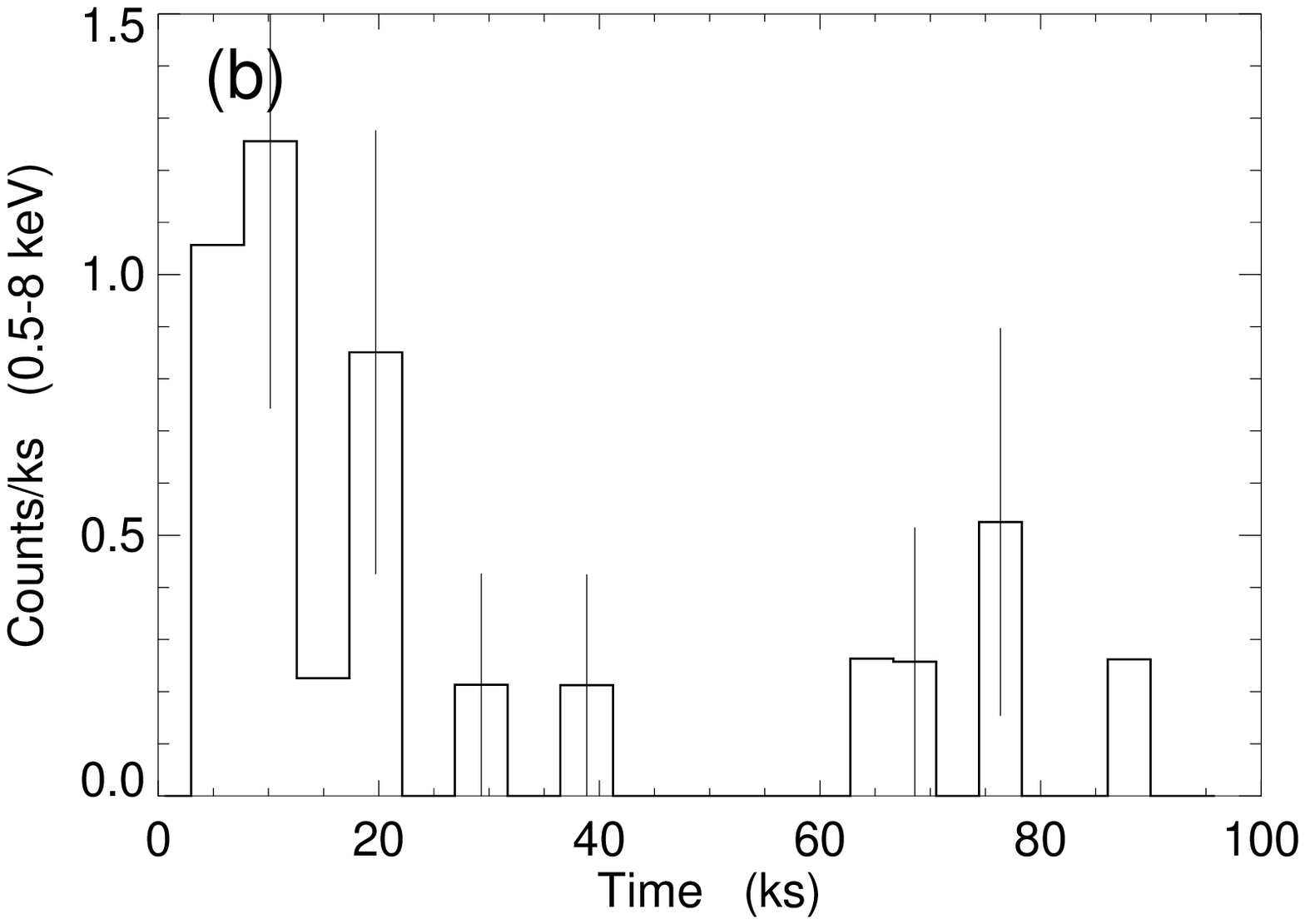}
  \end{minipage} \\ [-0.1in]
  \begin{minipage}[t]{1.0\textwidth}
  \centering
  \includegraphics[scale=0.4]{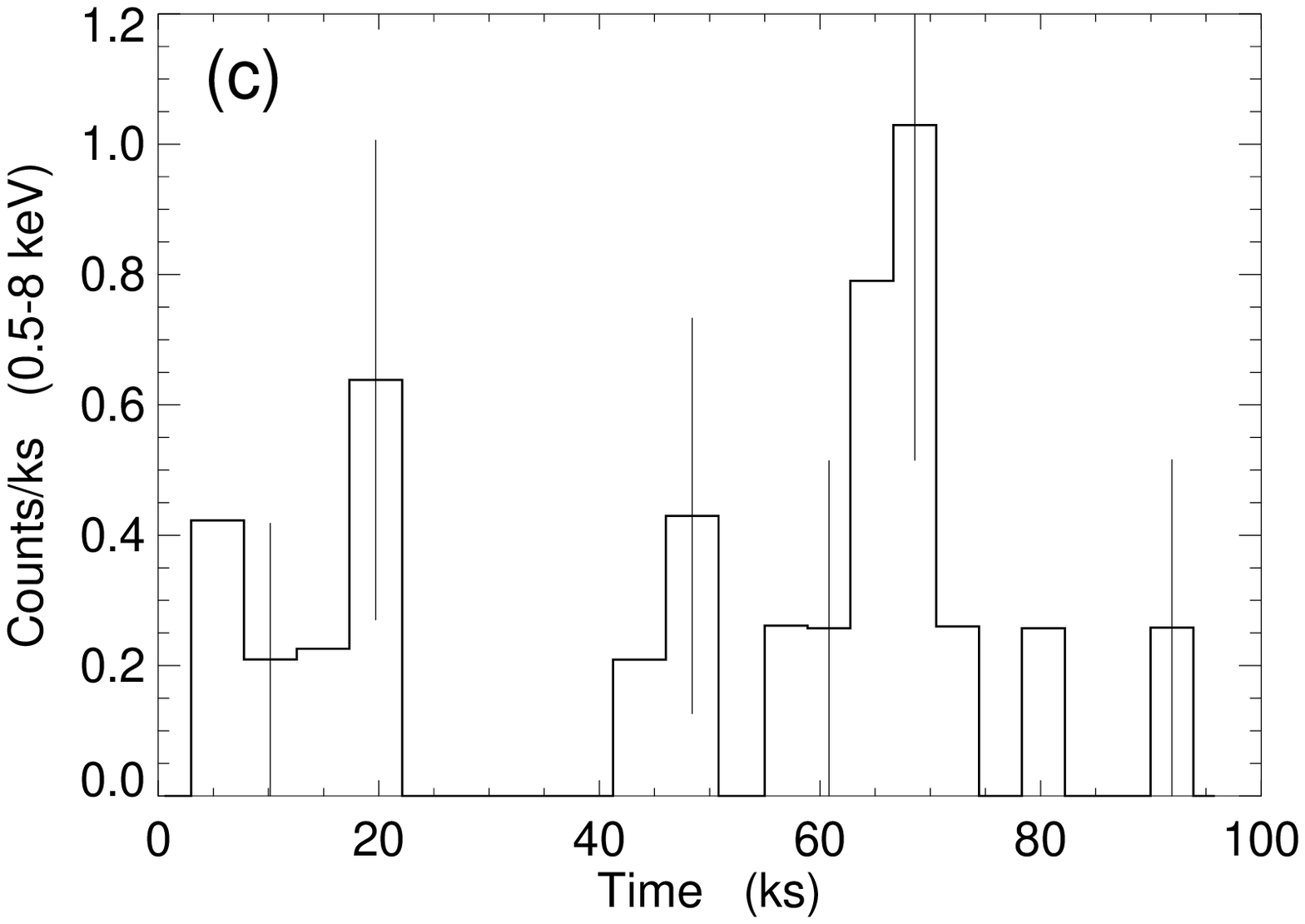}
  \includegraphics[scale=0.4]{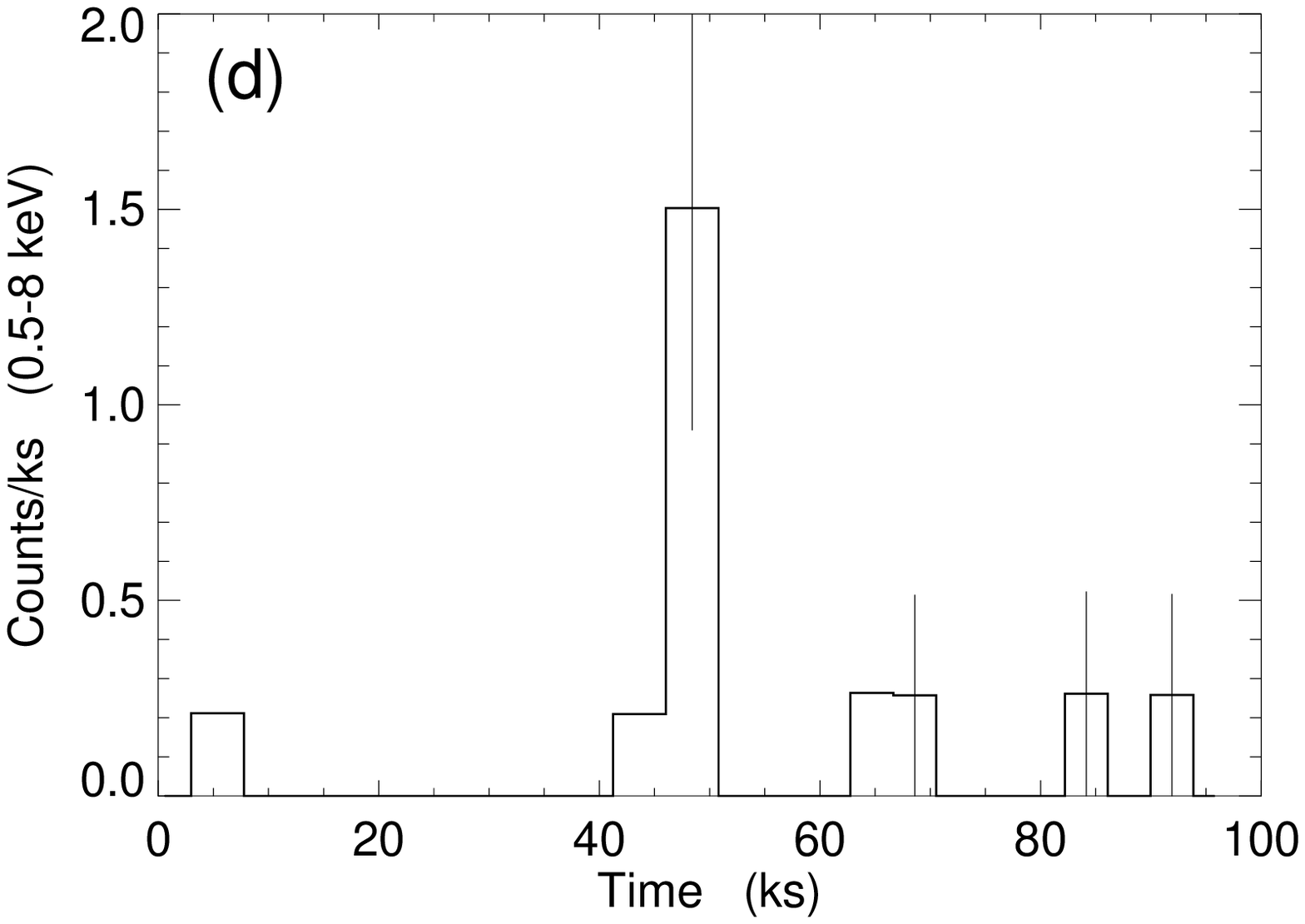}
  \end{minipage} \\ [-0.1in]  
  \begin{minipage}[t]{1.0\textwidth}
  \centering
  \includegraphics[scale=0.4]{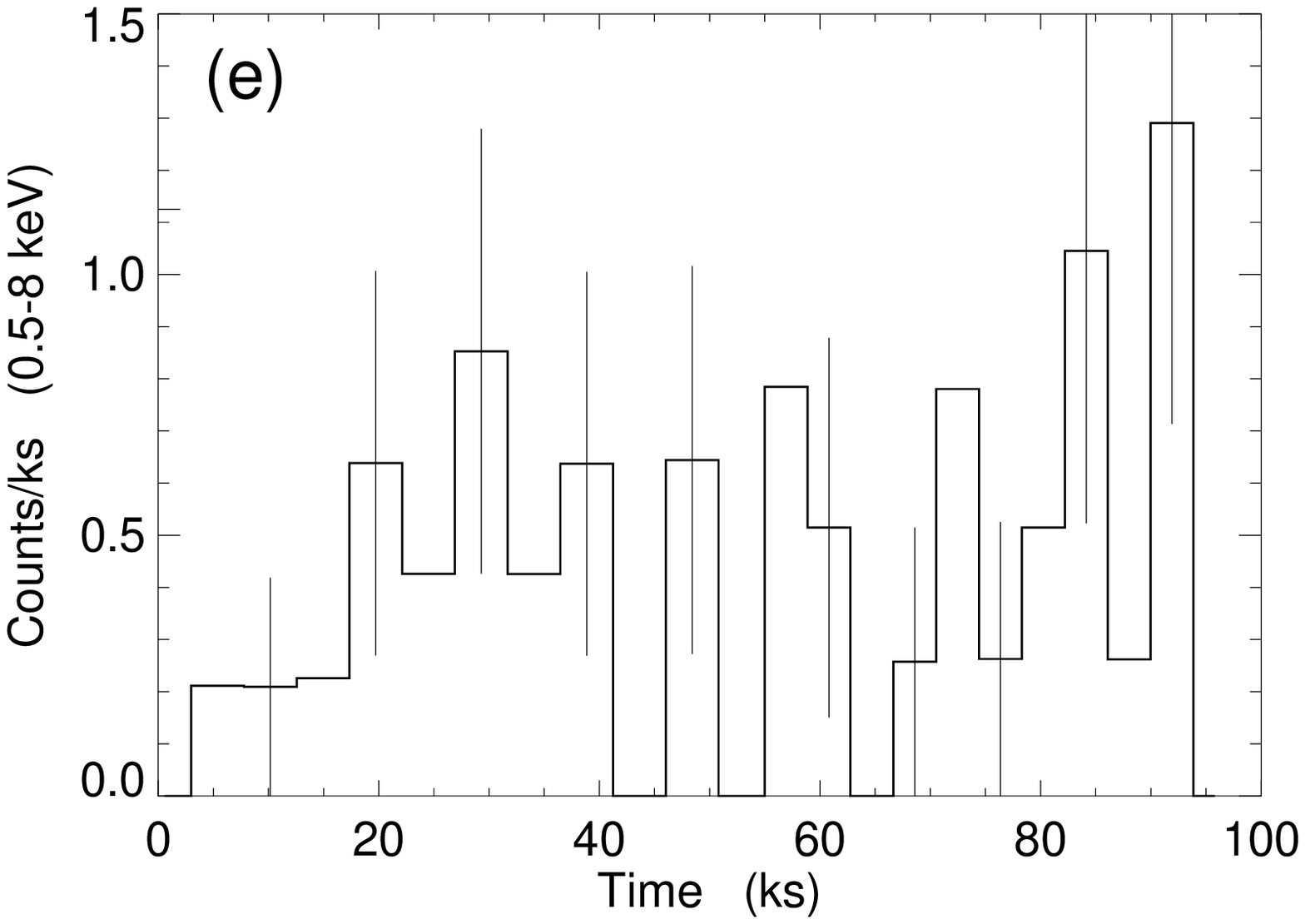}
  \includegraphics[scale=0.4]{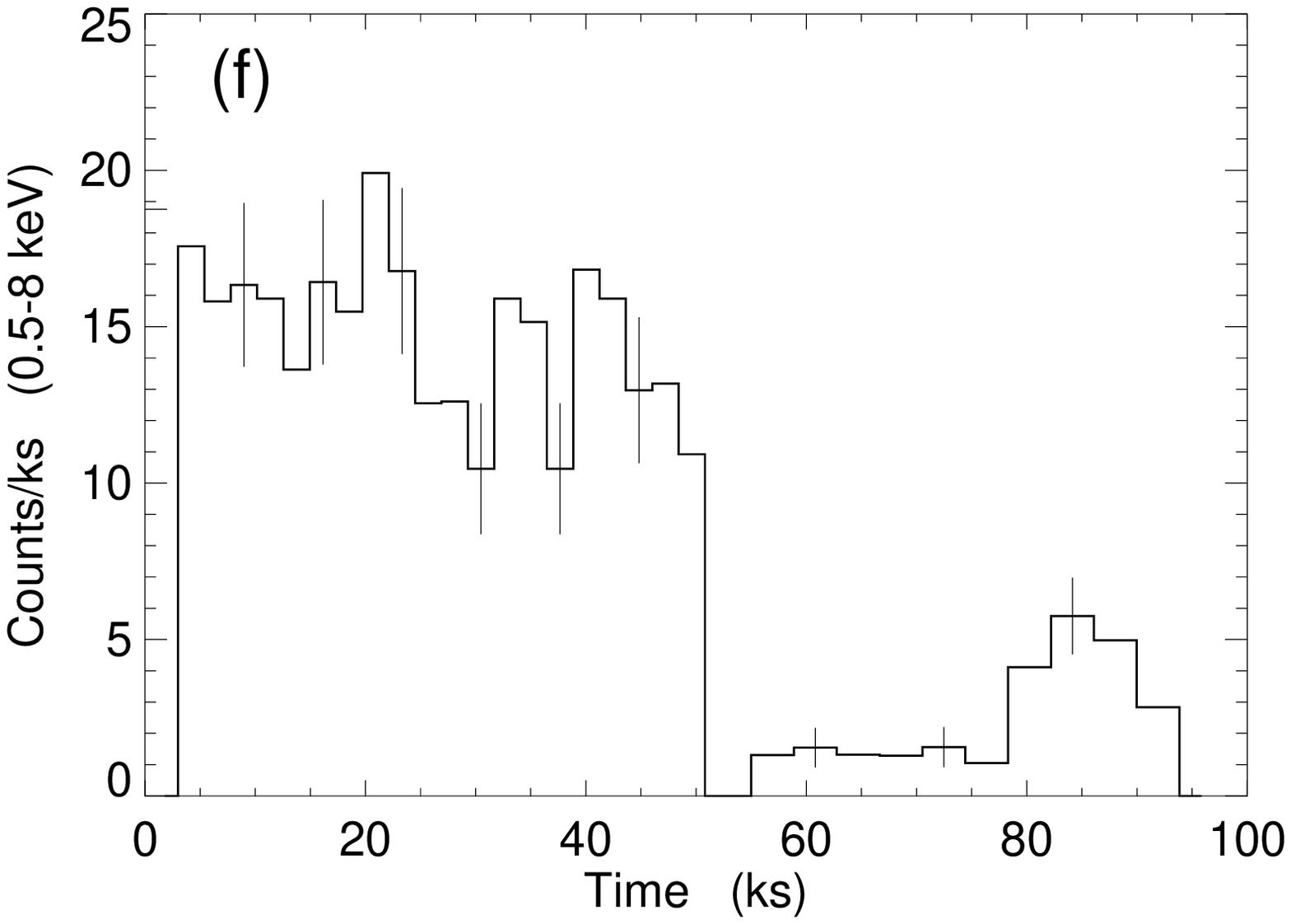}
  \end{minipage} \\ [-0.1in]
  \begin{minipage}[t]{1.0\textwidth}
  \centering
  \includegraphics[scale=0.4]{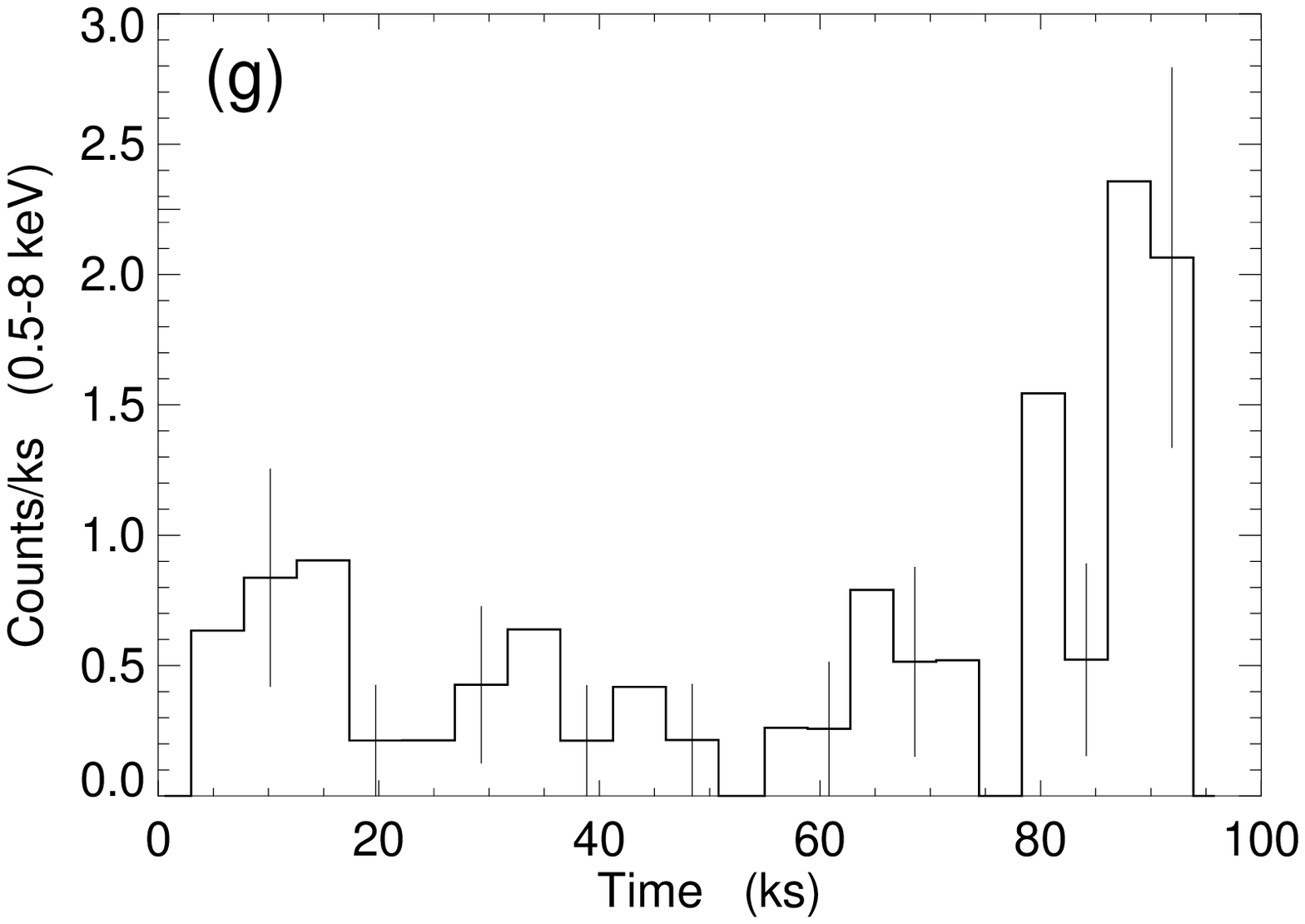}
\caption{Lightcurves of very low mass ONC objects with `Flare' or
`Possible flare' variability classifications:  (a) CHS 7273 ($M \simeq
0.07$ M$_\odot$); (b) HC 741 ($M \simeq 0.02$ M$_\odot$); (c) PSH 298
($M \simeq 0.08$ M$_\odot$); (d) PSH 116 ($M \simeq 0.04$ M$_\odot$);
(e) CHS 9480 ($M \simeq 0.06$ M$_\odot$); (f) HC 756 ($M \simeq 0.07$
M$_\odot$); and (g) CHS 11663 ($M \simeq 0.03$ M$_\odot$).  For
graphical convenience, the two observations are plotted consecutively 
separated by 5 ks, though in fact they are separated by $\simeq 6$ months.
\label{bd_var_fig}}
  \end{minipage} 
\end{figure}

\clearpage
\newpage

\begin{deluxetable}{cccccccc}
\tablewidth{0pt}
\tablecolumns{8}
\tablecaption{ACIS-I observations of the Orion Nebula Cluster 
\label{obs_table}}
\tablehead{
\multicolumn{4}{c}{Exposures} && 
\multicolumn{3}{c}{Background\tablenotemark{a} 
(cts arcsec$^{-2}$)} \\ \cline{1-4} \cline{6-8}
\colhead{\it Chandra} &
\colhead{T} &
\colhead{Start/stop times} &
\colhead{Eff.\ exp.} &&
\colhead{Soft} &
\colhead{Hard} &
\colhead{Total} \\
\colhead{ObsID} &
\colhead{($^\circ$C)} &
\colhead{(UT)} &
\colhead{(ks)} &&
\colhead{($0.5-2$ keV)} &
\colhead{($2-8$ keV)} &
\colhead{($0.5-8$ keV)} 
}
\startdata
  18   & -110    & 1999 Oct 12.43-13.04 & 45.3 && 0.025 & 0.049 & 0.074 \\
1522   & -120    & 2000 Apr  1.73-2.20  & 37.5 && 0.014 & 0.027 & 0.042 \\
Merged & \nodata &  \nodata             & 82.8 && 0.039 & 0.076 & 0.116 \\
\enddata

\tablenotetext{a}{These values apply $\theta > 3$\arcmin\/ away from
the bright Trapezium star $\theta^1$C Ori.  See \S \ref{extract_sec}
for the elevated background values in the inner region of the field.}

\end{deluxetable}

\clearpage
\newpage

\begin{deluxetable}{rrrrrcccrcrrrrrrcrcc}
\rotate
\tabletypesize{\scriptsize}
\tablewidth{680pt}
\tablecolumns{20}
\tablecaption{ACIS ONC sources and stellar counterparts (sample page) 
\label{src_list_table}}
\tablehead{
\multicolumn{6}{c}{X-ray source} && \multicolumn{2}{c}{Stellar ID} &&
\multicolumn{9}{c}{Stellar properties} & \multicolumn{1}{c}{Note} 
\\ \cline{1-6} \cline{8-9} \cline{11-19}
\colhead{Src}& 
\colhead{CXOONC J} &
\colhead{R.A.} &
\colhead{Dec.} &
\colhead{$\theta$} &
\colhead{Prev.} &&
\colhead{ID} & 
\colhead{$\phi$} &&
\colhead{log $T_{eff}$} &
\colhead{log $L_{bol}$} &
\colhead{$A_V$} & 
\colhead{log $M$} &
\colhead{log $t$} &
\colhead{$\Delta K$} &
\colhead{Other} &
\colhead{$P$} &
\colhead{$P$ ref.} &
\\
\colhead{}& 
\colhead{} &
\multicolumn{2}{c}{(J2000)} &
\colhead{\arcmin} &
\colhead{} &&
\colhead{} & 
\colhead{\arcsec} &&
\colhead{($^\circ$K)} &
\colhead{($L_\odot$)} &
\colhead{(mag)} & 
\colhead{($M_\odot$)} &
\colhead{(yr)} &
\colhead{(mag)} &
\colhead{} &
\colhead{(day)} &
\colhead{} &
}
\startdata

367 & 053513.5-052757 & 83.80637 & -5.46596 & 4.6 & \nodata &  & JW 418 & 
0.7 &  & \nodata & \nodata & \nodata & \nodata & \nodata & \nodata & 
\nodata & \nodata & \nodata & \nodata \\

368 & 053513.5-053057 & 83.80653 & -5.51605 & 7.6 & abe &  & JW 421 & 0.8 
&  &3.643 & 0.51 & 0.51 & -0.26 & 5.35 & 0.80 & \nodata & \nodata & 
\nodata & \nodata \\

369 & 053513.5-052355 & 83.80662 & -5.39866 & 0.9 & ef &  & HC 192 & 0.1 &  
& \nodata & \nodata & \nodata & \nodata & \nodata & \nodata & L & \nodata 
& \nodata & \nodata \\

370 & 053513.5-052400 & 83.80663 & -5.40000 & 0.9 & \nodata &  & HC 178 & 
0.7 &  & \nodata & \nodata & \nodata & \nodata & \nodata & \nodata & L & 
\nodata & \nodata & \nodata \\

371 & 053513.6-052425 & 83.80669 & -5.40711 & 1.3 & ce &  & JW 417 & 0.1 &  
& 3.568 & -0.16 & 1.58 & -0.48 & 5.72 & 0.56 & \nodata & 7.4 & H & \nodata 
\\

372 & 053513.6-052031 & 83.80676 & -5.34212 & 2.9 & e &  & PSH 47 & 0.4 &  
& 3.535 & -0.25 & 5.26 & -0.66 & 5.27 & -0.91 & \nodata & \nodata & 
\nodata & \nodata \\

373 & 053513.6-051745 & 83.80679 & -5.29593 & 5.7 & e &  & CHS 8393 & 1.0 
&  & \nodata & \nodata & \nodata & \nodata & \nodata & \nodata & \nodata & 
\nodata & \nodata & \nodata \\

374 & 053513.6-051954 & 83.80694 & -5.33188 & 3.5 & be &  & JW 413 & 0.7 &  
& 3.633 & -0.84 & 0.00 & -0.16 & 7.65 & \nodata & L & 10.1 & H & \nodata 
\\

375 & 053513.6-052846 & 83.80695 & -5.47960 & 5.4 & c &  & JW 422 & 0.4 &  
& \nodata & \nodata & \nodata & \nodata & \nodata & \nodata & \nodata & 
6.4 & S & \nodata \\

376 & 053513.6-052255 & 83.80703 & -5.38221 & 0.8 & e &  & \nodata & 
\nodata & & \nodata & \nodata & \nodata & \nodata & \nodata & \nodata & 
\nodata & \nodata & \nodata & \nodata \\

377 & 053513.6-051832 & 83.80704 & -5.30914 & 4.9 & e &  & CHS 8404 & 1.0 
&  & \nodata & \nodata & \nodata & \nodata & \nodata & \nodata & \nodata & 
\nodata & \nodata & \nodata \\

378 & 053513.7-052135 & 83.80724 & -5.35998 & 1.9 & e &  & HC 602 & 0.1 &  
& \nodata & \nodata & \nodata & \nodata & \nodata & \nodata & L & \nodata 
& \nodata & \nodata \\

379 & 053513.7-052221 & 83.80732 & -5.37278 & 1.2 & e &  & JW 420 & 0.3 &  
& 3.546 & 0.35 & 3.09 & -0.72 & 4.20 & \nodata & L pd & \nodata & \nodata 
& \nodata \\

380 & 053513.7-053024 & 83.80733 & -5.50682 & 7.1 & e &  & JW 428 & 0.8 &  
& 3.524 & -0.39 & 0.65 & -0.70 & 5.42 & 0.00 & \nodata & \nodata & \nodata 
& \nodata \\

381 & 053513.7-052217 & 83.80743 & -5.37151 & 1.3 & e &  & HC 499 & 0.1 &  
& \nodata & \nodata & \nodata & \nodata & \nodata & \nodata & \nodata & 
\nodata & \nodata & \nodata \\

382 & 053513.7-051743 & 83.80746 & -5.29550 & 5.7 & \nodata &  & \nodata & 
\nodata &  & \nodata & \nodata & \nodata & \nodata & \nodata & \nodata & 
\nodata &\nodata & \nodata & \nodata \\

383 & 053513.8-052207 & 83.80754 & -5.36862 & 1.4 & e &  & JW 423 & 0.2 &  
& 3.535 & 0.11 & 0.00 & -0.72 & 4.54 & 1.31 & pd & \nodata & \nodata & 
\nodata \\

384 & 053513.8-052202 & 83.80756 & -5.36746 & 1.5 & e &  & JW 424 & 0.2 &  
& \nodata & \nodata & \nodata & \nodata & \nodata & \nodata & \nodata & 
\nodata & \nodata & \nodata \\

385 & 053513.8-051925 & 83.80759 & -5.32377 & 4.0 & e &  & JW 419 & 0.5 &  
& \nodata & \nodata & \nodata & \nodata & \nodata & \nodata & \nodata & 
\nodata & \nodata & \nodata \\

386 & 053513.8-052209 & 83.80767 & -5.36921 & 1.4 & e &  & HC 525 & 0.2 &  
& \nodata & \nodata & \nodata & \nodata & \nodata & \nodata & \nodata & 
\nodata & \nodata & \nodata \\

387 & 053513.8-052425 & 83.80781 & -5.40722 & 1.2 & e &  & HC 130 & 0.2 &  
& \nodata & \nodata & \nodata & \nodata & \nodata & \nodata & \nodata & 
\nodata & \nodata & \nodata \\

388 & 053513.9-052229 & 83.80795 & -5.37495 & 1.1 & \nodata &  & HC 451 & 
0.7 &  & \nodata & \nodata & \nodata & \nodata & \nodata & \nodata & FIR & 
\nodata & \nodata & id \\

389 & 053513.9-052701 & 83.80803 & -5.45034 & 3.7 & e &  & CHS 8453 & 0.5 
&  & \nodata & \nodata & \nodata & \nodata & \nodata & \nodata & \nodata & 
\nodata & \nodata & x \\

390 & 053513.9-052319 & 83.80806 & -5.38888 & 0.6 & e &  & HC 314 & 0.2 &  
& \nodata & \nodata & \nodata & \nodata & \nodata & \nodata & pd & \nodata 
& \nodata & \nodata \\

391 & 053513.9-051853 & 83.80810 & -5.31482 & 4.5 & ce &  & JW 425 & 0.7 &  
& \nodata & \nodata & \nodata & \nodata & \nodata & \nodata & \nodata & 
\nodata & \nodata & \nodata \\

392 & 053513.9-052123 & 83.80826 & -5.35647 & 2.1 & ae &  & PSH 53 & 0.8 &  
& \nodata & \nodata & \nodata & \nodata & \nodata & \nodata & \nodata & 
\nodata & \nodata & \nodata \\

393 & 053514.0-052520 & 83.80837 & -5.42229 & 2.0 & e &  & PSH 93 & 1.1 &  
& \nodata & \nodata & \nodata & \nodata & \nodata & \nodata & pd & \nodata 
& \nodata & \nodata \\

394 & 053514.0-052636 & 83.80842 & -5.44337 & 3.3 & \nodata &  & JW 434 & 
0.9 &  & \nodata & \nodata & \nodata & \nodata & \nodata & \nodata & 
\nodata & \nodata & \nodata & x \\

395 & 053514.0-052338 & 83.80854 & -5.39400 & 0.7 & ef &  & JW 431 & 0.1 &  
& 3.513 & 0.22 & 0.68 & -0.82 & 4.10 & \nodata & r & \nodata & \nodata & 
\nodata \\

396 & 053514.0-052012 & 83.80862 & -5.33689 & 3.2 & \nodata &  & \nodata & 
\nodata &  & \nodata & \nodata & \nodata & \nodata & \nodata & \nodata & 
\nodata & \nodata & \nodata & \nodata \\

397 & 053514.0-051951 & 83.80863 & -5.33110 & 3.6 & ace &  & JW 429 & 0.4 
&  & 3.753 & 0.57 & 2.67 & 0.20 & 6.85 & \nodata & pd & \nodata & \nodata 
& \nodata \\

398 & 053514.0-052236 & 83.80867 & -5.37682 & 1.0 & ef &  & JW 432 & 0.0 &  
& 3.513 & 0.31 & 1.94 & -0.82 & 3.94 & 0.93 & r & \nodata & \nodata & 
\nodata \\

399 & 053514.0-052222 & 83.80873 & -5.37289 & 1.2 & ef &  & BN Obj & 0.6 &  
& \nodata & \nodata & \nodata & \nodata & \nodata & \nodata & FIR r & 
\nodata & \nodata & \nodata \\

400 & 053514.2-052613 & 83.80924 & -5.43699 & 2.9 & e &  & \nodata & 
\nodata &  & \nodata & \nodata & \nodata & \nodata & \nodata & \nodata & 
\nodata & \nodata & \nodata & \nodata \\

\enddata

\tablecomments{The full table of 1075 sources is available only on-line as a
machine-readable table} 

\end{deluxetable}

\clearpage
\newpage

{\bf Notes to Table \ref{src_list_table} (sample page)}

053513.9-052229
This is one of the high-mass, high-luminosity ($L_{FIR} \sim 1000$ 
L$_\odot$) deeply embedded stars in the BN/KL region.  Other designations 
include IRc 3 and ``Source i''.

053513.9-052701
This source lies near a readout trail.

053514.0-052636
This faint source lies near a readout trail.

053514.0-052222
This X-ray source is associated with the Becklin-Neugebauer Object,
which coincides with K-band source HC 705 and radio source B.  The
1.1\arcsec\/ offset between the ACIS position and radio position reported
by Garmire et al.\ (2000) is now 0.6\arcsec, consistent with a true
coincidence.

\clearpage
\newpage

\begin{deluxetable}{rrrrrrcrrccrrcccrrrrc}
\rotate
\tabletypesize{\scriptsize}
\tablewidth{650pt}
\tablecolumns{21}
\tablecaption{X-ray properties of ONC sources (sample, 1075 total) 
\label{src_prop_table}}
\tablehead{
\multicolumn{6}{c}{X-ray extraction} &&
\multicolumn{3}{c}{Variability} &&
\multicolumn{4}{c}{Spectrum} && 
\multicolumn{4}{c}{Luminosity} & 
\multicolumn{1}{c}{Note} \\  \cline{1-6} \cline{8-10} \cline{12-15} 
\cline{17-20}

\colhead{Src} & 
\colhead{CXOONC} &
\colhead{$C_{xtr}$} &
\colhead{$B_{xtr}$} &
\colhead{$R_{xtr}$} &
\colhead{$f_{PSF}$} &&
\colhead{$CR_1$} &
\colhead{$CR_2$} &
\colhead{Var Cl} &&
\colhead{log$N_H$} &
\colhead{$kT$} &
\colhead{$kT_1/kT_2$} &
\colhead{Feat} &&
\colhead{log$L_s$} &
\colhead{log$L_h$} &
\colhead{log$L_t$} &
\colhead{log$L_c$} &
\colhead{} \\

\colhead{} & 
\colhead{} &
\multicolumn{2}{c}{(counts)} &
\colhead{(\arcsec)} &
\colhead{} &&
\multicolumn{2}{c}{(ct ks$^{-1}$)} &
\colhead{} &&
\colhead{(cm$^{-2}$)} &
\multicolumn{2}{c}{(keV)} &
\colhead{} &&
\multicolumn{4}{c}{(erg s$^{-1}$)} &
\colhead{} 
}

\startdata

367 & 053513.5-052757 &   17 & 4 & 3.2 & 0.95 &  & 0.2 & 0.2 & Const  &  
&$<$20.0 & 2.9  & \nodata & \nodata &  & \nodata & \nodata & 28.3 & \nodata & f s \\

368 & 053513.5-053057 & 1350 & 15& 6.5 & 0.92 &  &11.6 & 24.0& Flare  &  &   
21.5 & 2.6  & \nodata & \nodata &  & 30.1 & 30.2 & 30.5 & 30.6 & v \\

369 & 053513.5-052355 &  340 & 3 & 1.8 & 0.92 &  & 7.3 & 0.7 & Flare  &  &   
22.7 & 5.0  & \nodata & \nodata &  & 28.7 & 30.4 & 30.4 & 30.8 & v \\

370 & 053513.5-052400 &   21 & 3 & 1.8 & 0.93 &  & 0.4 & 0.1 & LT Var &  &   
22.4 & 1.8  & \nodata & \nodata &  & \nodata & \nodata & 28.9 & \nodata & f 
\\

371 & 053513.6-052425 & 2086 & 5 & 2.3 & 0.96 &  & 9.9 & 45.1& Pos fl &  &   
21.3 & 2.6  & \nodata & \nodata &  & 30.3 & 30.3 & 30.6 & 30.7 & \nodata 
\\

372 & 053513.6-052031 &   51 & 2 & 2.2 & 0.95 &  & 0.7 & 0.4 & Const  &  &   
21.5 & 1.8  & \nodata & \nodata &  & 28.8 & 28.7 & 29.0 & 29.2 & \nodata 
\\

373 & 053513.6-051745 &   37 & 1 & 1.7 & 0.54 &  & 0.9 & 0.5 & Const  &  &   
22.6 & 1.3  & \nodata & \nodata &&$<$28.0 & 29.0 & 29.2 & 30.1 & \nodata 
\\

374 & 053513.6-051954 & 1178 & 6 & 4.0 & 0.97 &  &21.6 & 5.2 & Flare  &  &   
22.1 & 2.5  & \nodata & \nodata &  & 30.0 & 30.5 & 30.7 & 31.0 & v \\

375 & 053513.6-052846 &  108 & 6 & 4.1 & 0.94 &  & 0.1 & 2.8 & LT Var &  &   
21.6 & 4.0  & \nodata & \nodata &  & 29.1 & 29.5 & 29.6 & 29.8 & \nodata 
\\

376 & 053513.6-052255 &   28 & 4 & 1.8 & 0.93 &  & 0.5 & 0.1 & LT Var &  &   
22.7 & 1.9  & \nodata & \nodata && \nodata& \nodata & 29.0 & \nodata & f 
\\

377 & 053513.6-051832 &   32 & 5 & 3.6 & 0.95 &  & 0.6 & 0.0 & LT Var &  &   
20.1 & 3.6  & \nodata & \nodata &  & 28.5 & 28.5 & 28.8 & 28.8 & \nodata 
\\

378 & 053513.7-052135 &  146 & 2 & 1.9 & 0.93 &  & 1.8 & 1.9 & Const  &  &   
22.7 & 2.3  & \nodata & \nodata &  & 28.5 & 29.9 & 29.9 & 30.5 & \nodata 
\\

379 & 053513.7-052221 &   30 & 3 & 1.8 & 0.94 &  & 0.4 & 0.3 & Const  &  &   
21.3 & 6.4  & \nodata & \nodata &  & 28.4 & 28.7 & 28.9 & 29.0 & \nodata 
\\

380 & 053513.7-053024 &  151 & 14& 6.3 & 0.95 &  & 1.4 & 2.2 & Const  &  &   
20.7 & 1.3  & \nodata & \nodata &  & 29.1 & 28.6 & 29.3 & 29.3 & \nodata 
\\

381 & 053513.7-052217 &   26 & 3 & 1.8 & 0.93 &  & 0.3 & 0.4 & Const  &  &   
22.4 & 2.8  & \nodata & \nodata && \nodata& \nodata & 29.0 & \nodata & f 
\\

382 & 053513.7-051743 &   10 & 0 & 1.1 & 0.34 &  & 0.6 & 0.0 & Const  &  &   
22.4 &$>$10 & \nodata & \nodata && \nodata& \nodata & 28.8 & \nodata & f 
\\

383 & 053513.8-052207 &  634 & 0 & 1.0 & 0.89 &  & 8.2 & 8.6 & Const  &  &   
21.5 & 2.3  & \nodata & \nodata &  & 29.7 & 29.9 & 30.1 & 30.3 & \nodata 
\\

384 & 053513.8-052202 &   15 & 2 & 1.9 & 0.94 &  & 0.3 & 0.1 & Const  &  &   
22.0 & 3.0  & \nodata & \nodata &  & \nodata & \nodata & 28.6 & \nodata & f 
\\

385 & 053513.8-051925 &   18 & 2 & 2.5 & 0.93 &  & 0.3 & 0.1 & Const  &  
&$<$20.0 & 1.9  & \nodata & \nodata &  & \nodata & \nodata & 28.5 & \nodata & 
f \\

386 & 053513.8-052209 &  137 & 0 & 1.0 & 0.87 &  & 1.1 & 2.6 & Pos fl &  &   
22.2 & 1.7  & \nodata & \nodata &  & 29.0 & 29.5 & 29.6 & 30.1 & \nodata 
\\

387 & 053513.8-052425 &   93 & 3 & 1.8 & 0.93 &  & 1.4 & 0.9 & Pos fl &  
&\nodata&\nodata& 2/$>$10 & \nodata &  & 28.4 & 29.5 & 29.6 & \nodata & 
\nodata \\

388 & 053513.9-052229 &   13 & 3 & 1.8 & 0.92 &  & 0.1 & 0.3 & Const  &  &   
22.8 & 2.8  & \nodata & \nodata && \nodata& \nodata & 28.9 & \nodata & f 
\\

389 & 053513.9-052701 &   42 & 1 & 1.2 & 0.68 &  & 0.9 & 0.6 & Const  &  &   
22.2 &$>$10 & \nodata & \nodata &  & 28.1 & 29.3 & 29.3 & 29.5 & \nodata 
\\

390 & 053513.9-052319 &   36 & 5 & 1.9 & 0.95 &  & 0.5 & 0.4 & Const  &  
&$<$20.0 & 4.5  & \nodata & \nodata &  & 28.4 & 28.6 & 28.8 & 28.8 & 
\nodata \\

391 & 053513.9-051853 &   87 & 4 & 3.3 & 0.94 &  & 1.1 & 0.9 & Const  &  &   
21.7 & 2.9  & \nodata & \nodata &  & 28.9 & 29.2 & 29.4 & 29.5 & \nodata 
\\

392 & 053513.9-052123 &  250 & 2 & 1.9 & 0.93 &  & 2.7 & 3.7 & Const  &  &   
22.2 & 3.1  & \nodata & \nodata &  & 29.2 & 29.9 & 30.0 & 30.3 & \nodata 
\\

393 & 053514.0-052520 &   41 & 2 & 1.9 & 0.96 &  & 0.4 & 0.5 & Const  &  
&$<$20.0 & 1.4  & \nodata & \nodata &  & 28.6 & 28.1 & 28.7 & 28.7 & 
\nodata \\

394 & 053514.0-052636 &    9 & 2 & 2.6 & 0.96 &  & 5.0 & 7.1 & LT Var &  
&$<$20.0 & 3.0  & \nodata & \nodata && \nodata&\nodata&27.9 & \nodata & f s \\

395 & 053514.0-052338 &  359 & 4 & 1.9 & 0.96 &  & 3.9 & 4.9 & Pos fl &  &   
21.0 & 1.4  & \nodata & $\star$ &  & 29.5 & 29.1 & 29.7 & 29.8 & \nodata 
\\

396 & 053514.0-052012 &   25 & 2 & 2.3 & 0.93 &  & 0.1 & 0.6 & Flare  &  &   
23.1 & 1.1  & \nodata & \nodata && \nodata& \nodata & 29.2 & \nodata & f v \\

397 & 053514.0-051951 &  485 & 2 & 2.5 & 0.96 &  & 5.0 & 7.2 & LT Var &  
&$<$20.0 & 1.5  & \nodata & $\star$ &  & 29.8 & 29.3 & 29.9 & 29.9 & 
\nodata \\

398 & 053514.0-052236 &  218 & 3 & 1.8 & 0.96 &  & 1.2 & 4.4 & Pos fl &  &   
21.3 & 1.3  & \nodata & \nodata &  & 29.4 & 29.0 & 29.6 & 29.8 & \nodata 
\\

399 & 053514.0-052222 &   45 & 3 & 1.8 & 0.92 &  & 0.5 & 0.6 & Const  &  &   
22.6 &$>$10 & \nodata & \nodata && $<$28.0& 29.4 & 29.4 & 29.6 & \nodata 
\\

400 & 053514.2-052613 &   17 & 2 & 2.1 & 0.93 &  & 0.2 & 0.2 & Const  &  &   
22.6 & 2.8  & \nodata & \nodata && \nodata& \nodata & 28.8 & \nodata & f 
\\

\enddata

\tablecomments{The full table of 1075 sources is available only on-line as a
machine-readable table} 

\end{deluxetable}

\clearpage
\newpage

{\bf Notes to Table \ref{src_prop_table} (sample page)}

053513.5-052757
The spectral fit does not account for most of the emission below 1 keV.

053513.5-053057
Flares seen during both observations.  First, the source rose from $CR_q
\sim 0.005$ cts s$^{-1}$ to $CR_p = 0.018$ cts s$^{-1}$, remaining
high for $> 6$ hours.  Second, a short-duration (2 hours) event is seen
with $CR_p = 0.037$ cts s$^{-1}$ superposed on a quiescent level
$CR_q = 0.02$ cts s$^{-1}$.

053513.5-052355
High amplitude flare towards the end of the observation with $CR_q =
0.003$ cts s$^{-1}$ rising to $CR_p > 0.03$ cts s$^{-1}$ in $\simeq 3$
hours.

053513.6-051954
A very short flare with $CR_p = 0.03$ cts s$^{-1}$ and duration $<1$
hour appears to be superposed on the $>12$ hour decay of a larger flare.

053514.0-052636
The spectral model was obtained by fixing the plasma energy at 3 keV.

053514.0-052012
The observation begins during a flare with $CR_p > 0.0024$ cts s$^{-
1}$ and decays to $CR_q \sim 0.0003$ cts s$^{-1}$ over 2 hours.

053514.2-052004
Flare lasting several hours with $CR_p \simeq 0.0055$ cts s$^{-1}$
follows a quiescent level $CR_q \simeq 0.0015$ cts s$^{-1}$.

053514.2-052304
The spectral fit for this source is poor.

053514.3-052308
The spectra fit may have overestimated the soft absorption for this
source.

053514.3-052317
Dramatic flare at the end of an observation from $CR_q = 0.002$ cts
s$^{-1}$ to $CR_p > 0.016$ cts s$^{-1}$ with a rise of 3 hours.  The
spectral fit has greatly overestimated the soft band absorption for this
source.

053514.3-052232
The emission rises from $CR_q \sim 0.002$ cts s$^{-1}$ to $CR_p \simeq
0.008$ cts s$^{-1}$ with rise time 2 hours.  The broad-band spectrum of
the source is unusual with a nearly flat distribution across the
{\it Chandra} band. It can be interpreted as a very hard component with 
high absorption ($\log N_H \sim 22.5$ cm$^{-2}$) plus a soft unabsorbed
component.  The luminosity of the latter component is underestimated in
the spectral fit and tabulated $L_s$ value.

053514.3-052219
The spectral model was obtained by fixing the plasma energy at 3 keV.

053514.3-052333
Rise to $CR_p \simeq 0.009$ cts s$^{-1}$ from $CR_q \simeq 0.002$
cts s$^{-1}$ over several hours.

053514.4-052903
A hard spectral component not included in the spectral fit above 2 keV
may be present.

\clearpage
\newpage

\begin{deluxetable}{rrrcrcrc}

\tabletypesize{\scriptsize}
\tablewidth{347pt}
\tablecolumns{8}

\tablecaption{X-ray sources without stellar identifications  
\label{noid_table}}

\tablehead{

\colhead{Src} &
\colhead{CXOONC} &
\colhead{$C_{xtr}$} &
\colhead{Var Cl} &
\colhead{$\log N_H$} &
\colhead{kT} &
\colhead{$\log L_t$} &
\colhead{Suggested} \\

&
&
&
&
\colhead{(cm$^{-2}$)} &
\colhead{(keV)} &
\colhead{(erg s$^{-1}$)} &
\colhead{class} \\
}

\startdata

3  & 053439.6-052458 & 45 & Const & 21.6 & $>$10 & 29.4~~~ & ONC\\
24 & 053446.8-052342 & 29 & Const & $<$20.0 & $>$10 & 28.7~~~ & ONC\\
27 & 053447.9-052054 & 35 & Const & 22.6 & $>$10 & 29.1~~~ & Embd/Bk  \\
35 & 053450.2-052323 & 36 & Const & 22.0 & $>$10 & 29.2~~~ & Embd/Bk  \\
54 & 053452.9-052617 & 30 & Const & 21.4 & $>$10 & 28.9~~~ & ONC\\
57 & 053453.4-052650 & 44 & Const & 21.7 & $>$10 & 29.1~~~ & ONC\\
58 & 053453.6-051354 & 43 & Const & \nodata & 5/$>$10 & 29.2~~~ & ONC\\
71 & 053455.5-051529 & 87 & Const & \nodata & 0.1/$>$10 & 29.4~~~ & ONC\\
73 & 053455.8-052337 & 44 & Const & 22.6 & $>$10 & 29.5~~~ & Embd/Bk  \\
79 & 053456.2-052228 & 123 & Const & 22.3 & $>$10 & 29.8~~~ & Embd/Bk  \\
86 & 053457.0-051500 & 59 & LT Var & \nodata & 0.1/$>$10 & 29.0~~~ & ONC\\
89 & 053457.7-052223 & 23 & Const & 22.1 & $>$10 & 29.0~~~ & Embd/Bk  \\
100 & 053459.4-052615 & 27 & Const & 21.9 & $>$10 & 29.0~~~ & ONC\\
105 & 053500.1-052549 & 13 & Const & 22.3 & 2.7 & 28.6~~~ & Embd/Bk  \\
122 & 053501.7-052512 & 21 & Const & 22.3 & $>$10 & 29.0~~~ & Embd/Bk  \\
141 & 053503.8-052941 & 170 & Pos fl & 22.2 & $>$10 & 29.9~~~ & Embd/Bk  \\
151 & 053504.4-051951 & 28 & Const & 22.0 & $>$10 & 29.0~~~ & Embd/Bk  \\
186 & 053506.3-052335 & 16 & Const & 22.4 & $>$10 & 29.1~~~ & Embd/Bk  \\
193 & 053506.5-052734 & 10 & Const & 22.1 & 4.2 & 28.3~~~ & Embd/Bk  \\
195 & 053506.6-051622 & 31 & LT Var & 22.3 & $>$10 & 29.0~~~ & Embd/Bk  \\
201 & 053507.3-052253 & 13 & Const & 23.2 & 1.8 & 28.9~~~ & Embd/Bk  \\
202 & 053507.3-052547 & 14 & Const & 21.1 & 0.7 & 28.2~~~ & ONC\\
206 & 053507.4-052301 & 8 & Const & 22.0 & 3.0 & 28.3~~~ & Embd/Bk  \\
212 & 053507.8-052029 & 14 & Const & 23.2 & 2.1 & 28.8~~~ & Embd/Bk  \\
227 & 053508.6-052022 & 23 & Pos fl & 22.0 & 2.9 & 28.8~~~ & Embd/Bk  \\
250 & 053510.1-052004 & 11 & Const & 23.0 & 0.9 & 28.6~~~ & Embd/Bk  \\
258 & 053510.4-052223 & 28 & Pos fl & 22.5 & $>$10 & 29.3~~~ & Embd/Bk  \\
288 & 053511.6-052729 & 57 & Pos fl & 22.1 & 4.1 & 29.3~~~ & Embd/Bk  \\
312 & 053512.2-052424 & 26 & LT Var & 23.1 & 6.7 & 29.4~~~ & Embd/Bk  \\
315 & 053512.3-052241 & 12 & Const & 22.3 & 1.1 & 28.4~~~ & OMC 1\\
324 & 053512.6-052205 & 11 & Const & 22.5 & 1.4 & 28.6~~~ & Embd/Bk  \\
336 & 053512.9-052351 & 17 & Const & $<$20.0 & $>$10 & 28.6~~~ & ONC\\
337 & 053512.9-052354 & 27 & LT Var & 23.2 & $>$10 & 29.5~~~ & OMC 1S\\
351\tablenotemark{a} & 053513.2-052254 & 121 & LT Var & 23.4~~~ & $>$10 & 
30.2~~~ & OMC 1\\
354 & 053513.2-052239 & 74 & Pos fl & 22.8 & $>$10 & 30.0~~~ & OMC 1\\
362 & 053513.4-052354 & 5 & Const & 22.0 & 3.0 & 28.2~~~ & OMC 1S\\
376 & 053513.6-052255 & 28 & LT Var & 22.7 & 1.9 & 29.0~~~ & OMC 1\\
382 & 053513.7-051743 & 10 & Const & 22.4 & $>$10 & 28.4~~~ & Embd/Bk  \\
396 & 053514.0-052012 & 25 & Flare & 23.1 & 1.1 & 29.2~~~ & Embd/Bk  \\
400 & 053514.2-052613 & 17 & Const & 22.6 & 2.8 & 28.8~~~ & Embd/Bk  \\
406 & 053514.3-052317 & 221 & Flare & 23.3 & $>$10 & 30.3~~~ & Embd/Bk  \\
410 & 053514.3-052219 & 9 & Const & 23.1 & 3.0 & 28.6~~~ & OMC 1\\
417 & 053514.5-052630 & 22 & LT Var & 23.2 & 1.1 & 29.2~~~ & Embd/Bk  \\
419 & 053514.5-052315 & 70 & Const & \nodata & 0.3/3 & 29.5~~~ & ONC\\
422 & 053514.5-052407 & 131 & Flare & 23.2 & $>$10 & 30.1~~~ & OMC 1S\\
428 & 053514.6-052210 & 74 & LT Var & \nodata & 4/$>$10 & 29.9~~~ & OMC 1\\
431 & 053514.7-052412 & 50 & Pos fl & \nodata & 0.7/$>$10 & 29.7~~~ & OMC 
1S\\
435 & 053514.8-052057 & 11 & Const & 22.0 & 4.9 & 28.5~~~ & OMC 1N\\
436 & 053514.8-052406 & 94 & LT Var & 22.7 & $>$10 & 29.8~~~ & OMC 1S\\
442 & 053514.9-052225 & 625 & Pos fl & 23.1 & $>$10 & 30.7~~~ & OMC 1\\
450 & 053515.0-052336 & 9 & Const & $<$20.0 & 3.0 & 28.2~~~ & ONC\\
453\tablenotemark{b} & 053515.1-052229 & 35 & Const & \nodata & 0.1/$>$10 
& 29.3~~~ & OMC 1\\
454 & 053515.1-052238 & 8 & Const & $<$20.0 & 3.0 & 28.3~~~ & ONC\\
455 & 053515.1-052201 & 41 & Const & 20.6 & 1.9 & 28.8~~~ & ONC\\
456 & 053515.1-052217 & 222 & Pos fl & \nodata & 0.5/6 & 30.2~~~ & OMC 1\\
469 & 053515.3-052218 & 71 & Pos fl & \nodata & 0.9/$>$10 & 29.8~~~ & OMC 1\\
476 & 053515.4-051934 & 18 & Const & 22.1 & $>$10 & 28.9~~~ & OMC 1N\\
483 & 053515.6-052126 & 378 & Const & 23.3 & $>$10 & 30.5~~~ & OMC 1N\\
494 & 053515.7-051808 & 24 & LT Var & 22.3 & 2.1 & 29.0~~~ & Embd/Bk  \\
500 & 053515.8-052318 & 171 & Const & 20.5 & 2.0 & 29.4~~~ & ONC\\
509\tablenotemark{c} & 053515.9-052319 & 39 & Const & $<$20.0 & 2.0 & 28.7~~~ 
& ONC\\
515 & 053516.0-051944 & 38 & Const & 23.1 & $>$10 & 29.5~~~ & OMC 1N\\
532 & 053516.2-052306 & 11 & Const & $<$20.0 & $>$10 & 28.4~~~ & ONC\\
546 & 053516.5-052054 & 12 & Const & 22.6 & $>$10 & 28.8~~~ & OMC 1N\\
579 & 053517.1-052129 & 64 & LT Var & 21.5 & 3.9 & 29.1~~~ & ONC\\
580 & 053517.1-051813 & 53 & Const & 22.5 & 4.2 & 29.3~~~ & OMC 1N\\
587 & 053517.3-052051 & 8 & Const & 22.6 & 3.0 & 28.5~~~ & OMC 1N\\
595 & 053517.4-052315 & 17 & Const & $<$20.0 & 0.8 & 28.4~~~ & ONC\\
610 & 053517.7-051833 & 317 & LT Var & 22.4 & 2.5 & 29.9~~~ & OMC 1N\\
635 & 053518.0-052056 & 17 & Const & 22.9 & 2.3 & 28.9~~~ & Embd/Bk  \\
660 & 053518.5-052232 & 17 & Const & $<$20.0 & 5.1 & 28.6~~~ & ONC\\
665 & 053518.7-051905 & 597 & Flare & 23.2 & $>$10 & 30.7~~~ & OMC 1N\\
683 & 053519.1-052112 & 16 & Const & 22.2 & $>$10 & 29.1~~~ & Embd/Bk  \\
686 & 053519.1-052118 & 28 & Pos fl & 23.5 & 0.9 & 29.3~~~ & Embd/Bk  \\
701 & 053519.7-052110 & 22 & Const & 22.8 & $>$10 & 29.3~~~ & Embd/Bk  \\
702 & 053519.7-052155 & 9 & Const & 22.0 & 3.0 & 28.3~~~ & Embd/Bk  \\
715 & 053520.0-052038 & 131 & LT Var & 22.9 & 1.4 & 29.8~~~ & Embd/Bk  \\
756 & 053520.9-052234 & 36 & Const & 22.6 & 3.4 & 29.3~~~ & Embd/Bk  \\
776 & 053521.5-051752 & 17 & Const & 22.0 & $>$10 & 28.6~~~ & Embd/Bk  \\
779 & 053521.6-051952 & 57 & Flare & 23.1 & $>$10 & 29.7~~~ & Embd/Bk  \\
839 & 053523.4-051957 & 44 & Pos fl & 23.0 & 1.2 & 29.2~~~ & Embd/Bk  \\
842 & 053523.4-052001 & 72 & Pos fl & 22.4 & 1.4 & 29.1~~~ & Embd/Bk  \\
862 & 053524.0-052125 & 10 & Const & 22.6 & 1.1 & 28.4~~~ & Embd/Bk  \\
872 & 053524.3-052206 & 12 & Const & 22.5 & $>$10 & 28.6~~~ & Embd/Bk  \\
881 & 053524.6-052759 & 3703 & Const & 22.2 & $>$10 & 31.3~~~ & Embd/Bk  \\
888 & 053525.0-052326 & 26 & Const & 22.7 & 1.6 & 29.0~~~ & Embd/Bk  \\
896 & 053525.3-052720 & 23 & Const & 22.0 & $>$10 & 28.9~~~ & Embd/Bk  \\
899 & 053525.4-052012 & 24 & Const & 23.2 & 1.2 & 29.0~~~ & Embd/Bk  \\
902 & 053525.5-052136 & 282 & Const & 21.6 & 2.2 & 29.7~~~ & ONC\\
915 & 053526.3-051950 & 19 & Const & 22.8 & 2.4 & 28.9~~~ & Embd/Bk  \\
935 & 053527.6-052038 & 14 & Const & 22.0 & 5.9 & 28.6~~~ & Embd/Bk  \\
966 & 053529.8-052859 & 64 & Const & 22.1 & $>$10 & 29.4~~~ & Embd/Bk  \\
1006 & 053532.4-052822 & 52 & LT Var & 21.9 & $>$10 & 29.2~~~ & ONC\\
1014 & 053533.3-051508 & 95 & Const & 21.9 & $>$10 & 29.9~~~ & ONC\\
1015 & 053533.4-052702 & 34 & Const & 22.3 & 5.0 & 29.1~~~ & Embd/Bk  \\
1016 & 053533.5-051651 & 40 & Const & 22.2 & $>$10 & 29.3~~~ & Embd/Bk  \\
1031 & 053536.5-051628 & 58 & LT Var & 22.1 & $>$10 & 28.8~~~ & Embd/Bk  \\
1043 & 053539.2-052856 & 45 & LT Var & 21.8 & $>$10 & 28.8~~~ & ONC\\
1045 & 053540.1-053016 & 102 & LT Var & 21.9 & $>$10 & 29.4~~~ & ONC\\
1050 & 053541.7-052015 & 40 & Const & 22.3 & $>$10 & 29.1~~~ & Embd/Bk  \\
1059 & 053543.7-052400 & 25 & Const & 21.8 & $>$10 & 28.7~~~ & ONC\\
\enddata

\tablenotetext{a}{Radio source Q, 2.5 mJy at 2 cm \citep{Felli93}}

\tablenotetext{b}{This X-ray source lies 1.5\arcsec\/ from IRc 14,
a luminous mid-infrared member of the OMC 1 cluster \citep{Gezari98}.}

\tablenotetext{c}{This X-ray source coincides with a SIMBAD listing for
Parenago 1867 (V=15.8).  However, examination of the original charts of
\citet{Parenago54} indicates that the SIMBAD position is probably
incorrect:  Parenago 1867 corresponds to HC 304 $\simeq 3$\arcsec to
the south.  The X-ray position is resolvable from other stars in the
region and has no counterpart with $V<20$ or $K<18$.}

\end{deluxetable}

\clearpage
\newpage

\begin{deluxetable}{rcccrcrrrrcrcrrc}

\rotate
\tabletypesize{\scriptsize}
\tablewidth{625pt}
\tablecolumns{16}

\tablecaption{High- and intermediate-mass ONC stars 
\label{highmass_table}}

\tablehead{
\colhead{Src} &
\colhead{CXOONC} &
\multicolumn{6}{c}{Optical properties} &&
\multicolumn{7}{c}{X-ray properties} \\ \cline{3-8} \cline{10-16}

\colhead{} &
\colhead{} &
\colhead{ID} &
\colhead{Name} &
\colhead{V} &
\colhead{Sp.Ty.} &
\colhead{$\log L_{bol}$} &
\colhead{$\log M$} &&
\colhead{$C_{xtr}$} &
\colhead{Var Cl} &
\colhead{$\log N_H$} &
\colhead{kT} &
\colhead{$\log L_s$} &
\colhead{$\log L_t$} &
\colhead{$\log L_t/L_{bol}$} \\

&
&
&
&
\colhead{(mag)} &
&
\colhead{(L$_\odot$)} &
\colhead{(M$_\odot$)} &&
&
&
\colhead{(cm$^{-2}$)} &
\colhead{(keV)} &
\multicolumn{2}{c}{(erg s$^{-1}$)} &
\\
}

\startdata
542     &053516.4-052322&P 1891&$\theta^1$C Ori&5.13   &O6 pe   &5.38 
&1.65&& 21596 &LT Var &\nodata&0.3/2    &33.2 &33.3 &-5.8 \\
828     &053522.8-052457&P 1993&$\theta^2$A Ori&5.08   &O9.5 Vpe&5.02 
&1.49&& 16525 &Flare  &\nodata&0.2/6    &31.2 &31.6 &-7.4 \\
498     &053515.8-052314&P 1865&$\theta^1$A Ori&6.73   &O7      &4.49 
&1.27&& 13676 &LT Var &\nodata&1.0/5    &31.0 &31.5 &-7.1 \\
584     &053517.2-052316&P 1889&$\theta^1$D Ori&6.71   &B0.5 Vp &4.33 
&1.21&&   724 &Pos fl &20.47  &0.6      &30.1 &30.1 &-7.8 \\
996     &053531.4-051602&P 2074&NU Ori         &6.87   &B1 V    &4.33 
&1.21&&    490 &Const  &21.54  &7.0      &28.7 &29.3 &-9.2 \\
916     &053526.4-052500&P 2031&$\theta^2$B Ori&5.02   &B1 V    &3.96 
&1.08&&   242 &Flare  &21.20  &1.4      &29.3 &29.5 &-8.2 \\
\nodata &\nodata        &P 1772&LP Ori         &8.43   &B1.5 Vp &3.26 
&0.85&&\nodata&\nodata&\nodata&\nodata  &$<$28.4 &$<$28.7 &$<$-8.5~~ \\
519     &053516.0-052307&P 1863a&BM Ori        &7.96   &B0.5    &3.23 
&0.84&&  1427 &Pos fl &21.48  &2.5      &30.0 &30.4 &-6.8 \\
746     &053520.7-052144&JW 660&V1230 Ori      &9.66   &B8 IV   &3.06 
&0.79&&  6020 &LT Var &21.64  &2.5      &30.7 &31.1 &-6.0 \\ 
728     &053520.2-052057&JW 640&TU Ori         &\nodata&G9      &2.26 
&0.70&&  1895 &Pos fl &21.80  &2.2      &30.2 &30.6 &-5.7 \\ 
995     &053531.4-052516&P 2085&$\theta^2$C Ori&8.24   &B4 V    &2.86 
&0.70&&  3261 &Flare  &\nodata&0.1/2    &30.5 &30.7 &-5.9 \\ 
495     &053515.7-052309&P 1864&\nodata        &11.10  &\nodata &2.48 
&0.63&&  14968 &Const  &\nodata&0.9/$>$10&30.9 &31.7 &-5.1 \\ 
164     &053505.3-051449&JW 260&V1230 Ori      &11.07  &G5      &1.83 
&0.61&&  1833 &Const  &21.44  &2.1      &30.5 &30.8 &-4.9 \\ 
1046    &053540.3-051728&JW 945&\nodata        &14.60  &B6      &2.35 
&0.61&&   484 &Pos fl &21.94  &2.5      &29.5 &30.1 &-6.4 \\
\nodata &        \nodata&JW 108&\nodata        &10.27  &A2 Vp   &1.98 
&0.55&&\nodata&\nodata&\nodata&\nodata  &$<$28.6 &$<$28.9 &$<$-7.0~~ \\
670     &053518.8-051728&JW 599&\nodata        &\nodata&A9      &1.84 
&0.54&&    23 &Const  &20.96  &3.6      &\nodata &28.6 &-7.3 \\ 
291     &053511.6-051657&JW 364&LT Ori         &\nodata&K0      &1.37 
&0.50&&  2883 &Pos fl &21.73  &1.8      &30.4 &30.8 &-4.5 \\
508     &053515.9-052349&JW 499&$\theta^1$E Ori&13.79  &K0      &1.30 
&0.50&&  3965 &Flare  &21.70  &6.6      &30.6 &31.1 &-4.3 \\ 
1004    &053532.3-053111&JW 887&\nodata        &11.77  &\nodata &1.38 
&0.50&&  1319 &LT Var &\nodata&0.2/1    &30.2 &30.3 &-4.8 \\ 
663     &053518.6-052033&JW 595&MV Ori         &\nodata&\nodata &1.26 
&0.46&&  3504 &LT Var &21.73  &2.6      &30.5 &30.9 &-4.4 \\
70      &053455.2-053022&JW 153&\nodata        &9.01   &B9      &1.86 
&0.46&&    93 &Pos fl &20.29  &1.0      &29.0 &29.1 &-6.4 \\
651     &053518.3-052237&JW 589&V1229 Ori      &13.38  &M0      &1.25 
&0.45&&  5352 &Pos fl &\nodata&0.9/4    &30.8 &31.2 &-4.0 \\
914     &053526.3-052540&JW 799&AK Ori         &\nodata&G5      &1.16 
&0.44&&  4161 &Flare  &\nodata&0.8/3    &30.5 &31.0 &-4.2 \\
484     &053515.6-052256&JW 479&V348 Ori       &\nodata&K0      &1.18 
&0.43&&  5954 &LT Var &21.50  &2.5      &30.7 &31.1 &-4.1 \\ 
\nodata &        \nodata&P 1892&\nodata        &11.50  &B8      &1.85 
&0.45&&\nodata&\nodata&\nodata&\nodata  &$<$28.9 &$<$29.2 &$<$-6.5~~ \\
103     &053500.0-052515&JW 197&KS Ori         &10.19  &A0      &1.66 
&0.40&&   665 &LT Var &20.74  &1.6      &29.8 &30.0 &-5.4 \\
910     &053526.1-052737&JW 795&V1232 Ori      &11.59  &K0      &1.03 
&0.40&&  7891 &Flare  &21.98  &0.2/2    &30.9 &31.1 &-3.7 \\
347     &053513.1-052455&JW 401&\nodata        &\nodata&K1      &1.06 
&0.40&&    94 &Const  &21.80  &2.6      &28.8 &29.3 &-5.9 \\ 
\nodata &        \nodata&JW 531&MR Ori         &10.30  &A2 Vp   &1.68 
&0.42&&\nodata&\nodata&\nodata&\nodata  &$<$28.2 &$<$28.5 &$<$-7.1~~ \\ 
1051    &053541.9-052813&JW 959&AN Ori         &\nodata&K11 Ve  &1.05 
&0.39&&  8141 &Pos fl &\nodata&0.8/3    &31.0 &31.2 &-3.7 \\ 
81      &053456.4-053136&JW 165&KO Ori         &\nodata&\nodata &1.47 
&0.38&&  1256 &Const  &20.91  &1.7      &30.5 &30.7 &-4.6 \\
461     &053515.2-052256&JW 468&\nodata        &13.22  &G7      &0.97 
&0.38&&  5401 &Pos fl &21.39  &2.1      &30.7 &31.0 &-3.9 \\ 
952     &053528.4-052621&JW 831&V1073 Ori      &9.52   &B9.5 V  &1.52 
&0.38&&   125 &Const  &$<$20.0~&1.0     &29.1 &29.2 &-6.0 \\
261     &053510.4-052618&JW 348&LR Ori         &11.90  &\nodata &0.94 
&0.36&&  2499 &Pos fl &21.40  &1.8      &30.4 &30.7 &-4.1 \\
760     &053521.0-052349&JW 669&V1399 Ori      &12.30  &\nodata &1.15 
&0.36&&  7476 &Pos fl &\nodata&0.8/3    &30.8 &31.2 &-3.9 \\
722     &053520.1-052639&JW 641&V1338 Ori      &\nodata&\nodata &0.93 
&0.36&&  4954 &Flare  &21.20  &2.8      &30.6 &31.0 &-3.9 \\ 
104     &053500.1-052301&JW 193&KR Ori         &\nodata&K0 e    &0.89 
&0.35&&   844 &Const  &20.14  &2.0      &29.8 &30.1 &-4.6 \\
173     &053505.6-052519&JW 273&LL Ori         &10.70  &K0 e    &1.15 
&0.32&&  2449 &Pos fl &\nodata&0.7/3    &30.4 &30.6 &-4.4 \\
320     &053512.5-052343&JW 385&LV Ori         &12.10  &\nodata &1.01 
&0.32&&   413 &Pos fl &\nodata&0.6/2    &28.7 &28.5 &-5.9 \\
993     &053531.4-051533&JW 866&V1294 Ori      &\nodata&K1 IV   &1.01 
&0.32&&  1611 &Pos fl &21.44  &2.0      &30.4 &30.7 &-4.2 \\ 
576     &053517.0-052334&JW 538&\nodata        &\nodata&K1      &0.99 
&0.32&&  1830 &Pos fl &22.23  &3.7      &30.0 &31.0 &-4.5 \\
513     &053516.0-052353&JW 503&AC Ori         &12.50  &\nodata &1.14 
&0.30&&   302 &LT Var &22.41  &4.1      &29.0 &30.1 &-5.7 \\
573     &053517.0-052233&JW 536&V1333 Ori      &14.60  &K1      &0.78 
&0.28&&  3555 &Pos fl &21.45  &2.1      &30.5 &30.8 &-3.9 \\ 
6       &053439.8-052642&JW 46&\nodata         &12.57  &K3e var &0.75 
&0.27&&  1674 &LT Var &21.11  &1.3      &30.5 &30.6 &-3.8 \\
14      &053443.3-051828&JW 64&\nodata         &11.13  &F2 IV   &1.00 
&0.27&&   608 &Const  &21.32  &2.8      &30.1 &30.5 &-4.5 \\
853     &053523.7-053048&JW 747&V358 Ori       &12.20  &G8 V    &0.72 
&0.26&&  5651 &Pos fl &\nodata&0.8/2    &30.8 &31.0 &-3.5 \\
\nodata &\nodata        &JW 608&\nodata        &11.89  &A5      &0.98 
&0.25&&\nodata&\nodata&\nodata&\nodata  &$<$28.2 &$<$28.5 &$<$-6.4~~ \\ 
131     &053502.4-051547&JW 221&V403 Ori       &\nodata&\nodata &1.44 
&0.24&& 10505 &Flare  &\nodata&0.9/3    &30.9 &31.4 &-4.1 \\
17      &053445.1-052503&JW 75&\nodata         &\nodata&K2      &0.92 
&0.22&&  9461 &Flare  &\nodata&0.9/3    &31.0 &31.3 &-3.5 \\
133     &053502.9-053001&JW 232&KZ Ori         &\nodata&\nodata &0.84 
&0.20&&  2544 &Const  &21.31  &1.6      &30.4 &30.6 &-4.0 \\
397     &053514.0-051951&JW 429&\nodata        &\nodata&G       &0.57 
&0.20&&   485 &LT Var &$<$20.0~&1.5     &29.7 &29.8 &-4.4 \\
77      &053455.9-052312&JW 157&KM Ori         &\nodata&K1      &1.22 
&0.18&& 16231 &Flare  &\nodata&0.8/3    &31.2 &31.5 &-3.6 \\
769     &053521.2-052457&JW 678&V377 Ori       &12.80  &\nodata &0.58 
&0.17&&  2580 &Flare  &21.09  &2.1      &30.4 &30.6 &-3.8 \\

\enddata
\end{deluxetable}

\clearpage
\newpage

\begin{deluxetable}{rrcrrrrrrcrccc}

\rotate
\tabletypesize{\scriptsize}
\tablewidth{0pt}
\tablecolumns{14}

\tablecaption{X-ray detections of very low mass ONC objects 
\label{bd_table}}

\tablehead{
\colhead{Src} &
\colhead{CXOONC} &
\multicolumn{5}{c}{Optical-IR properties} &&
\multicolumn{6}{c}{X-ray properties} \\ \cline{3-7} \cline{9-14}

\colhead{} &
\colhead{} &
\colhead{ID} &
\colhead{K} &
\colhead{$H-K$} &
\colhead{$M_K$} &
\colhead{Mass} &&
\colhead{$C_{xtr}$} &
\colhead{Var Cl} &
\colhead{$\log N_H$} &
\colhead{kT} &
\colhead{$\log L_t$} &
\colhead{$\log L_t/L_{bol}$} \\

&
&
&
\colhead{(mag)} &
\colhead{(mag)} &
\colhead{(mag)} &
\colhead{(M$_\odot$)} &&
&
&
\colhead{(cm$^{-2}$)} &
\colhead{(keV)} &
\colhead{(erg s$^{-1}$)} &
}

\startdata

34~  & 053450.1-051959 & JW 110 & 13.82 & 0.40 & 5.39 & 0.07 &  & 25 & 
Const & \nodata & 0.3/$>$10 & 28.4 & -3.7\\

55~  & 053453.1-052400 & CHS 6221 & 13.76 & 0.43 & 5.28 & 0.07 &  & 41 & 
Const & 20.6 & 2.3 & 28.7 & -3.4 \\

56~  & 053453.3-052627 & CHS 6210 & 13.70 & 0.44 & 5.20 & 0.07 &  & 30 & 
Const & \nodata & 0.2/5 & 29.0 & -3.2 \\

65~  & 053454.5-052300 & CHS 6351 & 13.86 & 0.42 & 5.40 & 0.07 &  & 30 & 
Const & $<$20.0 & $>$10 & 28.8 & -3.3 \\

74~  & 053455.9-053113 & JW 159 & 13.24 & 0.14 & 4.88 & 0.09 &  & 57 & 
Const & $<$20.0 & 1.4 & 29.2 & -3.1 \\

112~ & 053501.2-052144 & JW 205 & 11.68 & 0.75 & 2.42 & 
0.65\tablenotemark{a} &  & 26 & Const & $<$20.0 & 3.6 & 28.6 & -4.7 \\

142~ & 053503.9-052809 & CHS 7273 & 14.05 & 0.53 & 5.43 & 0.07 &  & 91 & 
Pos fl & 22.4 & 2.7 & 29.5 & -2.6 \\

169~ & 053505.4-052230 & H 5096 & 13.31 & 0.81 & 4.20 & 
0.13\tablenotemark{b} &  & 15 & Const & 20.9 & $>$10 & 28.7 & -3.9 \\

196* & 053506.6-052243 & HC 741 & 16.26 & \nodata & $<$7.90& $>$0.02 &  & 
23 & Pos fl & 22.6 & 2.4 & 28.9 & $<$-2.2~~~\\

197* & 053506.8-052209 & PSH 298 & 13.66 & 0.45 & 5.15 & 0.08 &  & 22 & 
Pos fl & 20.8 & 0.4 & 28.5 & -3.7 \\

198* & 053506.9-052501 & HC 64 & 14.52 & 0.73 & 5.63 & 0.06 &  & 8 & Const 
& 22.0 & 3.0 & 28.3 & -3.7 \\

199~ & 053507.1-051828 & H 5064 & 13.44 & 0.31 & 5.08 & 
0.08\tablenotemark{c} &  & 17 & Const & $<$20.0 & 6.1 & 28.3 & -3.9 \\

241* & 053509.7-052152 & HC 748 & 16.22 & 1.96 & 5.61 & 0.06 &  & 10 & 
Const & 22.3 & 2.6 & 28.5 & -3.5 \\

309* & 053512.1-052447 & HC 95 & 15.92 & \nodata & $<$7.56 & $>$0.03 &  & 
16 & Const & 22.8 & 1.7 & 28.9 & $<$-2.3~~ \\

339~ & 053513.0-051547 & CHS 8315 & 13.77 & 0.48 & 5.22 & 0.07 &  & 52 & 
LT Var & $<$20.0 & 5.2 & 28.8 & -3.4 \\

413~ & 053514.4-052903 & JW 446 & 13.56 & -0.04 & 5.20 & 0.07 &  & 42 & LT 
Var & $<$20.0 & 0.8 & 28.5 & -3.7 \\

503* & 053515.8-052431 & PSH 116 & 12.55 & 0.87 & 3.21 & 
0.35\tablenotemark{d} &  & 13 & Pos fl & 23.3 & 1.5 & 28.8 & -4.2 \\

561* & 053516.8-052307 & PSH 153 & 13.19 & 0.12 & 4.83 & 0.08 &  & 37 & 
Const & $<$20.0 & 9.2 & 28.8 & -3.5 \\

643* & 053518.2-052346 & PSH 215 & 14.46 & 0.71 & 5.61 & 0.06 &  & 19 & 
Const & $<$20.0 & $>$10 & 28.5 & -3.5 \\

755~ & 053520.9-053005 & CHS 9480 & 14.45 & 0.57 & 5.77 & 0.06 &  & 116 & 
Flare & 21.2 & 2.7 & 29.2 & -2.7 \\

774~ & 053521.4-051710 & CHS 9558 & 13.99 & \nodata & $<$5.63 & $>$0.06 &  
& 143 & LT Var & 22.8 & 1.6 & 29.9 & $<$-2.1~~~\\

806~ & 053522.1-051857 & CHS 9657 & 13.96 & \nodata & $<$5.60 & $>$0.06 &  
& 32 & Const & 21.7 & $>$10 & 28.9 & $<$-3.1~~ \\

838~ & 053523.4-052038 & CHS 9819 & 14.57 & \nodata & $<$6.21 & $>$0.04 && 
11 & Const & 22.7 & 1.6 & 28.6 & $<$-3.2~~ \\

852~ & 053523.7-052804 & H 5131 & 13.75 & \nodata & $<$5.39 & $>$0.07 &  & 
20 & Const & \nodata & 0.2/$>$10 & 28.6 & $<$-3.5~~ \\

869~ & 053524.3-052647 & CHS 9924 & 13.83 & 0.45 & 5.33 & 0.07 &  & 17 & 
Const & $<$20.0 & 1.2 & 28.2 & -3.9 \\

873* & 053524.4-052440 & HC 756 & 14.29 & 0.77 & 5.33 & 0.07 &  & 790 & 
Pos fl & 22.1 & 3.3 & 30.4 & -2.7 \\

886* & 053525.0-052438 & HC 114 & 13.94 & 0.49 & 5.37 & 0.07 &  & 16 & 
Const & $<$20.0 & 1.1 & 28.2 & -3.9 \\

936~ & 053527.6-053109 & CHS  10299 & 14.61 & 0.39 & 6.18 & 0.05 &  & 55 & 
LT Var & 21.9 & $>$10 & 29.1 & -2.7\\

1047~ & 053540.8-052707 & CHS 11663 & 14.97 & 0.32 & $<$6.61 & $>$0.03 &  
& 56 & Pos fl & \nodata & 0.2/$>$10 & 28.9 & $<$-2.7~~ \\

1052~ & 053542.1-052005 & JW 958 & 13.41 & 0.24 & 5.05 & 0.08 &  & 66 & 
Const & \nodata & 0.2/$>$10 & 29.0 & -3.2 \\

\enddata

\tablenotetext{a}{Spectroscopic mass estimate of 0.04 M$_\odot$}

\tablenotetext{b}{Spectroscopic mass estimate of 0.05 M$_\odot$}

\tablenotetext{c}{Spectroscopic mass estimate of 0.07 M$_\odot$}

\tablenotetext{d}{Spectroscopic mass estimate of 0.04 M$_\odot$}

\end{deluxetable}

\clearpage
\newpage

\end{document}